%% file: main.tex
\documentclass[preprintnumbers,amsmath,amssymb,prd, 
notitlepage,nofootinbib,twocolumn,superscriptaddress]{revtex4-2}

\input{preamble}

\begin{document}

\title{Subtracting the kinetic Sunyaev-Zeldovich effect from the cosmic
microwave background with surveys of large-scale structure} 

\newcommand{\perimeter}{Perimeter Institute for Theoretical Physics, 
31 Caroline Street North, Waterloo, ON N2L 2Y5, Canada}
\newcommand{\penn}{Department of Physics and Astronomy, University of Pennsylvania, 209 South 33rd Street, Philadelphia, PA 19104, USA}
\newcommand{\princeton}{Department of Astrophysical Sciences, Princeton University, 4 Ivy Lane, Princeton, NJ 08544, USA}
\newcommand{\drao}{Dominion Radio Astrophysical Observatory, 
Herzberg Astronomy \& Astrophysics Research Centre, \\
National Research Council Canada, P.O.~Box 248, Penticton, BC V2A 6J9, Canada}
\newcommand{\jhu}{William H.\ Miller III Department of Physics and Astronomy, 
Johns Hopkins University, Baltimore, MD 21218, USA}
\newcommand{\asu}{School of Earth and Space Exploration, 
Arizona State University, Tempe, AZ 85287, USA}
\newcommand{\mitkavli}{MIT Kavli Institute for Astrophysics and Space Research, \\
Massachusetts Institute of Technology, 77 Massachusetts Ave, Cambridge, MA 02139, USA}

\author{Simon Foreman}
\affiliation{\perimeter}
\affiliation{\drao}
\affiliation{\mitkavli}
\author{Selim~C.~Hotinli}
\affiliation{\jhu}
\author{Mathew S. Madhavacheril}
\affiliation{\perimeter}
\affiliation{\penn}
\author{Alexander van Engelen}
\affiliation{\asu}
\author{Christina D. Kreisch}
\affiliation{\princeton}

\date{\today}

\begin{abstract}
The kinetic Sunyaev-Zeldovich (kSZ) effect will be an important source of cosmological and astrophysical information in upcoming surveys of the cosmic microwave background (CMB). However, the kSZ effect will also act as the dominant source of noise for several other measurements that use small angular scales in CMB temperature maps, since its blackbody nature implies that standard component separation techniques cannot be used to remove it from observed maps. In this paper, we explore the idea of ``de-kSZing": constructing a template for the late-time kSZ effect using external surveys of large-scale structure, and then subtracting this template from CMB temperature maps in order to remove some portion of the kSZ signal. After building intuition for general aspects of the de-kSZing procedure, we perform forecasts for the de-kSZing efficiency of several large-scale structure surveys, including BOSS, DESI, Roman, MegaMapper, and PUMA. We also highlight potential applications of de-kSZing to cosmological constraints from the CMB temperature power spectrum, CMB lensing reconstruction, and the moving-lens effect. While our forecasts predict achievable de-kSZing efficiencies of 10-20\% at best, these results are specific to the de-kSZing formalism adopted in this work, and we expect that higher efficiencies are possible using improved versions of this formalism.
\end{abstract}

\maketitle

%--------------------------------------------------------------------------------------
% Introduction
%--------------------------------------------------------------------------------------

\section{Introduction}

The ``primary" fluctuations of the cosmic microwave background (CMB), which encode the properties of the universe during the recombination era, have played a key role in the establishment of our current cosmological model. ``Secondary" fluctuations, created by processes occurring after recombination, are now attracting significant theoretical and observational interest, as they can in principle be used to probe numerous aspects of astrophysics and fundamental physics.

In particular, much recent interest has been focused on the various incarnations of the Sunyaev-Zeldovich (SZ) effect (\cite{Zeldovich:1969ff,Sunyaev:1972eq,Sunyaev:1980vz,Sazonov:1999zp}; see Ref.~\cite{Mroczkowski:2018nrv} for a recent review), by which CMB photons scatter off of free electrons associated with massive dark matter halos. 
The thermal SZ (tSZ) effect, generated by thermal motions of electrons within a halo, distorts the CMB blackbody spectrum and probes the pressure of those electrons. The kinetic SZ (kSZ) effect, generated by bulk motion of the electrons, leads to additional small-scale CMB anisotropies and is sensitive to a combination of large-scale velocity flows and the spatial distribution of electrons. The polarized SZ (pSZ) effect, generated by a quadrupole in the radiation field observed by the scatterer (and therefore sensitive to the various effects that can contribute to that quadrupole), induces linear polarization in the scattered photons. Higher-order corrections to these effects also have distinctive signatures and a variety of applications (e.g.~\cite{Itoh:1997ks,Coulton:2019ign,Hotinli:2022wbk}). 

These effects are interesting signals in many contexts, but in other contexts, they represent sources of bias and/or noise that degrade our ability to make useful measurements. A significant portion of the tSZ effect can be removed from observed CMB maps thanks to its non-blackbody spectral shape, but the (blackbody) kSZ effect cannot be removed in this way, and is expected to dominate the observed temperature fluctuations at $\ell \gtrsim 4000$ after non-blackbody signals have been filtered out. For sufficiently high-resolution observations, reconstruction of gravitational lensing of the CMB will make use of temperature fluctuations at these scales, and the kSZ effect can significantly bias this reconstruction~\cite{Ferraro:2017fac,Cai:2021hnb}, affecting its ability to constrain neutrino mass, dark energy, and dark matter (e.g.~\cite{Nguyen:2017zqu,Han:2021vtm}). Furthermore, kSZ fluctuations at small scales add noise to measurements of the Silk-damped regime of the primary CMB, which can be used to probe the effective number of free-streaming species $N_{\rm eff}$, the primordial helium abundance $Y_{\rm p}$, and the spectrum of primordial scalar perturbations. The kSZ effect also acts as noise for measurements of other CMB secondaries, such as the moving-lens effect~\cite{Hotinli:2018yyc,Hotinli:2020ntd,Hotinli:2021hih}.

This situation is analogous to gravitational lensing of the CMB, which is both a physically interesting signal and a source of bias and noise for other measurements (including primordial gravitational waves, acoustic oscillations in the primary CMB, and lensing itself). In the lensing case, methods to ``de-lens" CMB temperature and polarization maps have been extensively investigated~\cite{2012JCAP...06..014S,Green:2016cjr,Sehgal:2016eag,Manzotti:2017oby,Hotinli:2021umk}, applied to recent CMB observations~\cite{Larsen:2016wpa,Carron:2017vfg,Planck:2018lbu,SPTpol:2020rqg,ACT:2020goa},  and  integrated into the core analysis pipelines that are being developed for upcoming surveys~\cite{Namikawa:2021gyh}. ``Internal" de-lensing uses lensing maps reconstructed from a CMB survey to remove lensing effects from temperature and polarization maps from the same survey,
 while ``external" de-lensing uses an external dataset to estimate a lensing map that is used to de-lens the observed CMB.

In this work, we investigate the prospects for external ``de-kSZing" of CMB temperature maps, whereby an external galaxy survey is used to construct a template for the kSZ effect, which is then subtracted from the observed maps. Specifically, we consider the portion of the kSZ effect sourced by the post-reionization universe (the ``late-time" kSZ effect), since this is expected to either dominate over the kSZ signal from reionization or be comparable to it, depending on the models that are assumed~\cite{Cai:2021hnb}.
Following Refs.~\cite{Ho:2009iw,Shao:2010md}\footnote{See also Ref.~\cite{Munshi:2015anr} for a fully 3-dimensional formalism for kSZ templates.}, we construct a template by using a galaxy survey to reconstruct the large-scale velocity field, and multiplying by a linear reconstruction of the electron density field; this forms a template for the electron momentum field, which can then be projected along the line of sight to yield an estimate of the kSZ-induced CMB temperature fluctuations. While Refs.~\cite{Ho:2009iw,Shao:2010md} explored the prospect of cross-correlating this template with the CMB in order to measure the kSZ effect for probing baryons in the warm-hot intergalactic medium, our motivation here is to use the template to remove some portion of the kSZ effect from CMB observations.

The bulk of our paper is concerned with assessing the performance of kSZ templates constructed from galaxy surveys with different properties, and performing an initial exploration of several possible applications of de-kSZing with external data: namely, constraining cosmological parameters using the CMB damping tail, improving the reconstruction of gravitational lensing, and measuring halo profiles using the moving-lens effect.
While our overall conclusion is that it will be challenging to obtain significant benefits from de-kSZing within the formalism adopted in this work, it's likely that modifications to this formalism could yield substantial improvements, and we discuss this further in the final section of the paper.

Note that we will use the term ``kSZ effect" to refer only to the {\em late-time} kSZ effect in the body of the paper; we will briefly discuss the implications of de-kSZing the late-time kSZ effect on the measurement of the kSZ signal from reionization at the end of Sec.~\ref{sec:dampingtail}.

This paper is organized as follows. 
In Sec.~\ref{sec:theory}, we review the main theoretical expressions used to describe the kSZ effect, as well as our formalism for constructing kSZ templates from galaxy surveys and how we model the relevant quantities. 
In Sec.~\ref{sec:intuition}, we build intuition for the ability of different kSZ templates to capture a given fraction of the true kSZ signal, by enumerating the relevant redshifts and halo masses (\ref{sec:intuition-ranges}),  the separate impacts of the velocity and electron density templates (\ref{sec:templates-ev}), and the influence of redshift uncertainties in the input galaxy survey (\ref{sec:redshifts}).
In Sec.~\ref{sec:forecasts}, we forecast the usefulness
of specific galaxy surveys 
for de-kSZing, including BOSS, DESI, the Roman Space Telescope, the MegaMapper proposal, and the PUMA \tcm intensity mapping proposal. 
In Sec.~\ref{sec:applications}, we discuss the potential of de-kSZing to improve measurements of cosmological parameters, CMB lensing, and the moving lens effect.
Finally, we discuss plausible improved versions of de-kSZing and conclude in Sec.~\ref{sec:conclusion}. 

The appendices discuss the impact of shot noise in the galaxy survey used to construct the template (Appendix~\ref{app:shotnoise}), the details of our halo model approach to modelling the signal and templates (Appendix~\ref{app:halomodel}), the halo occupation distribution models we assume in our computations (Appendix~\ref{app:hod}), our approach to modelling \tcm intensity mapping surveys (Appendix~\ref{app:21cm}), and further details of our CMB forecasts (Appendices~\ref{sec:CMB_forecasts}-\ref{app:movinglens}).

%--------------------------------------------------------------------------------------
% Theory
%--------------------------------------------------------------------------------------
\section{Theory}
\label{sec:theory}

%----------------------
% Review of the kSZ effect
%----------------------
\subsection{Review of the kSZ effect}

The kSZ effect adds the following contribution to the observed CMB temperature at sky location $\hvn$ (e.g.~\cite{Ma:2001xr}):
\beq
\TkSZ(\hvn) = \int_0^{\chi_*} d\chi\, \tilde{K}(z[\chi]) \, \qr(\hvn,\chi;z[\chi])\ .
\label{eq:TkSZ}
\eeq
In this expression, $\chi$ is comoving distance, $\chi_*$ corresponds to the earliest epoch we wish to include in our calculations of the kSZ effect, $\qr$ is the radial (i.e.\ line-of-sight) component of the momentum of the free electrons at location $(\hvn,\chi)$ and redshift $z[\chi]$, and $\tilde{K}(z)$ is a radial weight function that captures the scattering of CMB photons by these electrons. In this work, we are only concerned with the post-reionization kSZ effect, so we take $\chi_*$ to be the comoving distance to the end of reionization, at $z\approx 6$. 

The electron momentum is well approximated by
\beq
\qr(\vx; z) \approx \velr(\vx; z) \deltae(\vx; z)\ ,
\eeq
where $\velr$ is the radial component of the velocity field and $\deltae$ is the electron density contrast. At the large scales where the velocity field has the majority of its power, it can be treated as curl-free, such that
\beq
\velr(\vk; z) = i\mu v(\vk; z)\ ,
\eeq
where $\mu$ is the cosine of the angle between $\vk$ and the line of sight.
Furthermore, at large scales, $v$ is linearly related to the matter density contrast $\deltam$ via
\beq
\deltam(\vk) = \frac{k}{faH} v(\vk)\ ,
\eeq
omitting the redshift arguments for brevity.
Note that the electron momentum is given in full by $\qr = \velr + \velr \deltae$, but the $\velr$ term is far subdominant on nonlinear scales (where $\deltae \gg 1$)~\cite{Ma:2001xr}. The radial weight function is
\beq
\tilde{K}(z) \equiv - T_{\rm CMB}\, \bar{n}_{\rm e,0}\, \sigma_{\rm T}\, (1+z)^2 e^{-\tau(z)}\ ,
\label{eq:kSZweight}
\eeq
where $T_{\rm CMB}$ is the mean CMB temperature, $\bar{n}_{\rm e,0}$ is the mean electron number density at redshift zero, $\sigma_{\rm T}$ is the cross section for Thomson scattering, and $\tau(z)$ is the spatially-averaged optical depth to redshift $z$. We take $\tau(z)\approx 0$ after reionization.

In the Limber approximation~\cite{Limber:1954zz,LoVerde:2008re}, the angular power spectrum of $T_{\rm kSZ}$ evaluates to
\beq
C_\ell^{\rm kSZ} = \int_0^{\chi_*} \frac{d\chi}{\chi^2} \tilde{K}(z[\chi])^2 
	\, P_{\qr \qr} \! \lp \frac{\ell+1/2}{\chi}; z[\chi] \rp\ ,
	\label{eq:clksz}
\eeq
where $P_{\qr \qr}$ is the 3d power spectrum of $\qr$, given by
\begin{align}
\label{eq:Pqrqr-vec}
P_{\qr \qr}(\kS; z)
	&\approx \int \frac{d^3\vkL}{(2\pi)^3} P_{\velr \velr}(\kL; z) \Pee(\kS; z) \\
&=\frac{1}{6\pi^2}
	\lp \int d\kL\, \kL^2 \Pvv(\kL; z) \rp \Pee(\kS; z)\ .
	\label{eq:Pqrqr}
\end{align}
In writing Eq.~\eqref{eq:Pqrqr-vec}, we have neglected the connected four-point function $\langle \velr \deltae \velr \deltae \rangle$, and only retained the dominant ``squeezed" contribution to the power spectrum, arising from the large-scale ($\kL$) velocity power and small-scale ($\kS$) electron density power. The accuracy of these approximations has been argued both analytically~\cite{Ma:2001xr,Smith:2018bpn}  and using simulations~\cite{Ma:2001xr,Shaw:2011sy,Giri:2020pkk}.\footnote{Often in the literature, the squeezed limit ($\kL \ll \kS$) is not taken when writing Eq.~\eqref{eq:Pqrqr-vec}, but $\kL$ is still assumed to be in the linear regime, in which case Eq.~\eqref{eq:Pqrqr-vec} becomes
\begin{align*}
P_{\qr \qr}(\kS) &\approx (faH)^2 \int \frac{d^3\vkL}{(2\pi)^3} P_{\rm mm}(\kL) \Pee(|\vkS-\vkL|) \\
&\qquad\qquad\quad \times \frac{\kS (\kS-2\kL \muL) (1-\muL^2)}{\kL^2 (\kS^2+\kL^2-2\kS\kL\muL)}\ .
	\numberthis
	\label{eq:Pqrqr-unsqueezed}
\end{align*}
If Eq.~\eqref{eq:Pqrqr-unsqueezed} is used when computing $C_\ell^{\rm kSZ}$ instead of Eq.~\eqref{eq:Pqrqr-vec}, the results differ by 30\% at $\ell=1000$ and by less than 5\% at $\ell>4000$. We use Eq.~\eqref{eq:Pqrqr-vec} in our computations, to enable us to compare the results with the templates described in Sec.~\ref{sec:templates}, which also assume the squeezed limit.}

%----------------------
% Constructing templates from galaxy surveys
%----------------------
\subsection{Constructing templates from galaxy surveys}
\label{sec:templates}

Given a galaxy density contrast $\deltags$ observed (in redshift space, hence the superscript ``$s$") by a galaxy survey, one can construct a template for the expected kSZ contribution to the observed CMB temperature.\footnote{Similar formalisms for kSZ templates were previously presented in Refs.~\cite{Ho:2009iw,Shao:2010md,Smith:2018bpn}. Ref.~\cite{Ho:2009iw} computed $\Pge$ and $\Pgg$ using a linear bias model and Ref.~\cite{Shao:2010md} effectively assumed that $\Pge=1$, while we use a more detailed halo model approach based on Ref.~\cite{Smith:2018bpn}, described in Sec.~\ref{sec:modelling} and Appendix~\ref{app:halomodel}. Also, in contrast with Refs.~\cite{Ho:2009iw,Smith:2018bpn}, we account for linear redshift-space distortions in the galaxy density when constructing our electron-density and velocity templates.}
First, one forms the following estimates for $\velr$ and $\deltae$ from~$\deltags$:
\begin{align}
\label{eq:eta-def}
\eta(\vk;z) &\equiv i\mu \frac{\Pgv(k, \mu;z)}{\Pggtot(k, \mu;z)} \deltags(\vk;z)\ , \\
\label{eq:eps-def}
\epsilon(\vk;z) &\equiv \frac{\Pge(k, \mu;z)}{\Pggtot(k, \mu;z)} \deltags(\vk;z)\ .
\end{align}
The redshift-space galaxy power spectrum $\Pggtot(k, \mu;z)$ is defined to include the effect of shot noise (assumed to be Poissonian in this work), 
\beq
\label{eq:Pggtot}
\Pggtot(k, \mu;z) \equiv \Pgg(k, \mu;z) + \frac{1}{\bar{n}_{\rm g}(z)}\ ,
\eeq
such that $\eta$ and $\epsilon$ are Wiener-filtered quantities that downweight noise-dominated modes of $\deltags$. 
We will discuss our approach to modelling the power spectra needed for Eqs.~\eqref{eq:eta-def}-\eqref{eq:eps-def} in Sec.~\ref{sec:modelling}.

One then forms an estimate for the line-of-sight electron momentum field,
\beq
\hqr(\vx,z) = \eta(\vx,z) \epsilon(\vx,z)\ ,
\label{eq:qhatr}
\eeq
and computes a line-of-sight projection using the kSZ radial weight function, analogous with Eq.~\eqref{eq:TkSZ}:
\beq
\hTkSZ(\hvn) = \int_0^{\chi_*} d\chi\, \tilde{K}(z[\chi]) \, \hqr(\hvn,\chi;z[\chi])\ .
\eeq

The cross-correlation between the template $\hTkSZ$ and the true signal $\TkSZ$ will depend on the 3d cross power spectrum between $\hqr$ and $\qr$. Under the same assumptions applied to Eq.~\eqref{eq:Pqrqr-vec}, this evaluates to
\begin{align}
\nonumber
&P_{\hqr \qr}(\kS, \muS; z) \\
\label{eq:Phqrqr1}
&\qquad\approx \int \frac{d^3\vkL}{(2\pi)^3} P_{\velr \eta}(\kL, \muL; z) \Peeps(\kS, \muS; z) \\
\nonumber
&\qquad=\frac{1}{4\pi^2}
	\lp \int d\kL\, \kL^2 \int_{-1}^1 d\muL\, \muL^2 \frac{\Pgv(\kL, \muL; z)^2}{\Pggtot(\kL, \muL; z)} \rp \\
&\qquad\qquad\times
	\frac{\Pge(\kS, \muS; z)^2}{\Pggtot(\kS, \muS; z)}\ .
	\label{eq:Phqrqr}
\end{align}
Redshift-space distortions in the observed galaxy density contrast~$\deltags$ imply that $P_{\velr \eta}$ and $\Peeps$ both depend on the angle with respect to the line of sight, through $\muL$ and $\muS$ respectively. On the large scales that dominate the velocity contribution, redshift-distortions are approximately described by the Kaiser factor $\deltags(k, \mu) \propto (\bg + f\mu^2)$~\cite{Kaiser:1987qv}. On the other hand, at the scales relevant for the electron density contribution, there will be ``Finger of God" damping of large-$\muS$ power caused by velocity dispersions on small scales~\cite{Jackson:1971sky}. However, the angular power spectrum of the template is a line-of-sight projection weighted by the kSZ weight function from Eq.~\eqref{eq:kSZweight}, and the width and smoothness of this weight function will largely suppress the influence of large-$\muS$ modes on the final result (e.g.~\cite{Jalilvand:2019brk,GrasshornGebhardt:2020wsw}). This, it will not be necessary to include the Finger of God effect in our computations. We may then write the angular power spectrum derived from the $\hqr$-$\qr$ cross-correlation as
\beq
C_\ell^{\hqr \qr} = \int_0^{\chi_*} \frac{d\chi}{\chi^2} \tilde{K}(z[\chi])^2 
	\, P_{\hqr \qr} \! \lp \frac{\ell+1/2}{\chi}, 0; \, z[\chi] \rp\ ,
	\label{eq:clhqrqr}
\eeq
following the Limber approximation in setting $\muS=0$~\cite{LoVerde:2008re}.

A short calculation shows that the auto spectrum of the template $\hTkSZ$ is also given by Eqs.~\eqref{eq:Phqrqr}-\eqref{eq:clhqrqr}. However, in this case we must revisit the approximation that the connected four-point function can be neglected: the power spectrum of $\hqr$ is related to the trispectrum of $\deltags$, which has several contributions from the shot noise of the corresponding galaxy sample. We can write this as
\beq
C_\ell^{\hqr \hqr} = C_\ell^{\hqr \qr} + C_\ell^{\rm shot}\ ,
\eeq
with $C_\ell^{\rm shot}$ including all four-point shot noise terms.
 In the approximation of Poissonian shot noise, these contributions depend on the mean number density of the sample, along with the galaxy power spectrum and bispectrum (e.g.~\cite{Sugiyama:2019ike,Darwish:2020prn}). In Appendix~\ref{app:shotnoise}, we estimate the dominant such contributions, finding that the added power to a kSZ template  
can be as high as several tens of percents of the reconstructed kSZ power for a few of the surveys we consider.
We do not include these terms in the computations in the body of the paper, but note that they should be accounted for in detailed de-kSZing analyses, and we refer the reader to Appendix~\ref{app:shotnoise} for more details.

%----------------------
% Modelling
%----------------------
\subsection{Modelling}
\label{sec:modelling}

For modelling the various auto and cross power spectra defined in the previous two subsections, we follow the halo model approach of Ref.~\cite{Smith:2018bpn}. We briefly summarize this approach here, with further details contained in Appendix~\ref{app:halomodel}. We use the public \texttt{hmvec} code for our computations.\footnote{\url{https://github.com/simonsobs/hmvec}}

In the halo model, all matter is assumed to belong to bound, spherical halos. Correlation functions of matter may be computed once several ingredients are specified: the halo mass function, mass-dependent halo bias, and mass-dependent halo density profile. Correlation functions of galaxies additionally require models for how central and satellite galaxies occupy halos of a given mass, and the average spatial distribution of satellite galaxies with in a halo; similarly, predictions involving the electron number density require a model for the electron density profile within halos. Two-point correlations are composed of a ``two-halo term," describing correlations between points located in different halos, and a ``one-halo term," corresponding to two points located within the same halo.

In our calculations, we use the Sheth-Tormen halo mass function and halo bias~\cite{Sheth:1999su}, and a Navarro-Frenk-White (NFW) density profile~\cite{Navarro:1995iw} with the concentration-mass relation from Ref.~\cite{Duffy:2008pz}. Electron density power spectra are computed by replacing the NFW profile with an electron profile model based on hydrodynamical simulations from Ref.~\cite{Battaglia:2016xbi}; in particular, we use the model including AGN feedback as a baseline. Following Ref.~\cite{Bolliet:2022pze}, we truncate the electron profile at a maximum radius chosen so that the enclosed gas mass is the same as for a NFW profile truncated at $r_{\rm cut}=r_{\rm 200c}$. To describe galaxy clustering, we use the halo occupation distribution (HOD) formalism (e.g.~\cite{Berlind:2001xk}), with specific HOD models introduced when they are used later in the paper.

Finally, we use linear expressions for power spectra involving velocities, since only the large-scale velocity power will enter our results:
\begin{align}
\Pvv(k) &= \lp \frac{faH}{k} \rp^2 P_{\rm m}(k)\ , \\
\Pgv(k, \mu) &= \frac{faH}{k} (\bg + f\mu^2) P_{\rm m}(k)\ ,
\end{align}
where $\bg$ is computed within the halo model and the $z$-dependence is left implicit.

%--------------------------------------------------------------------------------------
% Intuition
%--------------------------------------------------------------------------------------
\section{Intuition}
\label{sec:intuition}

Our ability to remove the late-time kSZ signal from observed CMB maps will clearly depend on how much of the true kSZ signal is captured by a given template. In this section, we build intuition for the relationship between a template's properties and its performance in a de-kSZing procedure.

%----------------------
% Relevant redshifts and halo masses
%----------------------
\subsection{Relevant redshifts and halo masses}
\label{sec:intuition-ranges}

\begin{figure*}[t]
\includegraphics[width=\textwidth]{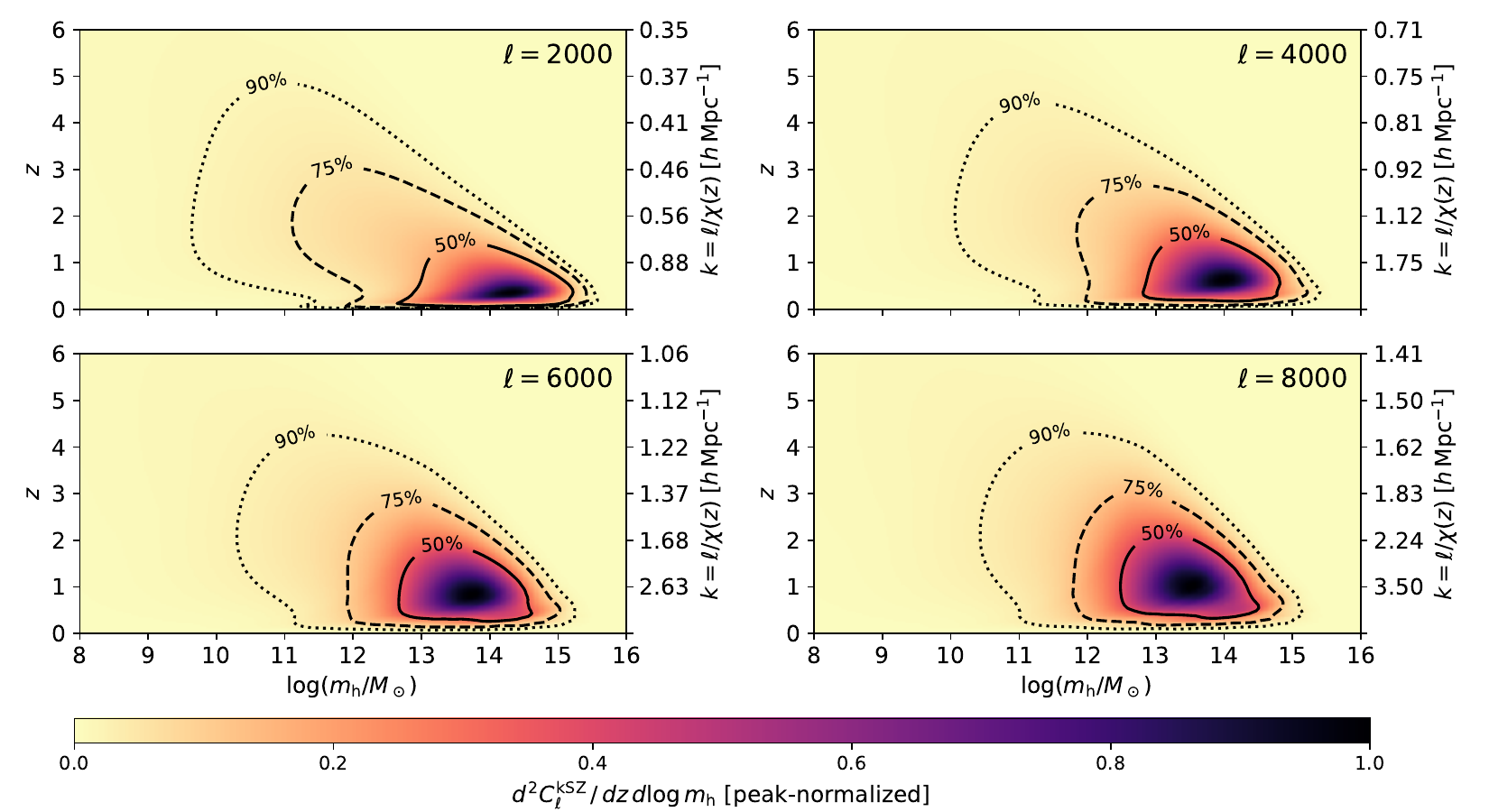}
\caption{%
Relative contribution to the kSZ angular power spectrum $C_\ell^{\rm kSZ}$ from different redshifts and halo masses, with each panel corresponding to the indicated $\ell$ value. The right-hand axis of each plot denotes the comoving wavenumbers that are relevant at a given redshift in the Limber approximation ($k \approx \ell / \chi[z]$). The color scale of each panel is normalized to have a maximum value of unity. The contours enclose regions than contribute 50\%, 75\%, and 90\% of the value of $C_\ell^{\rm kSZ}$ at the given~$\ell$. We find that 50\% the kSZ power generally comes from $z\lesssim 2$ and $12.5\lesssim \log_{10}(\mh/M_\odot) \lesssim 15$, implying that galaxy surveys focusing on these redshifts and halo masses are best suited for constructing kSZ templates.
}
\label{fig:cl_ksz_integrand}
\end{figure*}

We begin with Fig.~\ref{fig:cl_ksz_integrand}, which visualizes the relative contribution to the kSZ auto power from different redshifts and halo masses. Specifically, we rewrite Eq.~\eqref{eq:clksz}, Eq.~\eqref{eq:Pqrqr}, and the halo model expression for $\Pee$ as
\beq
C_\ell^{\rm kSZ} = \int_0^{z_*} dz \int d\log \mh \frac{d^2C_\ell^{\rm kSZ}}{dz\, d\log \mh}
\label{eq:clksz-integrand}
\eeq
and plot the integrand of this expression, normalized to unity at its peak, for several representative $\ell$ values. (See Appendix~\ref{sec:clksz-integrand} for the details of how this is computed.) We show contours enclosing regions of the $z-\mh$ plane that contribute 50\%, 75\%, and 90\% of the total signal. We also translate the redshift axis into the corresponding wavenumbers probed in the Limber approximation ($k\approx\ell/\chi(z)$). 

We find that 50\% of the signal is generally localized to $z\lesssim 2$ and $12.5\lesssim \log_{10}(\mh/M_\odot) \lesssim 15$, while 75\% of the signal arises from $z\lesssim 3$. The overall tilt of the contours stems from the fact that there are fewer higher-mass halos at higher redshift, so the contribution to $C_\ell^{\rm kSZ}$ from higher redshifts is naturally concentrated at lower halo masses.

The ranges of redshift and halo mass that dominate the kSZ signal depend strongly on the halo mass function. The electron-momentum power spectrum $P_{\qr\qr}$ is proportional to the electron density power spectrum $\Pee$ (recall Eq.~\ref{eq:Pqrqr}); in the regime where this is dominated by the one-halo term, 
 if we approximate $u
 _{\rm gas}(k, \mh,z) \approx 1$, Eq.~\eqref{eq:Pee-1h} shows that $\Pee \propto \int d\log \mh\, \mh^3 n(\mh, z)$. The integrand of this expression peaks in the range $13.5 < \log_{10}(\mh/M_\odot) < 14.5$ for $0.5<z<2$, with width $\Delta \log_{10}(\mh/M_\odot)$ of order unity; furthermore, this integral declines sharply with redshift, with roughly a factor of 40 difference between $z=1$ and $z=3$. The assumed electron density profile also has an important effect: in the model from Ref.~\cite{Battaglia:2016xbi}, the profile is strongly suppressed at $r \gtrsim 2R_{200}(\mh,z)$, implying that 
the one-halo kSZ power from a halo with mass $\mh$ will be suppressed at wavenumbers $k$ satisfying $k\gtrsim 2/R_{200}(\mh, z)$.\footnote{To see this mathematically, recall that the one-halo term depends on the Fourier transform of the density profile, $u_{\rm gas}(k) \propto \int dr\, r^2 \rho_{\rm gas}(k) \sin(kr) / kr$, and for a top-hat density profile with maximum radius $r_{\rm max}$, this evaluates to $u_{\rm gas}(k) \propto k^{-3} j_1(kr_{\rm max})$, which has its first zero at $k \approx 4.5 r_{\rm max}^{-1}$ and is strongly suppressed at higher $k$.} Thus, there is an effective (redshift-dependent) maximum $k$ to which a mass-$\mh$ halo will contribute significant kSZ power, and since $R_{200}$ grows with halo mass, this maximum $k$ will be smaller for more massive halos. We can see this in Fig.~\ref{fig:cl_ksz_integrand}: lower $\ell$ values probe a higher maximum halo mass.
In summary, the redshifts and halo masses that dominate the kSZ signal can mostly be understood in terms of (1) the abundance of halos of different masses at a given redshift, and (2) the spatial extent of the free electrons associated with each halo.

Overall, these results indicate that a galaxy survey (or combination of surveys) with broad coverage in redshift and halo mass will be necessary to construct a high-fidelity kSZ template. 
For example, a sample of emission-line galaxies (ELGs) would likely miss a significant part of the kSZ signal, since ELGs generally occupy halos with much lower mass (i.e.\ $\mh \sim 10^{12}\,M_\odot$~\cite{Gonzalez-Perez:2017mvf,Alam:2019pwr,Avila:2020rmp,Yuan:2022rsc}) than where the bulk of the kSZ contribution comes from. (We will explore specific galaxy samples, including ELGs, later in the paper.)
Also, it will likely be challenging to recover more than roughly 80\% of the kSZ signal unless one has access to a survey probing very low-mass ($\mh\lesssim 10^{11.5}M_\odot$) halos at higher redshifts (such as a \tcm intensity mapping survey~\cite{Villaescusa-Navarro:2018vsg,Modi:2019ewx}), and is able to combine it with measurements of higher-mass halos.

%----------------------
% Separate templates
%----------------------
\subsection{Importance of electron density and velocity templates}
\label{sec:templates-ev}

\subsubsection{Electron density template}
\label{sec:template-e}

The accuracies of the electron density template and velocity template both determine the fraction of true kSZ power captured by the kSZ template. We investigate the electron template in this subsection and the velocity template in Sec.~\ref{sec:template-v}.

\begin{figure}[t]
\includegraphics[width=\columnwidth, trim=0 20 0 0]{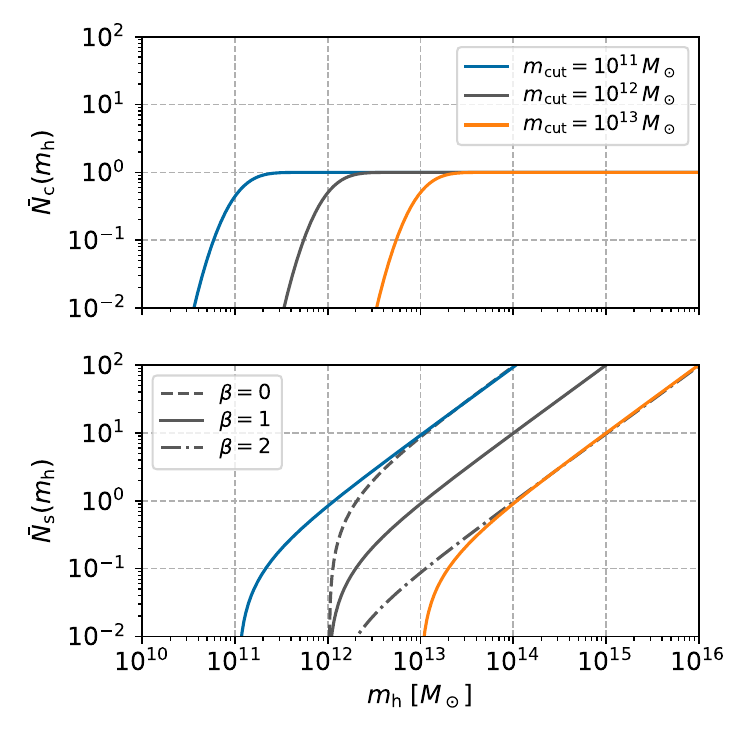}
\caption{%
Mean central occupation ({\em upper panel}) and satellite occupation ({\em lower panel}) for the HOD model used to investigate the electron density template in Sec.~\ref{sec:template-e}. In both panels, different colors denote different values of the lower mass cutoff $m_{\rm cut}$, while in the lower panel, different linestyles denote different values for $\beta$, which controls the relative amplitude of $\bar{N}_{\rm c}$ and $\bar{N}_{\rm s}$. This figure can be used in tandem with Fig.~\ref{fig:r_ge_squared_fixed_ngal} to gain intuition for how the properties of the HOD influence the fidelity of the electron density template that goes into a kSZ template.
}
\label{fig:zheng05_hod_variations}
\end{figure}

To assess the fidelity of the electron template, we compute the squared correlation coefficient between $\deltags$ and~$\deltae$, for purely-transverse ($\muS=0$) modes:
\beq
r_{\rm ge}^2(k; z) =
	\frac{\Pge(\kS, 0; z)^2}{\Pggtot(\kS, 0; z) \Pee(\kS; z)}\ .
\eeq
We compute this quantity for a family of fictional galaxy surveys which have the following HOD model:
\begin{align}
\label{eq:simpleHOD-Nc}
\bar{N}_{\rm c}(\mh) 
	&= \frac{1}{2} \lb 1 + {\rm erf}\!\lp \frac{\log \mh - \log m_{\rm cut}}{\sigma_{\log m}} \rp\rb\ , \\
\label{eq:simpleHOD-Ns}
\bar{N}_{\rm s}(\mh) 
	&= \lp \frac{\mh - m_{\rm cut}}{10^\beta\, m_{\rm cut}} \rp^\alpha\ .
\end{align}
This is a simplified version of the HOD from Ref.~\cite{Zheng:2004id}, with the amplitude of the satellite occupation parameterized relative to the lower mass cutoff $m_{\rm cut}$, to separate the effects of changing $m_{\rm cut}$ and this amplitude. For this exercise, we fix $\sigma_{\log m}=0.3$ and $\alpha=1$, and examine the effects of varying $m_{\rm cut}$ and $\beta$. The influence of these parameters on the central and satellite occupations is visualized in Fig.~\ref{fig:zheng05_hod_variations}.

\begin{figure*}[t]
\includegraphics[width=\textwidth, trim=0 15 0 0]{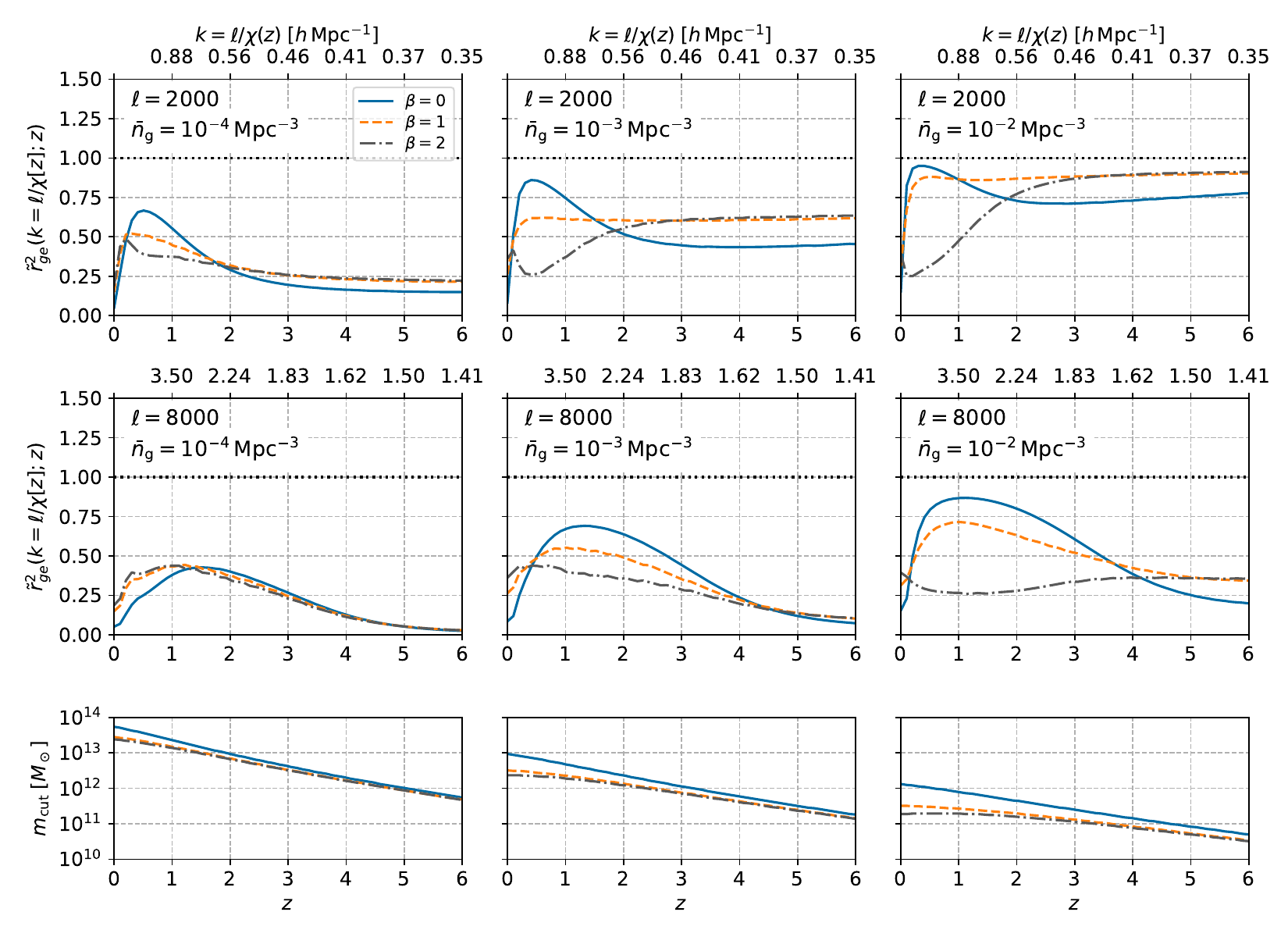}
\caption{%
An exploration of how accurate a template for the electron density $\deltae$ can be constructed by Wiener-filtering the observed galaxy density $\deltags$. We plot the squared correlation coefficient between $\deltags$ and $\deltae$,  assuming different parameters in the simplified HOD model from Eqs.~\eqref{eq:simpleHOD-Nc}-\eqref{eq:simpleHOD-Ns}. Curves in the {\em upper} and {\em middle rows} are evaluated at $k=\ell/\chi[z]$ for $\ell=2000$ or $8000$, while the {\em lower row} shows the redshift-dependent minimum mass $m_{\rm cut}$ corresponding to the galaxy number density indicated in each column. 
In both the two-halo-dominated regime (at sufficiently high $z$ and low $k$) and the one-halo-dominated regime (at low $z$ and high $k$), a high-fidelity electron template is achievable if the input galaxy survey is sufficiently dense.
}
\label{fig:r_ge_squared_fixed_ngal}
\end{figure*}

In Fig.~\ref{fig:r_ge_squared_fixed_ngal}, we plot $r_{\rm ge}^2$ as a function of $z$, at the $k$ values relevant for the Limber approximation ($k \approx \ell/\chi[z]$) at two different $\ell$ values. We vary $\beta$ between $0$ and $2$, noting that $\beta \approx 1$ is a typical value found in real or simulated galaxy samples (e.g.~\cite{Zheng:2004id,Smith:2017tzz,Walsh:2019luq,Yuan:2022rsc}). We also vary the galaxy number density $\bar{n}_{\rm g}$, taken to be redshift-independent, and derive the corresponding (redshift-dependent) lower mass cutoff $m_{\rm cut}$ by solving Eq.~\eqref{eq:ng-halomodel}. The resulting $m_{\rm cut}$ values are shown in the lower panels of the figure.

The shapes of the curves in Fig.~\ref{fig:r_ge_squared_fixed_ngal} can be interpreted in terms of whether the one-halo or two-halo term dominates. In particular, there are two noteworthy trends:
\begin{itemize}
%%%
\item At higher redshift, the g-e correlation improves with increasing $\bar{n}_{\rm g}$ (and decreasing $m_{\rm cut}$). This is because, in the angular power spectrum, higher $z$ corresponds to lower $k$, where the two-halo term is more important. At sufficiently low $k$ (i.e.\ high~$z$ and low~$\ell$), the two-halo term completely dominates and perfect cross-correlation can be achieved, although the (angular) kSZ power will also have some sensitivity to lower~$z$ where the correlation is worse. The correlation is noticeably worse for $\beta=0$ than for $\beta=1$ or $2$, because $\beta=0$ implies a particularly large amplitude for the one-halo term, moving the pure-two-halo regime to lower $k$.
%%%
\item In the one-halo regime (low $z$, high $k$), a high-fidelity electron template is achievable for sufficiently high galaxy number density. In this regime, lower values of $\beta$ result in a better template, because a higher satellite fraction implies a better measurement of the distribution of satellites within the halo, which will trace the distribution of electrons if the satellite profile and gas profile are similar.
%%%
\end{itemize}

\subsubsection{Velocity template}
\label{sec:template-v}

\begin{figure}[t]
\includegraphics[width=\columnwidth, trim= 0 20 0 0]{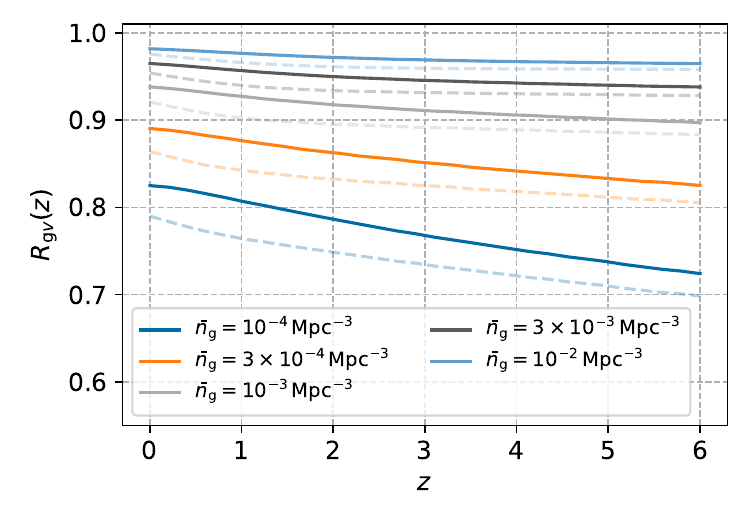}
\caption{%
Correlation-coefficient-like quantity, defined in Eq.~\eqref{eq:Rgv},  for the observed galaxy density and large-scale velocity field, reflecting the impact of the velocity template on the kSZ template. Different curves are computed as in Fig.~\ref{fig:r_ge_squared_fixed_ngal} with $\beta=1$, with faint lines ignoring the effects of RSD. For lower galaxy number density, the velocity template fidelity can degrade the kSZ template fidelity by a few tens of percents, while for higher number density, the electron template fidelity will mostly determine the fraction of true kSZ power that can be captured by the kSZ template.
}
\label{fig:R_gv2}
\end{figure}

Next, we assess the performance of the template for large-scale velocities. We do so by computing the
ratio of the velocity factors in Eq.~\eqref{eq:Phqrqr} and~\eqref{eq:Pqrqr}:
\beq
\Rgv(z) \equiv \frac{3}{2} 
	\frac{\int d\kL\, \kL^2 \int_{-1}^1 d\muL\, \muL^2 \frac{\Pgv(\kL,\muL; z)^2}{\Pggtot(\kL,\muL; z)}}
	{\int d\kL\, \kL^2 \Pvv(\kL; z)}\ .
\label{eq:Rgv}
\eeq
This is analogous to a correlation coefficient between $\deltags$ and $v$, but is more relevant to the kSZ template than a standard correlation coefficient, since it directly reflects the redshift-dependent suppression of kSZ template power due to the properties of the velocity template.

Fig.~\ref{fig:R_gv2} shows $\Rgv(z)$ computed using the same HOD model as in Sec.~\ref{sec:template-e}, showing only the $\beta=1$ case for simplicity. The inclusion of RSD slightly boosts the power in the velocity template thanks to the Kaiser factor increasing the signal-to-shot-noise ratio in $\Pggtot$. For $\bar{n}_{\rm g} = 10^{-4}\,{\rm Mpc}^{-3}$, the accuracy of the velocity template can degrade the fraction of the kSZ power captured by the kSZ template by tens of percents, but for higher number densities, a comparison with Fig.~\ref{fig:r_ge_squared_fixed_ngal} reveals that the electron template will be the dominant factor in the performance of the kSZ template.

\subsubsection{Combination}

\begin{figure}[t]
\includegraphics[width=\columnwidth, trim= 0 20 0 0]{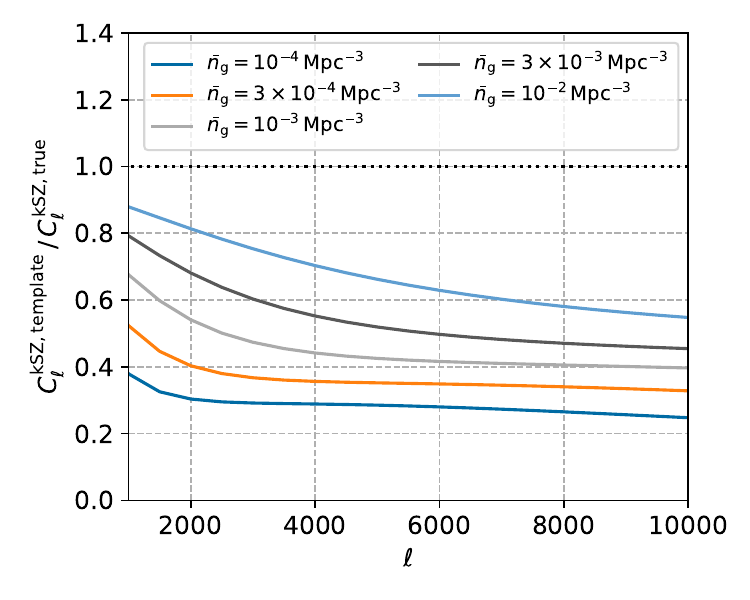}
\caption{%
Ratio of kSZ template power spectrum and true kSZ power spectrum, with the same line colors as Fig.~\ref{fig:R_gv2}. As expected, a template constructed from a denser galaxy survey captures a higher fraction of the true kSZ power, with better performance at lower multipoles that probe larger physical scales.
}
\label{fig:cl_ksz_ratio_beta1}
\end{figure}

Fig.~\ref{fig:cl_ksz_ratio_beta1} shows the auto power spectrum of the kSZ template constructed using the same HOD as the curves in Fig.~\ref{fig:R_gv2}, as a ratio to the true kSZ power. As expected, we find better performance at lower multipoles, because these multipoles probe larger physical scales and higher redshifts where the electron template is more accurate. The accuracy decreases at higher multipoles, but can still reach several tens of percents of the true kSZ power if a dense enough galaxy survey is used.

These general investigations provide context for survey-specific forecasts that we will present in Sec.~\ref{sec:forecasts}.

%----------------------
% Accuracy of model for galaxy-electron cross power spectrum
%----------------------
\subsection{Accuracy of model for galaxy-electron cross power spectrum}
\label{sec:Pge_error}

The kSZ template in Sec.~\ref{sec:templates} requires input models for $\Pgv$, $\Pge$, and $\Pggtot$. Since we only require $\Pgv$ on quasi-linear scales, it will be relatively straightforward to model, while $\Pggtot$ is directly measurable from the associated galaxy survey. On the other hand, the modelling of $\Pge$ is expected to be considerably more uncertain.  This uncertainty leads to what is known as the ``optical depth degeneracy" in kSZ studies (e.g.,~\cite{Madhavacheril:2019buy}) and is a limiting factor, for example, in using the reconstructed velocities from the kSZ effect as a probe of the growth rate of structure.  Here, we consider what impact uncertainties in the assumed form of $\Pge$ will have on a de-kSZing procedure.

Specifically, we are interested in the power spectrum of the kSZ effect that remains in a temperature map after a template has been subtracted off. This angular power spectrum, of $\hTdekSZ \equiv \TkSZ - \hTkSZ$, evaluates to
\beq
C_\ell^\text{de-kSZ} = C_\ell^{\rm kSZ} - 2 C_\ell^{\hqr \qr} + C_\ell^{\hqr \hqr}\ ,
\label{eq:cldeksz}
\eeq
where $C_\ell^{\rm kSZ}$ is given by Eq.~\eqref{eq:clksz}, $C_\ell^{\hqr \qr}$ is given by Eq.~\eqref{eq:clhqrqr}, and $C_\ell^{\hqr \hqr}$ is equal to $C_\ell^{\hqr \qr}$ if the assumed $\Pge$ is exactly correct.

We can relax this assumption by writing the assumed spectrum as a sum of the true spectrum and an error term:
\beq
\Pge^{\rm model}(k,\mu;z) = \Pge(k,\mu;z) + \Delta \Pge(k,\mu;z)\ .
\eeq
Repeating the derivations of $P_{\hqr \qr}$ (recall Eq.~\ref{eq:Phqrqr}) and $P_{\hqr\hqr}$, we find that using $\Pge^{\rm model}$ instead of $\Pge$ in the construction of the electron density templates will change each spectrum like so:
\begin{align*}
&\Delta P_{\hqr \qr}(\kS, \muS; z) \\
&\qquad=\frac{1}{4\pi^2}
	\lp \int d\kL\, \kL^2 \int_{-1}^1 d\muL\, \muL^2 \frac{\Pgv(\kL, \muL; z)^2}{\Pggtot(\kL, \muL; z)} \rp \\
\numberthis
&\qquad\qquad\times
	\frac{\Pge(\kS, \muS; z) \Delta \Pge(\kS, \muS; z)}{\Pggtot(\kS, \muS; z)}\ , \\
&\Delta P_{\hqr \hqr}(\kS, \muS; z) \\
&\qquad=\frac{1}{4\pi^2}
	\lp \int d\kL\, \kL^2 \int_{-1}^1 d\muL\, \muL^2 \frac{\Pgv(\kL, \muL; z)^2}{\Pggtot(\kL, \muL; z)} \rp \\
&\qquad\qquad\times
	\lb \frac{2\Pge(\kS, \muS; z) \Delta \Pge(\kS, \muS; z)}{\Pggtot(\kS, \muS; z)} \right. \\
\numberthis
&\qquad\qquad\qquad\left.
	+ \frac{\Delta \Pge(\kS, \muS; z)^2}{\Pggtot(\kS, \muS; z)} \rb\ .
\end{align*}
If we compute $\Delta C_\ell^\text{de-kSZ}$ by combining these expressions, the definitions of $C_\ell^{\hqr \qr}$ and $C_\ell^{\hqr \hqr}$, and Eq.~\eqref{eq:cldeksz}, we find that the term linear in $\Delta \Pge$ cancels out, such that the leading effect of $\Delta\Pge$ on $C_\ell^\text{de-kSZ}$ is of order $\Delta \Pge^2$. As a concrete example, if $\Delta\Pge \approx \alpha \Pge$ where $\alpha$ is a constant, then the fractional impact on $C_\ell^\text{de-kSZ}$ is
\beq
\frac{\Delta C_\ell^\text{de-kSZ}}{C_\ell^\text{de-kSZ}}
	\approx \frac{\alpha^2 C_\ell^{\hqr \hqr}}{ C_\ell^{\rm kSZ} -  C_\ell^{\hqr \hqr} }\ .
\eeq
This mild dependence on errors in $\Pge$, along with prospects for externally constraining it using e.g.\ dispersion measures of fast radio bursts~\cite{Madhavacheril:2019buy}, is a promising indication that modelling of the galaxy-electron cross power spectrum will not be a serious impediment to template-based de-kSZing.

%----------------------
% Redshift precision
%----------------------
\subsection{Redshift precision}
\label{sec:redshifts}

\begin{figure}[t]
\includegraphics[width=\columnwidth, trim= 0 20 0 0]{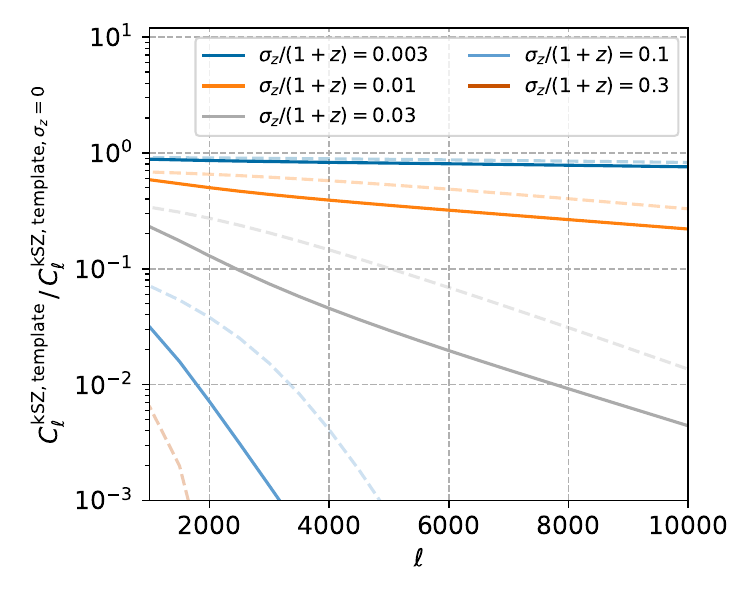}
\caption{%
Impact of galaxy redshift uncertainties on a kSZ template constructed using those galaxies. Each curve is computed with the HOD from  Sec.~\ref{sec:template-e} with $\beta=1$ and with the specified Gaussian redshift uncertainty, and has been divided by the corresponding template power spectrum assuming perfectly measured redshifts. {\em Solid curves} assume $\bar{n}_{\rm g} = 10^{-4}\,{\rm Mpc}^{-3}$, while {\em dashed curves} assume $\bar{n}_{\rm g} = 10^{-2}\,{\rm Mpc}^{-3}$. For $\sigma_z/(1+z) = 0.03$, which is the target for the Vera Rubin Observatory's LSST, and for which SPHEREx will obtain a galaxy sample with $\bar{n}_{\rm g} \sim \mathcal{O}(10^{-4})\,{\rm Mpc}^{-3}$ at $z\sim 1$, significant suppression of the kSZ template power is expected. This motivates our focus on spectroscopic and \tcm surveys in this work.
}
\label{cl_ksz_ratio_sigz_beta1}
\end{figure}

Thus far, we have assumed that the galaxy survey used for the template has negligible redshift uncertainty, but we can also consider surveys for which this uncertainty may be substantial. Specific examples include photometric surveys such as the Dark Energy Survey~\cite{DES:2005dhi} and the Vera Rubin Observatory's Legacy Survey of Space and Time (LSST;~\cite{LSSTScience:2009jmu}), or the SPHEREx satellite~\cite{Dore:2014cca} which is expected to acquire redshifts with a wide range of uncertainties~\cite{Dore:2014cca}.

We will not attempt detailed forecasts for such surveys in this work, but instead present a simple computation that indicates what we might expect. Under the assumption that the uncertainty on each measured redshift is Gaussian, the galaxy power spectrum is multiplied by two powers of a Gaussian kernel
\beq
W_{\Delta z}(k, \mu; z) = \exp\!\lb -\frac{\sigma_z^2 k^2 \mu^2 c^2}{2H(z)^2} \rb\ ,
\eeq
while cross spectra between galaxies and another field are multiplied by a single power (e.g.~\cite{Smith:2018bpn}):
\begin{align*}
\Pgg(k, \mu; z) &\to \Pgg(k, \mu; z) W_{\Delta z}(k, \mu; z)^2\ , \\
\Pge(k, \mu; z) &\to \Pge(k, \mu; z) W_{\Delta z}(k, \mu; z)\ , \\
\Pgv(k, \mu; z) &\to \Pgv(k, \mu; z) W_{\Delta z}(k, \mu; z)\ .
\numberthis
\end{align*}
Since the redshift kernel for the kSZ effect is quite broad, one might expect that redshift errors would not have a strong effect on a kSZ template, as is the case when constructing templates for the integrated Sachs-Wolfe effect (e.g.~\cite{Ferraro:2022twg}), for example. However, this intuition does not hold in this case, because our kSZ template involves a product of separate 3d templates for velocity and electron density, and low-$\mu$ modes of this product are affected by higher-$\mu$ modes of each component template. In particular, in the squeezed limit we have assumed for the kSZ effect in Eq.~\eqref{eq:Phqrqr}, the effect of redshift uncertainties on higher-$\mu$ modes of the velocity template propagates into the final projected kSZ template.

In Fig.~\ref{cl_ksz_ratio_sigz_beta1}, we show how different levels of redshift uncertainty affect $C_\ell^{\rm kSZ,template}$ computed using the HOD from Sec.~\ref{sec:template-e} with $\beta=1$ and $\bar{n}_{\rm g} = 10^{-4}\,{\rm Mpc}^{-3}$ (solid lines) or $10^{-2}\,{\rm Mpc}^{-3}$ (dashed lines). For $\sigma_z / (1+z) = 0.03$, which is the target for LSST~\cite{LSSTDarkEnergyScience:2018jkl}, the kSZ template power is suppressed at $\ell=2000$ by 85\% (70\%) for $\bar{n}_{\rm g} = 10^{-4}\,{\rm Mpc}^{-3}$ ($10^{-2}\,{\rm Mpc}^{-3}$), and more than 90\% for $\ell \gtrsim 2500$ ($5000$). SPHEREx is expected to achieve a galaxy number density of $\mathcal{O}(10^{-4})\,{\rm Mpc}^{-3}$ at $z\sim 1$ with $\sigma_z/(1+z) \sim 0.03$~\cite{Dore:2014cca}, implying a similar suppression of kSZ template power for a SPHEREx-based template. For this reason, we will focus on spectroscopic surveys, which have $\sigma_z / (1+z) \sim \mathcal{O}(0.001)$ (e.g.~\cite{DESI:2016fyo}), or \tcm surveys with equivalent redshift precision, for the remainder of this work.

%--------------------------------------------------------------------------------------
% Forecasts
%--------------------------------------------------------------------------------------
\section{Forecasts}
\label{sec:forecasts}

In this section, we investigate the amount of kSZ power that can be recovered when different large-scale structure surveys are used to construct kSZ templates using the procedure in Sec.~\ref{sec:templates}. As discussed in Sec.~\ref{sec:intuition}, we will find that the ranges of redshift and halo mass probed, along with the number density of the galaxy sample, play an important role in the performance of the templates.

%----------------------
% Spectroscopic surveys
%----------------------
\subsection{Spectroscopic surveys}
\label{sec:specsurveys}

\begin{figure}[t]
\includegraphics[width=\columnwidth, trim= 0 20 0 0]{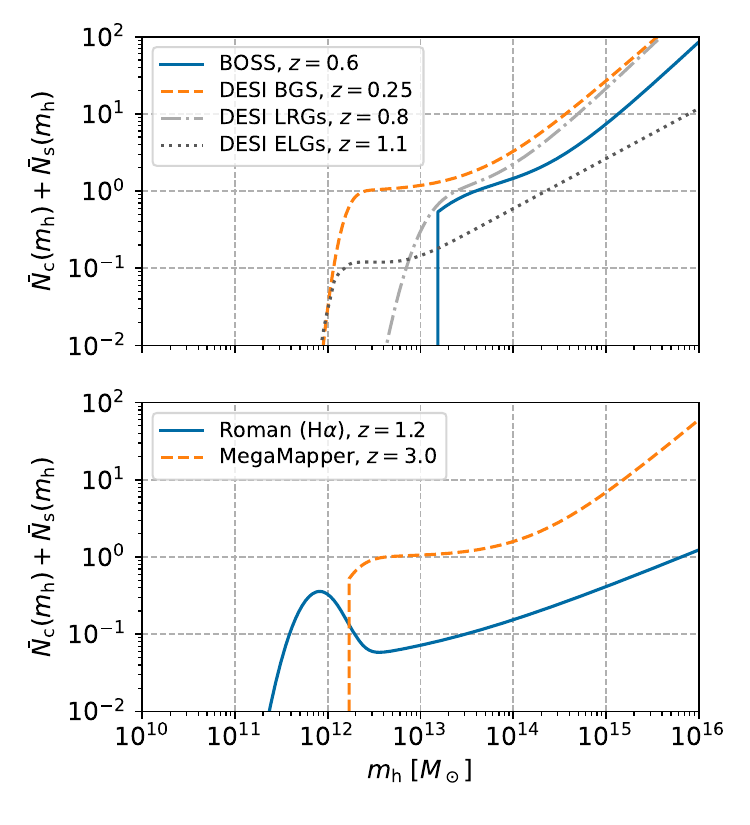}
\caption{%
HOD models for the spectroscopic surveys we forecast for in Sec.~\ref{sec:specsurveys}. Specifically, we show the sum of central and satellite occupations, evaluated at a representative redshift for each survey. For visual clarity, we plot the models in two separate panels. These models are described in detail in Sec.~\ref{sec:specsurveys} and Appendix~\ref{app:hod}.
}
\label{fig:spec_hods}
\end{figure}

\begin{figure}[t]
\includegraphics[width=\columnwidth, trim= 0 15 0 0]{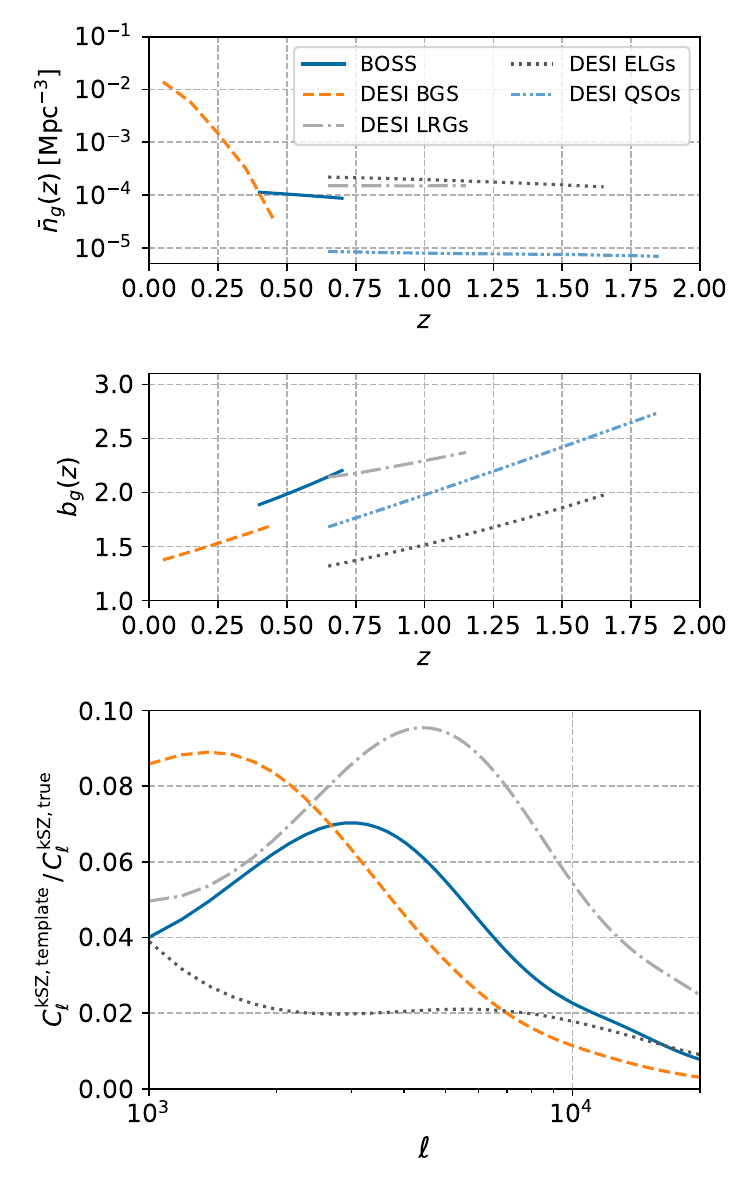}
\caption{%
Assumed number density ({\em upper panel}) and linear bias ({\em middle panel}) for BOSS and various clustering samples from DESI, along with forecasts for the ratio of kSZ template power spectrum and true kSZ power spectrum for templates constructed from each sample ({\em lower panel}). The DESI QSO template recovers less than 1\% of the true kSZ power, so we do not show it in the lower panel. Overall, the amount of kSZ power that can be recovered by these templates is limited by the halo masses and redshift range probed by each galaxy sample, along with galaxy shot noise that restricts access to smaller-scale modes.
}
\label{fig:desi_templates}
\end{figure}

First, we consider a representative set of spectroscopic galaxy surveys: BOSS, DESI, the High Latitude Spectroscopic Survey of the Roman Space Telescope, and the MegaMapper proposal. We state our assumptions about each survey, and their implications for kSZ reconstruction, in the following subsections. As a visual aid for these discussions, in Fig.~\ref{fig:spec_hods} we plot the HOD model assumed for each survey at a representative redshift. Further details about HOD modelling are included in Appendix~\ref{app:hod}. Our main results are summarized in Figs.~\ref{fig:desi_templates} and~\ref{fig:spec_templates}, where we show the angular power spectrum of each kSZ template normalized to the true kSZ power spectrum, along with the linear galaxy bias and number density assumed for each survey.

\subsubsection{BOSS}

BOSS (the Baryon Oscillation Spectroscopic Survey~\cite{BOSS:2012dmf}) measured the redshifts of 1.5 million luminous red galaxies (LRGs) over $0.4\lesssim z \lesssim 0.7$ and $10000\,{\rm deg}^2$. We use the HOD from Ref.~\cite{Walsh:2019luq}, which was jointly fit to the projected correlation function and void probability function measured from a subset of the BOSS CMASS sample spanning $0.46 \lesssim z \lesssim 0.57$. We use this HOD over the full BOSS redshift range, where it predicts a roughly constant number density of $\bar{n}_{\rm g} \approx 10^{-4}\,{\rm Mpc}^{-3}$ and a linear bias ranging from $1.9$ at $z=0.4$ to $2.2$ at $z=0.7$.

Fig.~\ref{fig:desi_templates} shows that the corresponding kSZ template can recover at most 7\% of the total kSZ power at $\ell\approx 3000$, and even less at lower and higher multipoles. This is consistent with our intuition from Sec.~\ref{sec:intuition}: Fig.~\ref{fig:cl_ksz_ratio_beta1} showed that a survey with $\bar{n}_{\rm g} \approx 10^{-4}\,{\rm Mpc}^{-3}$ over $0<z<6$ and a BOSS-like HOD can reproduce at most 30\% of the true kSZ power for $\ell\gtrsim 2000$, but BOSS covers a considerably smaller redshift range than what is significant for kSZ (recall Fig.~\ref{fig:cl_ksz_integrand}), so we would expect a correspondingly smaller fraction of the kSZ power to be captured by the template.

\subsubsection{DESI}

DESI (the Dark Energy Spectroscopic Instrument~\cite{DESI:2016fyo}) is an ongoing survey project with the goal of measuring roughly 30 million spectroscopic galaxy and quasar redshifts over $14000\,{\rm deg}^2$. We separately consider the planned Bright Galaxy Sample (BGS), which will have number density greater than BOSS for $z\lesssim 0.4$; the LRG sample over $0.6\lesssim z \lesssim 1$; the emission-line galaxy (ELG) sample over $0.6\lesssim z \lesssim 1.6$; and the quasar (QSO) clustering sample over $0.6\lesssim z \lesssim 1.8$. In some cases, we adjust a given HOD model or two-halo contribution to agree with the expected galaxy number density or linear bias, and we describe these adjustments below.

\paragraph*{\bf BGS:} 
We use the HOD from Ref.~\cite{Smith:2017tzz}, which is fit to a mock galaxy catalog with the same apparent magnitude threshold as the BGS sample. Specifically, the authors fit their 5 free HOD parameters to the clustering of galaxy subsamples corresponding to different absolute magnitudes. We use the parameter fits corresponding to ${}^{0.1} M_r - 5\log h = -20.5$, motivated by evidence from Ref.~\cite{Zarrouk:2021ctz} that the distribution of absolute magnitudes of BGS targets peaks at this value. However, we rescale the minimum halo mass and normalization of the satellite occupation by a redshift-dependent prefactor such that the predicted mean galaxy density matches the forecast values from Ref.~\cite{DESI:2016fyo} (see Appendix~\ref{app:hod-desibgs}). We also rescale the predicted two-halo term of the power spectrum to agree with the linear bias assumed for BGS in Ref.~\cite{DESI:2016fyo} in the low-$k$ limit.

\paragraph*{\bf LRGs:} We use the LRG HOD from Ref.~\cite{Yuan:2022rsc}, which was fit to galaxies selected from the IllustrisTNG simulation according to color cuts corresponding to the LRG sample. As with our BGS model, we rescale the HOD such that the mean galaxy density matches the LRG target density of $1.5\times10^{-4}\,{\rm Mpc}^{-3}$ quoted in Ref.~\cite{Yuan:2022rsc}. We use the linear bias predicted by this model, which ranges from $2.2$ at $z=0.6$ to $2.4$ at $z=1$. If we instead rescale the HOD to match the galaxy number density used for forecasting in 
Ref.~\cite{DESI:2016fyo}, or rescale the two-halo term to match the linear bias used in that work, the end result for $C_\ell^{\rm kSZ, template}$ can change by up to 75\% at lower multipoles, but never exceeds 6\% of the true kSZ power spectrum. We have also compared with results from using a different HOD fit to imaging of LRG targets in Ref.~\cite{Zhou:2020nwq}, and found that $C_\ell^{\rm kSZ, template}$ is around 20\% higher, which is not sufficient to qualitatively change our conclusions.

\paragraph*{\bf ELGs:} We use the ELG HOD from Ref.~\cite{Yuan:2022rsc}, which is based on the ``high-mass quenched" ELG model from Ref.~\cite{Alam:2019pwr} and which was also fit to galaxies selected from the IllustrisTNG simulation. We use the galaxy density ($\sim$$1.5\times10^{-4}{\rm Mpc}^{-3}$) and linear bias (ranging from $1.3$ at $z=0.65$ to $2.0$ at $z=1.65$) predicted by this model. In contrast to other HODs described above, the expected number of centrals in the ELG model peaks in lower-mass halos ($\mh \sim 10^{12} M_\odot$) and declines at higher halo masses\footnote{This feature of the central galaxy occupation for DESI ELGs is not visible in Fig.~\ref{fig:spec_hods} because we have plotted the sum of the central and satellite occupations, but a peak at $\mh \sim 10^{12} M_\odot$ is visible in the Roman (H$\alpha$) HOD plotted in the lower panel, for which the satellite occupation has lower amplitude.}, reflecting the expectation that star formation is quenched in higher-mass halos and therefore these halos are less likely to host ELGs as their central galaxy (e.g.~\cite{Gonzalez-Perez:2017mvf,Alam:2019pwr,Avila:2020rmp,Yuan:2022rsc}). Using the galaxy density and linear bias from Ref.~\cite{DESI:2016fyo} decreases $C_\ell^{\rm kSZ, template}$ by roughly 30\%, while using a parameterized version of the HOD from Ref.~\cite{Hadzhiyska:2020iri} changes it by around a factor of 2, but as with the LRGs, this does not alter the nature of our conclusions.

\paragraph*{\bf QSOs:} We use the error-function--based HOD from Ref.~\cite{Alam:2019pwr}, with the best-fit parameters fit to eBOSS quasar clustering but with the lower halo mass threshold and satellite fraction normalization adjusted to match the QSO number density given in Ref.~\cite{DESI:2016fyo}, and with the 2-halo term rescaled to match the linear bias from Ref.~\cite{DESI:2016fyo} in the low-$k$ limit. Due to the low expected QSO number density ($\lesssim 10^{-5}\,{\rm Mpc}^{-3}$), the QSO sample contributes negligibly to a reconstruction of the late-time kSZ signal, recovering less than 1\% of the true kSZ power, so we do not show the corresponding template in Fig.~\ref{fig:desi_templates} or discuss the QSO sample further.

\vspace{11pt}

Amongst the different DESI samples, Fig.~\ref{fig:desi_templates} shows that a template constructed from the LRGs captures the largest fraction of kSZ power, because its combination of halo mass range and redshift range is best matched to the ranges relevant for kSZ. However, for the same reasons as BOSS, DESI LRGs can still only recover a small fraction of kSZ power. The ranges of halo mass and/or redshift are less optimal for the other samples, which fare even worse. Finally, Fig.~\ref{fig:spec_templates} shows that a combination of all BOSS and DESI samples could recover around 10-20\% of the true kSZ power over a wide range of multipoles.

\subsubsection{Roman}

The Roman Space Telescope~\cite{Spergel:2015sza} is a planned multi-purpose satellite with cosmology among its scientific drivers. In particular, the Roman High Latitude Spectroscopic Survey will obtain redshifts for 12 million ELGs over $2000\,{\rm deg}^2$, using H$\alpha$ emission over $1<z<2$ and [OIII] emission over $1<z<3$~\cite{Wang:2021oec}. We only consider the H$\alpha$ sample in this work, due to its higher number density and more relevant redshift range for construction of a kSZ template.

For an HOD, we use a parameterized fit to the HOD measured from a mock ELG catalog generated from the {\sc Galacticus} semi-analytical model applied to the {\sc UNIT} N-body simulation~\cite{Zhai:2021pja}. This model has a similar form to the DESI ELG used above, but with a central occupation that peaks at slightly lower halo masses and with a lower satellite fraction. We rescale the amplitudes of both the central and satellite occupations to match the H$\alpha$ number densities from Ref.~\cite{Wang:2021oec} corresponding to fluxes $>10^{16}\,{\rm erg}\,{\rm s}^{-1} {\rm cm}^{-2}$ and dust attenuation parameter $A_V=1.65$, matching the catalog from Ref.~\cite{Zhai:2021pja}. We also rescale the 2-halo term such that the linear bias agrees with that in Ref.~\cite{Wang:2021oec} ($b(z)=1+0.5z$) in the low-$k$ limit.

Fig.~\ref{fig:spec_templates} shows that the corresponding kSZ template can recover 7\% of the kSZ power at $\ell \approx 1000$, declining to less than 1\% at $\ell\gtrsim3500$. Despite the relatively high ELG number density projected for the Roman H$\alpha$ sample, the corresponding halo masses are too low and the redshift range is too high to capture a significant part of the kSZ signal. The de-kSZing efficiency exhibits a minimum at $\ell\approx 6000$, which corresponds to spatial scales in the regime where neither the two-halo nor one-halo terms are dominant at the relevant redshifts.

In principle, one may also consider the Euclid satellite, which will perform its own spectroscopic survey of H$\alpha$ emitters. However, Table~3 of Ref.~\cite{Euclid:2019clj} indicates that this sample will have a lower number density than Roman and similar linear bias, so a kSZ template built from Euclid's H$\alpha$ sample is not expected to fare any better than what we have forecast for Roman.

\subsubsection{MegaMapper}

\begin{figure}[t]
\includegraphics[width=\columnwidth, trim= 0 15 0 0]{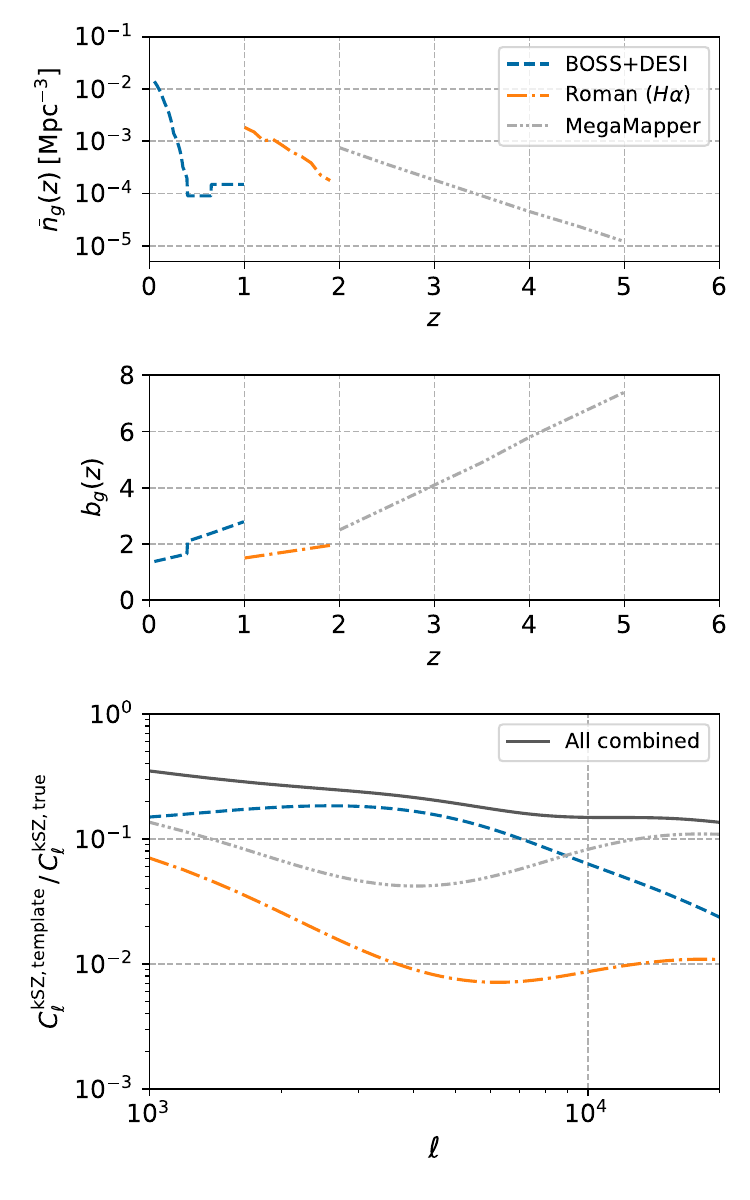}
\caption{%
Same as Fig.~\ref{fig:desi_templates}, but for kSZ templates constructed from the combination of BOSS and DESI ({\em blue dashed}), the H$\alpha$ emitter sample from the Roman Space Telescope ({\em orange dot-dashed}), the Lyman-break galaxy sample from MegaMapper ({\em grey dot-dot-dashed}), and an idealistic combination of all of the above ({\em solid black}) that assumes identical sky coverage for each survey. The first three samples are mostly limited to recovering 20\% or less of the true kSZ power, while the combination of all of them can recover 35\% at $\ell\approx 1000$ and between 15 and 25\% for $\ell\gtrsim 2000$.
}
\label{fig:spec_templates}
\end{figure}

The MegaMapper~\cite{Schlegel:2019eqc} is a proposed ground-based telescope outfitted with DESI fiber-positioning technology that would target Lyman-break galaxies (LBGs) and Lyman-alpha emitters over $2<z<5$ using imaging from the Vera Rubin Observatory's Legacy Survey of Space and Time. We perform a forecast using the HOD for LBGs obtained from early observations of roughly $6\times 10^5$ objects by the Hyper Suprime-Cam~\cite{Harikane:2017lcw}. In particular, we use their ``linear HOD model" fit at $z\approx 3.8$ using galaxies with threshold apparent magnitude $m_{\rm UV}^{\rm th}=24.5$, which is the same limiting magnitude assumed for the ``idealised sample" from Ref.~\cite{Ferraro:2019uce}. The galaxy number density and linear bias predicted by this model are very close to the values used for forecasting in Refs.~\cite{Ferraro:2019uce,Schlegel:2019eqc} (the number density ranges from $5\times 10^{-3}\,{\rm Mpc}^{-3}$ at $z=2$ to $10^{-5}\,{\rm Mpc}^{-3}$ at $z=5$, and the bias ranges from $2.5$ at $z=2$ to $7$ at $z=5$), so we use them as is.

Fig.~\ref{fig:spec_templates} shows that the corresponding kSZ template can recover roughly 15\% of kSZ power at $\ell\approx 1000$ and less than 10\% at $1500\lesssim\ell\lesssim 14000$. As with the Roman forecast discussed above, the minimum at $\ell\lesssim 4000$ is due to the transition between the two-halo and one-halo regimes, where the performance of the halo model framework is known to be particularly poor in general~\cite{Mead:2015yca,Mead:2020qdk}. The high linear bias of LBGs enhances the template's performance at lower $\ell$ (where the two-halo term is more important) compared to other samples, while the performance at higher $\ell$ (into the pure one-halo regime) also shows promise.

\subsubsection{Combination: BOSS + DESI + Roman + MegaMapper}

The solid black line in the lower panel of Fig.~\ref{fig:spec_templates} shows the kSZ template that results in combining the templates from BOSS, DESI, Roman, and MegaMapper, using only the dominant survey at a given redshift rather than optimally combining templates at redshifts where surveys overlap. (For these forecasts, only a small loss of information is incurred by using this sub-optimal combination.) Note that we assume identical sky coverage of each survey when forming this combination, which will not be true in practice. With this caveat, we find that even this combination can recover roughly 35\% of the late-time kSZ power at $\ell\sim 1000$, between 20 and 25\% for $2000\lesssim\ell\lesssim 6000$, and between 15 and 20\% at higher multipoles.

%----------------------
% 21cm intensity mapping
%----------------------
\subsection{\tcm intensity mapping}

Post-reionization large-scale structure can also be measured via \tcm emission from neutral hydrogen (e.g.~\cite{Bull:2014rha,CosmicVisions21cm:2018rfq,Liu:2019awk}). In this subsection, we consider the PUMA (Packed Ultra-wideband Mapping Array~\cite{PUMA:2019jwd,Castorina:2020zhz}) proposal for a next-generation \tcm intensity mapping project, intended to map LSS over $0.3<z<6$. We also comment on other \tcm projects below.

\begin{figure}[t]
\includegraphics[width=\columnwidth, trim= 0 20 0 0]{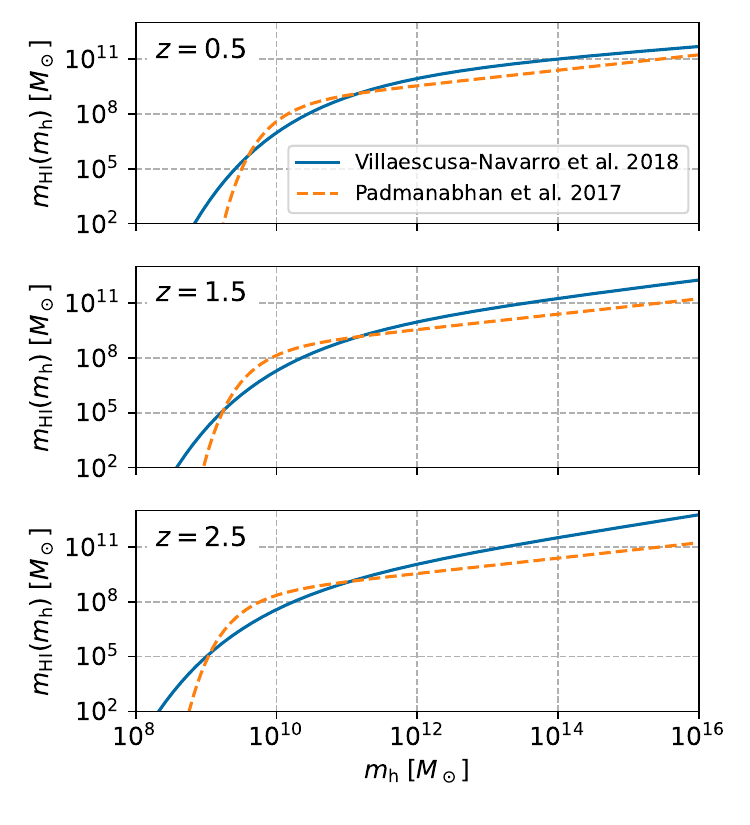}
\caption{%
The two HI mass-halo mass relations that we use in our forecasts for \tcm intensity mapping surveys: the relation from Ref.~\cite{Villaescusa-Navarro:2018vsg} ({\em blue solid lines}), which was fit to the IllustrisTNG simulation over $0<z<5$, and that from Ref.~\cite{Padmanabhan:2016fgy} ({\em orange dashed lines}), which was jointly fit to a variety of low- and high-redshift observations of neutral hydrogen. Halos with masses relevant to the kSZ effect ($\mh \sim 10^{13.5}M_\odot$) contain more HI if the former relation is true, such that a kSZ template constructed from a \tcm survey would have higher fidelity than if the latter relation is true.
}
\label{fig:mhi_mh}
\end{figure}

\begin{figure}[t]
\includegraphics[width=\columnwidth, trim= 0 20 0 0]{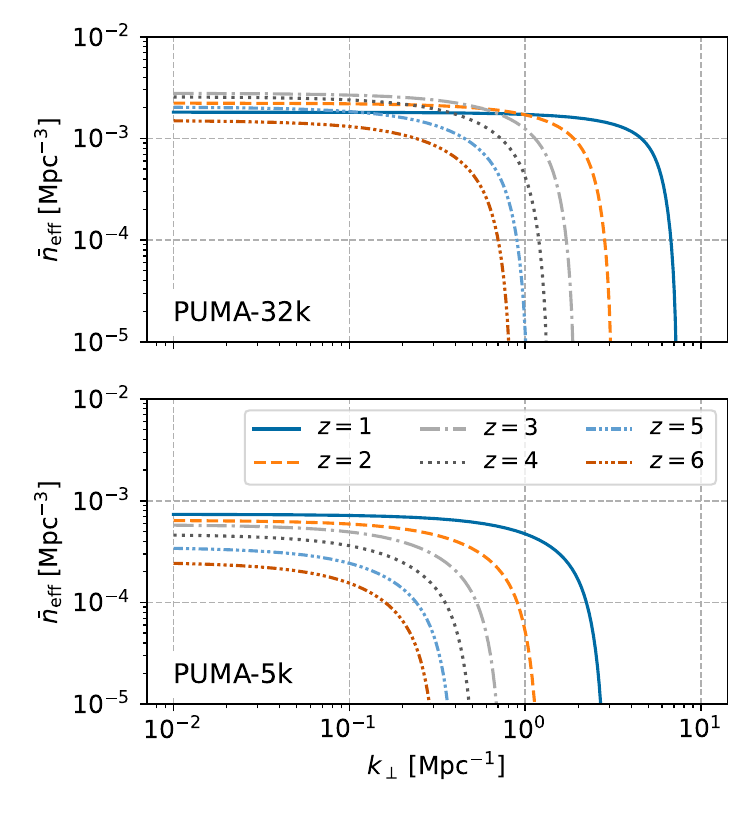}
\caption{%
For ease of comparison with our forecasts for spectroscopic galaxy surveys, we translate the expected noise in a \tcm survey, which has contributions from both the intrinsic shot noise of \tcm emitters and instrumental noise, into an ``effective shot noise" $\bar{n}_{\rm eff}$ which varies with redshift and transverse wavenumber $\kperp$. This decreases at high $\kperp$ due to the telescope's finite angular resolution. For PUMA-32k, the non-monotonicity with redshift at low $\kperp$ is due to the competing influence of intrinsic shot noise and instrumental noise, which scale oppositely with redshift, while for PUMA-5k, instrumental noise is dominant.
}
\label{fig:21cm_neff}
\end{figure}

Our forecasts follow a similar procedure to those for spectroscopic surveys, but with several distinctions (see Appendix~\ref{app:21cm} for more details). Our chosen ``HOD" model does not parameterize the occupation statistics of galaxies in halos, but instead makes use of a model for the relation between neutral hydrogen (HI) mass and halo mass in a given halo, following Refs.~\cite{Padmanabhan:2016fgy,Villaescusa-Navarro:2018vsg}. In particular, we perform forecasts that assume either the $m_{\rm HI}(\mh)$ relation fit to the IllustrisTNG simulations in Ref.~\cite{Villaescusa-Navarro:2018vsg}, or the parameterized function from Ref.~\cite{Padmanabhan:2016fgy} that was fit to observations of HI in resolved galaxies at $z\sim 0$, constraints on the mean HI density at $z\sim 1$, and properties of damped Lyman-$\alpha$ absorbers at $z\gtrsim 2$. We show these two options in Fig.~\ref{fig:mhi_mh}.
 We also assume the HI density profile within a halo is given by the fitting formula from Ref.~\cite{Villaescusa-Navarro:2018vsg}, again determined from IllustrisTNG; we show in Appendix~\ref{app:21cm-assumptions} that using the exponential HI profile from Ref.~\cite{Padmanabhan:2016fgy} changes our forecasts negligibly at low $\ell$ and by tens of percents at higher $\ell$, which is subdominant to the impact of the $m_{\rm HI}(\mh)$ relation. 
 
For the noise power of the HI auto spectrum, we add the (scale-dependent) instrumental noise power spectrum from Ref.~\cite{CosmicVisions21cm:2018rfq} to the intrinsic shot noise of \tcm emitters, computed using an updated version of the model from Ref.~\cite{Castorina:2016bfm}. To ease comparisons with the surveys from Sec.~\ref{sec:specsurveys}, we show an ``effective number density" $\bar{n}_{\rm eff}$, equal to the inverse of the \tcm instrumental+shot noise power spectrum, in Fig.~\ref{fig:21cm_neff}. We consider configurations of PUMA with 32000 or 5000 dishes (``PUMA-32k" and ``PUMA-5k" respectively, as considered in Refs.~\cite{CosmicVisions21cm:2018rfq,Castorina:2020zhz}).
The effective number density decreases at higher $\kperp$ due to the (redshift-dependent) finite angular resolution of the instrument. For PUMA-5k, the instrumental noise dominates over the shot noise, and the former increases with redshift due to the higher system temperature at lower frequencies, such that $\bar{n}_{\rm eff}$ decreases at higher redshift. For PUMA-32k, the instrumental and shot noise both contribute to the total noise, with the former increasing and the latter decreasing at higher redshifts, leading to the observed non-monotonic behavior of $\bar{n}_{\rm eff}$ with redshift.

Finally, Galactic and extragalactic foregrounds are expected to prevent certain Fourier modes of the HI distribution from being observed, and we account for this by assuming that modes with $k_\parallel < 0.03\invMpc$ or within a ``foreground wedge" (see e.g.~\cite{Morales:2012kf,Parsons:2012qh,Liu:2014bba}) defined by 3 times the width of a PUMA dish's primary beam (following Ref.~\cite{CosmicVisions21cm:2018rfq}) will be inaccessible. 
(We will comment on the impact of these assumptions below.)
Note that the loss of low-$\kpar$ modes in the \tcm maps does {\em not} imply that the kSZ template's angular power spectrum has negligible amplitude, because the template is constructed from a quadratic combination of \tcm maps; see Appendix~\ref{app:21cm-foregrounds} for more discussion on this point.

\begin{figure}[t]
\includegraphics[width=\columnwidth, trim= 0 15 0 0]{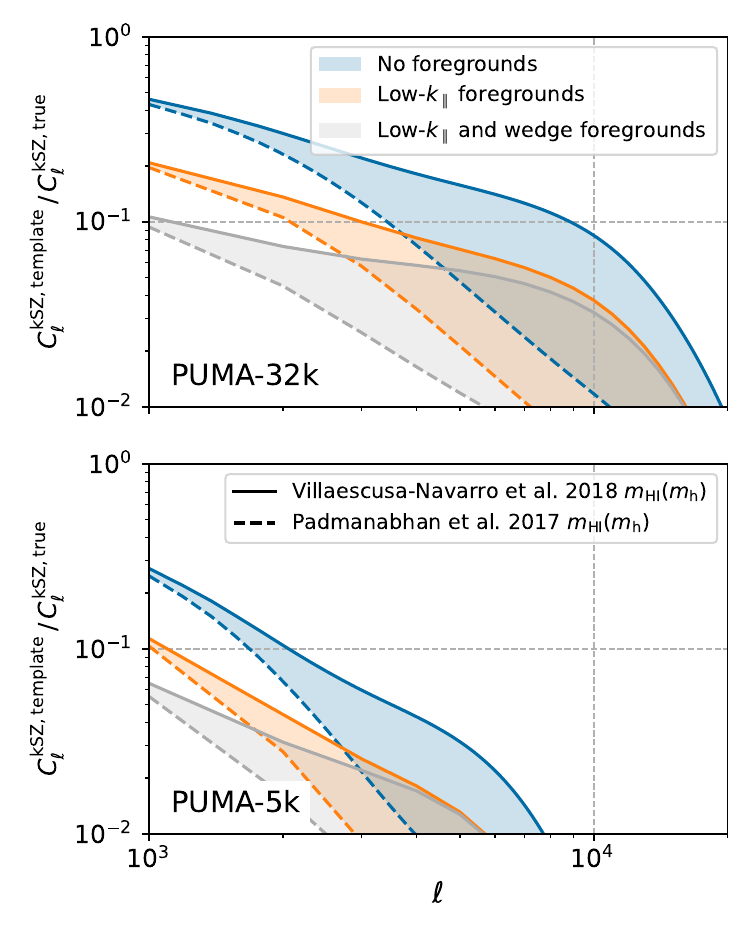}
\caption{%
Ratio of kSZ template power spectrum and true kSZ power spectrum for templates constructed from 32000-dish ({\em upper panel}) and 5000-dish ({\em lower panel}) versions of the proposed PUMA \tcm intensity mapping survey. Shaded bands denote uncertainty related to the choice of HI mass-halo mass relation in the forecasts, computed using either of the two options from Fig.~\ref{fig:mhi_mh}. In the absence of foregrounds ({\em blue}), the template performance at lower multipoles exceeds our forecasts for spectroscopic galaxy surveys, but when we include the loss of low-$\kpar$ modes due to spectrally-smooth foregrounds ({\em orange}) and leakage of these foreground into the ``foreground wedge" ({\em grey}), the template's performance is significantly degraded. Furthermore, as explained in the main text and Appendix~\ref{app:21cm-assumptions}, the amplitude of the full-foreground forecasts is sensitive to the precise assumptions made about foregrounds, so the grey bands should only be taken as a rough indication.
}
\label{fig:21cm_templates}
\end{figure}

We show the resulting kSZ template forecasts in Fig.~\ref{fig:21cm_templates}. In the complete absence of foregrounds, a template from PUMA-32k would be able to recover $\sim$45\% of the true kSZ power at $\ell\approx 1000$, $\sim$20\% at $\ell\approx2500$, and less at higher multipoles. This strong performance compared to the spectroscopic surveys from Sec.~\ref{sec:specsurveys} is due to the high sensitivity of PUMA-32k (equivalent to very low shot noise) shown in Fig.~\ref{fig:21cm_neff}. 
The results at higher multipoles are strongly dependent on the form of the $m_{\rm HI}(\mh)$ relation: the relation from Ref.~\cite{Villaescusa-Navarro:2018vsg} ascribes more HI to halo masses relevant for kSZ, and therefore results in a better kSZ template, than the relation from Ref.~\cite{Padmanabhan:2016fgy}. 

Accounting for the loss of low-$\kpar$ modes due to foregrounds, the fraction of recovered kSZ power drops to roughly 20\%, and also accounting for \tcm signal loss within the foreground wedge reduces the recovered kSZ power to less than 10\%.
Note that the detrimental effect of the foreground wedge can in principle be completely removed with sufficiently accurate characterization of the \tcm telescope and careful analysis (e.g.~\cite{Ghosh:2017woo}), or via machine learning methods~\cite{Gagnon-Hartman:2021erd}. Also, low-$\kpar$ modes can in principle be recovered with various reconstruction techniques (e.g.~\cite{Zhu:2016esh,Modi:2019hnu,Darwish:2020prn}). However, modes recovered in this way will have different noise properties than  those assumed in our forecasts, so we leave it to future work to assess their impact on a \tcm-based kSZ template.

As expected, PUMA-5k fares worse than PUMA-32k, due to its lower sensitivity and worse angular resolution (which prevents it from accessing higher-$\ell$ modes). At $\ell\approx 1000$, PUMA-5k could recover 25\% of the kSZ power in the absence of foregrounds, 10\% if low-$\kpar$ modes are lost, and 6\% if wedge modes are also lost. The performance of the kSZ template also degrades more quickly with $\ell$ than for PUMA-32k.

We have carried out forecasts for the HIRAX~\cite{Crichton:2021hlc} and CHORD~\cite{Vanderlinde:2019tjt} instruments, but found that neither could recover more than 5\% of the true kSZ power even in the absence of foregrounds, so we have not shown them in Fig.~\ref{fig:21cm_templates}.

In these forecasts, we have followed Ref.~\cite{CosmicVisions21cm:2018rfq} in our assumptions about \tcm foreground cleaning, but we note that different assumptions can lead to significant variations in the grey curves in Fig.~\ref{fig:21cm_templates}. In Appendix~\ref{app:21cm-assumptions}, we explore different choices for the minimum accessible $\kpar$ and the extent of the foreground wedge, and we find that the performance of the ensuing kSZ template can vary by as much as a factor of 10 (at low $\ell$) or 5 (at higher~$\ell$), mostly in the downward direction (in no case do the analogs of the grey curves exceed the orange curves plotted in Fig.~\ref{fig:21cm_templates}). The reader should thus bear in mind that the grey band shown in Fig.~\ref{fig:21cm_templates} is only roughly indicative of the template performance with a full treatment of foregrounds.

A further source of uncertainty in these forecasts is the stochasticity of the distribution of neutral hydrogen, which we have assumed is solely attributable to the discreteness of \tcm-emitting objects. Ref.~\cite{Obuljen:2022cjo} has recently used hydrodynamical simulations to show that the stochasticity of the HI density can far exceed the Poissonian approximation at $z=0$ but is of the same order as this approximation at $z=1$. We leave it to future work to incorporate more accurate estimates of HI stochasticity into our forecasts.

%--------------------------------------------------------------------------------------
% Applications
%--------------------------------------------------------------------------------------
\section{Applications}
\label{sec:applications}

\begin{figure*}[t]
\includegraphics[width=2\columnwidth, trim= 0 15 0 0]{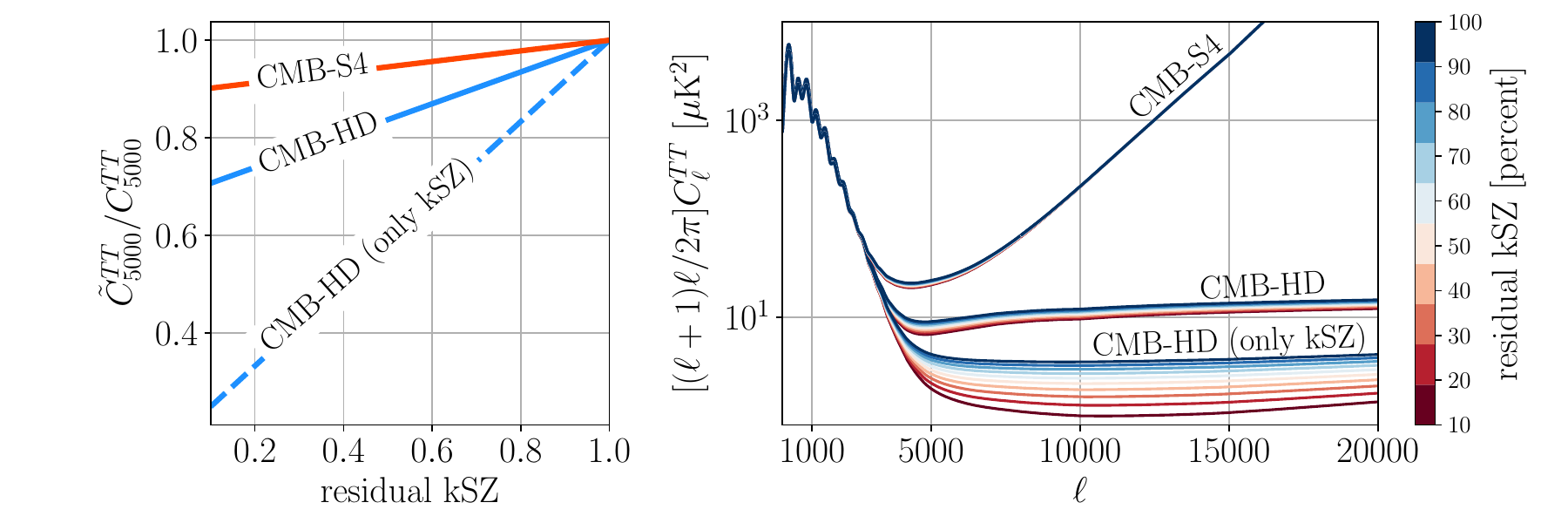}
\caption{Reducing the CMB variance via de-kSZing. {\em Left panel}: Reduction of the CMB temperature variance
 at multipole $\ell=5000$ for the CMB-S4 and CMB-HD surveys as a function of fractional residual kSZ power spectrum amplitude after de-kSZing. {The CMB variance is reduced by $5\%$ ($10\%$) for $50\%$ ($90\%$) kSZ removal for CMB-S4 and by $20\%$ ($60\%$) for 
$60\%$
($90\%$) kSZ removal for CMB-HD.} {\em Right panel}: The CMB variance 
as a function of~$\ell$
for varying levels of residual kSZ power spectrum amplitude from $10\%$ to $100\%$. The reduction of the CMB variance
is 
marginal  for CMB-S4, while future experiments such as CMB-HD can potentially significantly benefit from de-kSZing. Our forecasts for lines labelled CMB-S4 and CMB-HD include CMB foregrounds from CIB and tSZ after ILC-cleaning, as well as the reionization kSZ~\citep[e.g.][]{Park:2013mv}, which is not removed. {Forecasts for lines labelled ``CMB-HD (only kSZ)" include only the late-time and reionization kSZ as foregrounds for CMB-HD,}
and indicate that the benefit of de-kSZing would be much greater if frequency-dependent foregrounds could be cleaned much better than currently available ILC techniques.
}
\label{fig:residual_kSZ}
\end{figure*}

The kSZ effect constitutes the dominant contribution to the CMB black-body anisotropies on small scales ($\ell\gtrsim4000$) and, for the next-generation Stage-3 and Stage-4 CMB experiments~\citep{SimonsObservatory:2018koc,SimonsObservatory:2019qwx,CMB-S4:2016ple,Abazajian:2019eic}, is expected to be only within a factor $\sim2$ smaller than the frequency-dependent CMB foregrounds, the cosmic infrared background (CIB) and the thermal SZ (tSZ) effects, after standard harmonic-space internal linear combination (ILC) cleaning~\citep{Tegmark:2003ve}. For futuristic lower-noise CMB experiments like CMB-HD, the kSZ effect will likely dominate the CMB signal due to better  cleaning of frequency-dependent foregrounds~\citep{CMB-HD:2022bsz}. The removal of the kSZ effect via de-kSZing discussed in this paper hence may allow a significant reduction of the observed small-scale CMB variance.

In Fig.~\ref{fig:residual_kSZ}, we demonstrate the reduction of the total ILC-cleaned CMB variance by removing the (late-time) kSZ contribution. We describe our implementation of the ILC and the CMB forecasts in Appendix~\ref{sec:CMB_forecasts}. The left panel demonstrates the fractional reduction of the CMB variance at $\ell=5000$ for the CMB-S4 and CMB-HD surveys as a function of the fractional kSZ power spectrum amplitude that remains after de- kSZing (assuming that a scale-independent fraction of the amplitude is removed). The right panel demonstrates the same reduction for a range of CMB multipoles satisfying  
$\ell\in[1,20000]$.
In particular, for futuristic surveys such as CMB-HD, de-kSZing can lead to a 
$\sim$20\% reduction of the CMB variance if $\sim$60\%
of the kSZ signal can be removed.
We also show forecasts for CMB-HD that omit frequency-dependent foregrounds, as a proxy for a case where these foregrounds can be cleaned extremely efficiently; in this case, de-kSZing allows for a substantial reduction in temperature variance, particularly at higher multipoles.

In this section, we briefly highlight various ways in which de-kSZing may improve cosmological inference, leaving a more detailed analysis to future work. We envision a de-kSZing procedure in which a kSZ template $\hTkSZ$ is subtracted from an observed temperature map. Based on the results in Sec.~\ref{sec:forecasts}, we approximate this procedure as removing a scale-independent fraction of kSZ power, and examine the consequences of such a reduction. We find that significant gains in cosmological inference generally require de-kSZing to perform better than the detailed forecasts from Sec.~\ref{sec:forecasts}, and we take this as motivation to explore improvements to the de-kSZing formalism used in this paper.

\subsection{Improving parameter constraints from the CMB power spectrum}
\label{sec:dampingtail}

\begin{figure}[t]
\includegraphics[width=1\columnwidth, trim= 0 10 0 0]{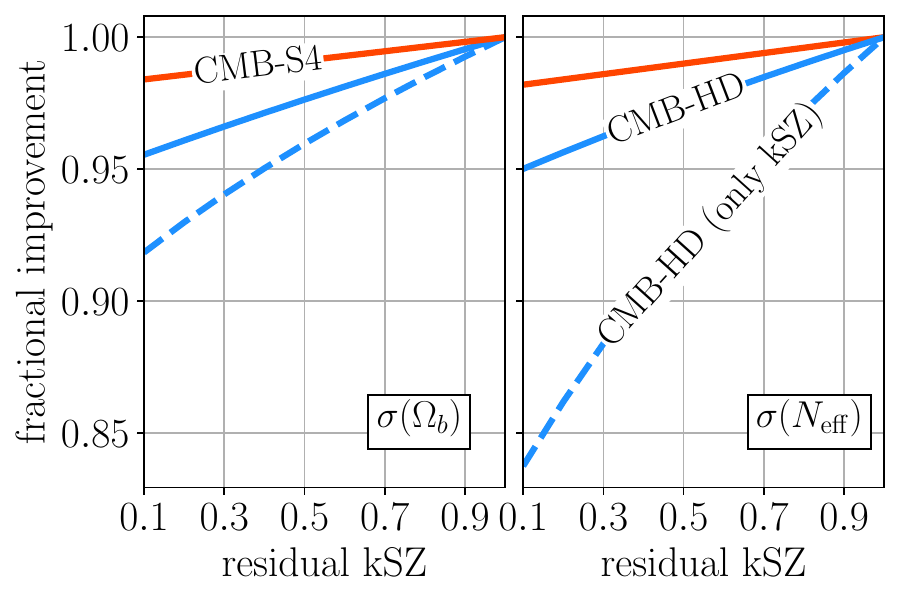}
\caption{{Improvement from de-kSZing on measurement errors on two example cosmological parameters: baryon density $\Omega_b$ and number of free-streaming species $N_{\rm eff}$. We find $50\%$ de-kSZing with CMB-S4 (\textit{solid orange}) may improve errors on these parameters by around a percent, while CMB-HD (\textit{solid blue}) can obtain 3 to $5\%$ improvement for similar de-kSZing efficiency. We also demonstrate the power de-kSZing in the absence of frequency-dependent foregrounds with the dashed blue curves.}}
\label{fig:imp_error}
\end{figure}

The small-scale primary CMB anisotropies are exponentially suppressed due
to Silk damping. The angular scale of this damping is determined by the expansion rate of the Universe and the free electron density before recombination, which are in turn sensitive to cosmological parameters such as the number of free-streaming species $N_{\rm eff}$, the baryon density $\Omega_b$, and the Helium abundance $Y_{\rm p}$. De-kSZing can improve the measurement precision of the CMB damping tail, in principle leading to improved parameter constraints. Unfortunately, for the cases we have examined{, including polarization information}, these improvements turn out to be quite mild: even assuming 90\% de-kSZing efficiency, we find that the uncertainty on determinations of $N_{\rm eff}$ and $\Omega_b$ decreases by no more than 10\% (see Fig.~\ref{fig:imp_error}).

For our forecasts in Fig.~\ref{fig:imp_error}, we used the publicly available forecasting software \texttt{FisherLens}\footnote{\url{https://github.com/ctrendafilova/FisherLens}}~\citep{Hotinli:2021umk}, and considered a cosmological model with parameters $\{\Omega_c h^2,\Omega_b h^2,\theta_s,\tau,A_s,n_s,N_{\rm eff}\}$ with fiducial values set to match the parameters determined by ~\textit{Planck}~\citep{Planck:2018lbu}. Here, $\Omega_c h^2$ is the physical cold dark matter density, $\Omega_b h^2$ is the physical baryon density, $\theta_s$ is the angle subtended by the acoustic scale, $\tau$ is the Thomson optical depth to recombination, $A_s$ is the primordial scalar fluctuation amplitude, and $n_s$ is the primordial scalar fluctuation slope. We define the information matrix with elements given by
\begin{equation}
    F_{ij} = \sum\limits_{\ell_1, \ell_2} \ \sum\limits_{W X Y Z} 
    \frac{\partial C_{\ell_1}^{XY}}{\partial \lambda^i} 
    \left[ \mathrm{Cov}_{\ell_1\ell_2}^{XY,WZ} \right]^{-1}
    \frac{\partial C_{\ell_2}^{WZ}}{\partial \lambda^j} \, .
    \label{eq:FisherMatrix}
\end{equation}
where $\lambda_i$ are the cosmological parameters. We include lensed $TT$, $TE$, $EE$, and $dd$ spectra, where $C_\ell^{dd} = \ell(\ell+1)C_\ell^{\phi\phi}$ is the lensing deflection spectrum. {We set the range of multipoles considered in our analysis to $\ell\in[30,10^4]$ and describe our modelling of the foregrounds and noise in Appendix~\ref{sec:CMB_forecasts}.} {The covariances include the lensing-induced non-Gaussian contributions as introduced in Ref.~\citep{Hotinli:2021umk}, which we calculate using \texttt{FisherLens}.
We set the sky fraction as $f_{\rm sky}=0.5$.} We also included a prior on $\tau$ with $\sigma_\tau = 0.007$, similar to what has been achieved by \textit{Planck}~\cite{Pagano:2019tci}.

Another observable of the CMB power spectrum is the kSZ effect from reionization.  In this paper we focussed our attention on the portion of the kSZ effect that is caused by the ionized gas in moving massive halos in the  relatively recent universe, at $z \lesssim 2$, but  
CMB maps also contain the signatures of the kSZ effect during reionization at $6 \lesssim z \lesssim 12$, originating mainly from the motions of ionized bubbles of the intergalactic medium as reionization progresses. The power spectrum of this effect has both the same spectral dependence (as a function of electromagnetic frequency) and a very similar power spectrum shape (as a function of $\ell)$ as that of the late-time kSZ. Its amplitude is also expected to be comparable. If a significant fraction of the late-time kSZ could be removed using the methods that we have explored, this could in principle help isolate the earlier effect and thereby reduce the resulting uncertainty on the duration of reionization, which is the property to which the kSZ power spectrum is most sensitive \citep[e.g.,][]{Gruzinov:1998un,Mesinger:2011aa,Zahn:2011vp}.  This would be achieved both by removing part of the highly degenerate signal power spectrum from low redshift, and, less significantly, by reducing the power spectrum variance.  Finally, removing some of  the late-time kSZ could help the search for signatures of reionization kSZ by using the four-point function \citep{Smith:2016lnt}.  Although one of the main advantages of this method is in helping to distinguish the late-time and reionization signals, reducing the late-time four-point signal by de-kSZing would only help in the use of this four-point method to study reionization in a new way.

\subsection{Measurement of other CMB secondaries}
\label{sec:secondaries}

\subsubsection{CMB lensing}
\label{sec:cmblensing_noise}

\begin{figure*}[t]\includegraphics[width=1.55\columnwidth, trim= 0 30 0 0]{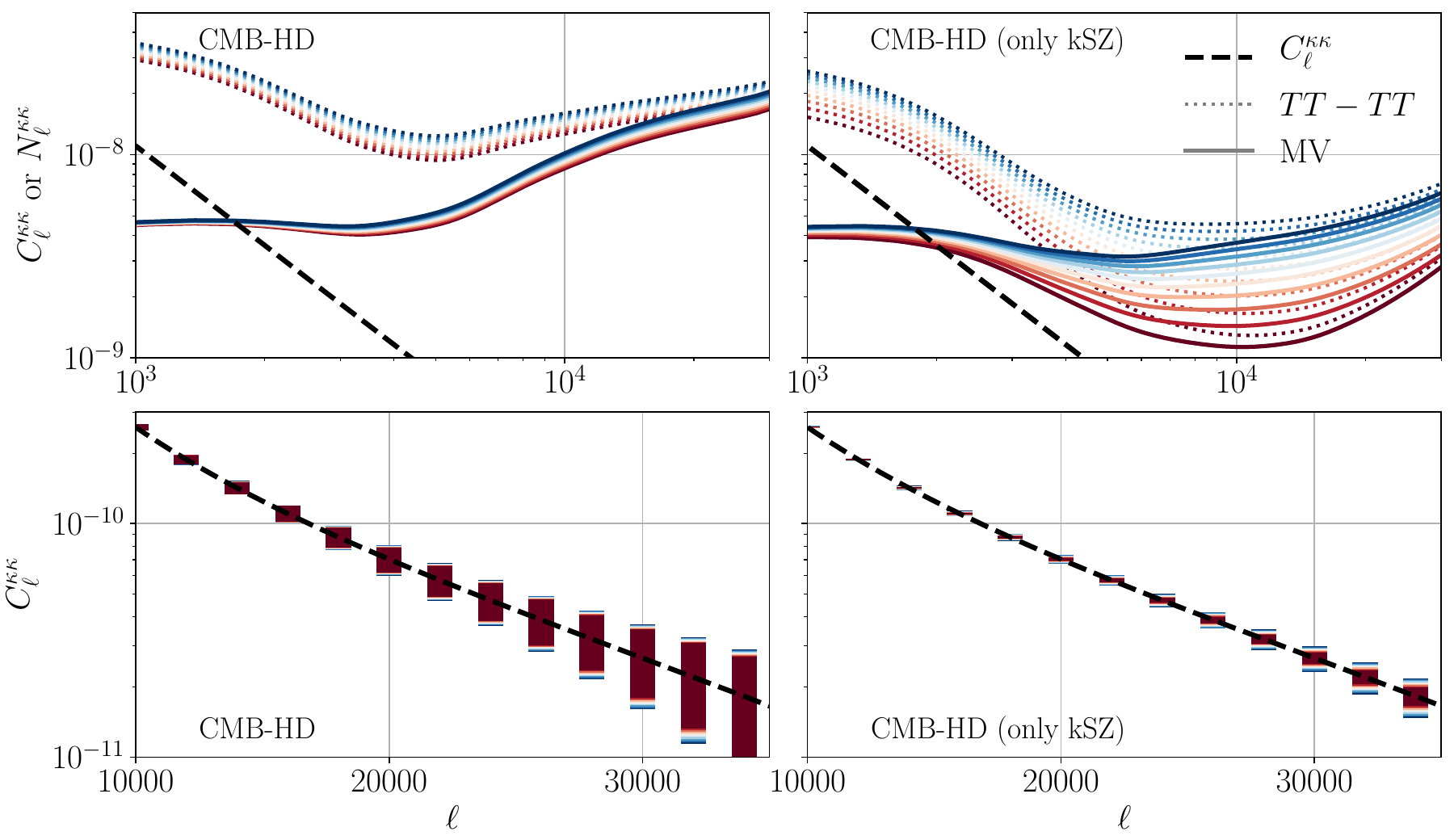} \includegraphics[width=0.15\columnwidth, trim= 0 -110 0 0]{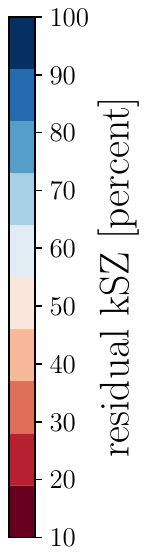}
\caption{%
{\textit{Upper panels:} Improvement of the lensing-reconstruction $TT$-$TT$ quadratic-estimator noise ({\em dotted curves}) together with the improvement on the minimum-variance noise from combining all temperature and polarization quadratic-estimators ({\em solid curves}). \textit{Lower panels:} Lensing convergence power spectra  ({\em solid black}). The error bars shown are from the diagonal terms of the minimum-variance lensing reconstruction noise. We use the \texttt{class$\_$delens} software~\citep{Hotinli:2020ntd} and consider a CMB-HD-like survey throughout. \textit{Left panels:} We include all significant foregrounds (CIB and tSZ, after ILC-cleaning, as well as the reionization kSZ) to demonstrate the power of de-kSZing in the {presence} of frequency dependent foregrounds. \textit{Right panels:} We include only the kSZ foreground (both late-time and reionization). De-kSZing improves the $TT$-$TT$ lensing reconstruction noise at all scales, while the minimum-variance noise is only improved at smaller scales where the $TT$-$TT$ estimator contributes significantly to the minimum-variance lensing reconstruction.}}
\label{fig:imp_lensing}
\end{figure*}

Reducing the CMB temperature variance induced by kSZ also improves lensing reconstruction. In Figure~\ref{fig:imp_lensing}, we demonstrate the reduction of the $TT$-$TT$ lensing quadratic-estimator noise ({\em upper panels, dotted curves}) together with the improvement on the minimum-variance noise from combining all lensing quadratic estimators ({\em solid curves}). Here, we use the \texttt{class$\_$delens} software~\citep{Hotinli:2020ntd} and assume a CMB-HD-like survey including CIB and tSZ foregrounds after ILC-cleaning, as well as the reionization kSZ. 
The $TT$-$TT$ lensing quadratic-estimator reconstruction noise is given by~\citep{Okamoto:2003zw}
\begin{equation}
    \label{eq:normalization}
    N^{TT{\rm -}TT}_{\ell} = (2\ell+1)\left[\sum\limits_{\ell_1\ell_2}\frac{|f^{TT}_{\ell_1\ell\ell_2}|^2}{2\tilde{C}_{\ell_1}^{TT}\tilde{C}_{\ell_2}^{TT}}\right]^{-1}\,,
\end{equation} 
where $f_{\ell_1\ell\ell_2}^{TT}$ is the optimal filter for the full-sky CMB lensing $TT$-$TT$ quadratic estimator~\citep{Okamoto:2003zw} and $\tilde{C}_\ell^{TT}$ is the observed CMB temperature spectrum. Performing lensing reconstruction on CMB maps after de-kSZing the CMB temperature would hence reduce the reconstruction noise, increasing the fidelity of lensing measurements. 

Temperature will be the dominant lensing channel for pre-S4 experiments, such as the Simons Observatory~\cite{SimonsObservatory:2018koc,SimonsObservatory:2019qwx}.
At CMB-S4 noise levels, however, the benefit of de-kSZing for reconstructing large-scale lensing modes will be much less, due to the sub-dominant contribution of the $TT$-$TT$ estimator to the minimum variance lensing estimator,
\be
N_\ell^{\rm mv}=\frac{1}{\sum_{\alpha\beta}(\boldsymbol{N}_\ell^{-1})^{\alpha\beta}}\,,
\ee
where $\boldsymbol{N}$ is the covariance of all quadratic estimators from temperature and polarization maps. In particular, on scales $\ell\lesssim10^4$, the minimum-variance lensing reconstruction is dominated by the $EB$-$EB$ quadratic estimator~\citep{Okamoto:2003zw}, which is not improved by de-kSZing. On the other hand, when reconstructing the 
smaller-scale ($\ell\gtrsim10^4$)
lensing modes, the $TT$-$TT$ estimator plays a more significant role, suggesting that de-kSZing can be important for improving small-scale CMB lensing reconstruction in the future.\footnote{Note also that de-kSZing can potentially reduce the kSZ-induced biases on the lensing reconstruction calculated in~Refs~\citep{Ferraro:2017fac,Darwish:2021ycf,Sailer:2020lal}, and the corresponding bias on de-lensed B-modes discussed in Ref.~\cite{BaleatoLizancos:2022vvr}.} We find that for a CMB-HD-like survey, the $TT$-$TT$ lensing quadratic-estimator noise decreases by {$\sim$15\%} on small scales if $\sim$30\% of the kSZ signal can be removed.

Small-scale CMB lensing reconstruction has applications including cosmological parameter inference, distinguishing between different dark-matter
models~\cite{Nguyen:2017zqu}, constraining high-redshift astrophysics, and validating galaxy weak-lensing
shear measurements~\citep{Hadzhiyska:2019cle}.

\subsubsection{Moving-lens effect}
\label{sec:movinglens}

\begin{figure}[t]
\includegraphics[width=1\columnwidth, trim= 0 30 0 0]{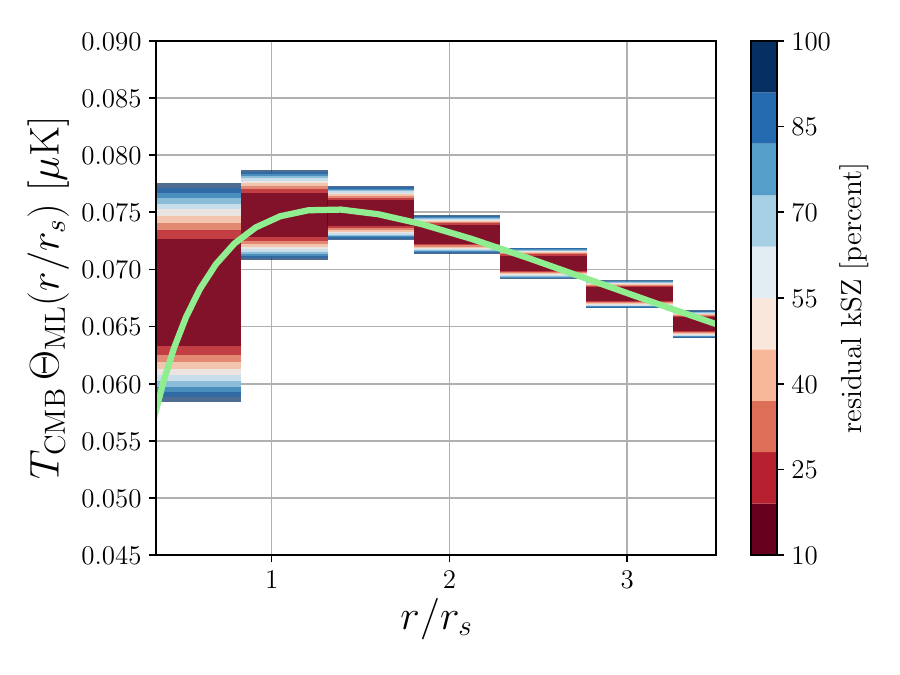}
\caption{Improvement on the measurement accuracy of the moving-lens profile. The error bars are scaled to match the net uncertainty anticipated from measurements of {$5 \times 10^4$} halos of the same mass ($M=10^{14}M_\odot$) at redshift $z=1$ using CMB-HD {(considering only the kSZ foreground)}. Different bins in $x=r/r_s$ are calculated by filtering the CMB around halos in radial bins with radial distance  $x$ from the halo center with width $\Delta x\simeq0.5$. The improvement in the measurement error of the halo profile at a given $x$-bin can be seen as a function of the residual kSZ (in percentage) in the CMB maps after de-kSZing.  
The degrading effect of different foregrounds on the detection and characterisation of the moving-lens effect could depend on the statistics used in the analysis, which may determine the benefit of de-kSZing in practice.
}
\label{fig:CMB_movinglens}
\end{figure}

Reducing the CMB variance on small scales can improve the measurement precision of cluster and halo profiles from several CMB secondaries. In Fig.~\ref{fig:CMB_movinglens}, we consider the moving-lens effect~\citep{Hotinli:2018yyc,Hotinli:2020ntd,Hotinli:2021hih} as a concrete example. In the moving-lens effect, a halo with peculiar motion transverse to the line of sight creates a small-scale dipolar temperature anisotropy centered on the halo's location on the sky. If we use $\vecr_\perp$ to denote the transverse proper distance from the halo center, and define $\vx_\perp \equiv \vecr_\perp / r_s$ where $r_s$ is the halo scale radius, the moving lens signal in the dimensionless CMB temperature $\Theta \equiv \Delta T/ T_{\rm CMB}$ is
\beq
\label{eq:ThetaML}
\Theta_{\rm ML}(\vx_\perp) = - a_0 \vv_{\rm b,\perp}\cdot\boldsymbol{\mathcal{M}}(\vx_\perp)\, ,
\eeq
where $\vv_{\rm b,\perp}$ is the halo's transverse velocity. The prefactor $a_0$ and moving-lens profile $\boldsymbol{\mathcal{M}}(\vx_\perp)$ depend on the halo density profile; assuming an NFW profile, they evaluate to~\citep{Hotinli:2020ntd}
\begin{align}
a_0 &\equiv  \frac{16\pi G\rho_s r_s^2}{c^3}\ , \\
\boldsymbol{\mathcal{M}}(\vx_\perp) 
	&\equiv \frac{\vx_\perp}{2x_\perp^2}
	\!\left[\Big|\frac{2\mathrm{sec}^{-1}(x_\perp)}{\sqrt{x_\perp^2-1}}\Big|
	+\ln\left(\frac{x_\perp^2}{4}\right)\right]\ ,
\end{align}
where
\beq
\rho_s = \frac{M}{4\pi r_s^3} 
	\left[-\frac{r_\mathrm{vir}}{r_s+r_\mathrm{vir}}-\ln\left(\frac{r_s+r_\mathrm{vir}}{r_s}\right)\right]\, .
	\label{eq:rhos}
\eeq

Here, we forecast the measurement precision on the angle-averaged moving-lens signal of a $10^{14}$ solar-mass halo, in $x_\perp$ bins with width $\Delta x_\perp = 0.5$ and taking the fiducial density profile to be NFW. 
For this forecast, we use the optimal matched filter derived in Ref.~\cite{Hotinli:2020ntd}, applied separately to each $x_\perp$ bin to enable a measurement of the radially-binned moving lens signal (see Appendix~\ref{app:movinglens} for details).

De-kSZing reduces the observed CMB variance $\tilde{C}_\ell^{TT}$ on scales where the moving-lens reconstruction noise gets its dominant contribution. The uncertainty bands in Fig.~\ref{fig:CMB_movinglens} are scaled to match the net uncertainty anticipated from measurements of {$5 \times 10^4$} halos of the same mass at redshift $z=1$, roughly representative of the number of cluster-size halos probed by recent and upcoming imaging surveys~\citep{Gao:2021tlp}.
The improvement in the uncertainty in a given bin can be seen as a function of the residual kSZ power remaining in the CMB maps after de-kSZing. We find that $\sim50\%$ removal of the kSZ effect can lead {up to} $\sim30\%$ improvement of the measurement of the moving-lens signal at a given distance from the halo center, { depending on the survey systematics and other foregrounds.} 

Unlike kSZ or other Sunyaev Zel'dovich effects, the moving-lens effect is purely gravitational and can be used to probe quantities such as the growth rate of cosmological structure~\citep{Hotinli:2021hih}, for example, without depending on the modelling of the electron gas. De-kSZing would not only boost the detection significance of the moving-lens effect but also potentially increase the prospects to perform cosmological and astrophysical inference with this promising observable.

%--------------------------------------------------------------------------------------
% Conclusion
%--------------------------------------------------------------------------------------
\section{Conclusion}
\label{sec:conclusion}

In this paper, we have explored the prospects for ``de-kSZing" the cosmic microwave background: constructing a template for the late-time kinetic Sunyaev-Zeldovich effect using a galaxy or \tcm survey, and subtracting this template from observed CMB temperature maps. The template is constructed following the procedure from Refs.~\cite{Ho:2009iw,Shao:2010md,Smith:2018bpn}: separate templates for the small-scale electron density and large-scale velocity field are formed by Wiener-filtering the observed galaxy density; these templates are combined into an estimate of the electron momentum field; and this combination is projected along the line of sight, using the theoretical kSZ redshift kernel, to form an estimate of kSZ-induced temperature fluctuations in the CMB. We have identified the properties of the input galaxy survey that most strongly determine the fidelity of the associated kSZ template: redshift and halo mass coverage (Sec.~\ref{sec:intuition-ranges}) and galaxy number density (Sec.~\ref{sec:templates-ev}). 

We have also carried out detailed forecasts for a number of recent, upcoming, or proposed surveys, assessing the ability of each survey to provide a template that could be used for de-kSZing. Since redshift uncertainties in photometric surveys will significantly degrade the usefulness of such a template (Sec.~\ref{sec:redshifts}), we have focused on spectroscopic and \tcm surveys in our forecasts.

The combination of BOSS and several of DESI's planned samples could in principle yield a template that captures 10-20\% of the total kSZ power, while templates derived from the H$\alpha$ emission-line galaxy sample from the Roman Space Telescope or the Lyman-break galaxy sample from the proposed MegaMapper telescope can capture no more than 10\% of the kSZ power at $\ell\gtrsim 1500$ (Fig.~\ref{fig:spec_templates}). Once foreground contamination is accounted for, our forecasts indicate that the proposed PUMA \tcm intensity mapping survey can do no better than the combination of BOSS and DESI, although uncertainty in modelling of the \tcm signal on small scales, along with the achievable level of foreground cleaning, makes this conclusion rather uncertain (Fig.~\ref{fig:21cm_templates}). Table~\ref{tab:summary} provides representative numbers for the surveys we have considered.

\begin{table}
\begin{ruledtabular}
\begin{tabular}{llll}
{\bf Survey} & \multicolumn{3}{c}{\bf De-kSZing Efficiency (\%)} \\ 
& $\ell=1000$ & $\ell=4000$ & $\ell=8000$ \\
\hline
BOSS + DESI		&	15	&	17	&	8 \\
Roman (H$\alpha$)	&	7	&	$<$1	&	$<$1 \\
MegaMapper		&	13	&	4	&	7 \\
All of the above		&	35	&	21	&	15 \\
\hline
PUMA-32k		&	10	&	4	&	$<$1 \\
\end{tabular}
\end{ruledtabular}
\caption{%
Representative de-kSZing efficiencies for kSZ templates constructed from the surveys considered in this work. See Sec.~\ref{sec:forecasts} for discussions of important caveats associated with these numbers; in particular, the ``all of the above" line only applies to the overlapping sky footprints of the surveys it combines, and the PUMA-32k numbers are highly uncertain due to modelling uncertainties in the HI distribution at small scales and the achievable level of foreground cleaning in \tcm surveys. De-kSZing procedures that improve upon that presented in this work could potentially improve these efficiencies substantially.
}
\label{tab:summary}
\end{table}

However, we emphasize that we have only performed an initial exploration of a specific de-kSZing procedure in this work, and we expect that alternative procedures can yield significant improvements to the results we have presented here. With this in mind, we have highlighted several applications of the idea of de-kSZing, all of which rely on the associated reduction of the small-scale variance in CMB temperature maps: better recovery of cosmological information from the CMB damping tail (Sec.~\ref{sec:dampingtail}); lower noise in measurements of CMB lensing at small scales, which can be used to test models for dark matter (Sec.~\ref{sec:cmblensing_noise}); and more precise measurements of the moving-lens effect, which can be used to probe the cosmic growth rate and the properties of dark matter halos (Sec.~\ref{sec:movinglens}).

Motivated by these applications, there are several pathways towards an improved de-kSZing procedure that would be worth pursuing:
\begin{itemize}
%%%%
\item Different surveys (or combinations of surveys) could be used for the velocity and electron 
density templates, which would make better use of the properties of each survey. In particular, 
photometric redshift errors would be tolerable if a photometric survey was only used for the 
electron density template. 
As a concrete example, the LSST Y10 Gold lens sample is estimated\footnote{This 
estimate is based on a total angular number density of $48\,{\rm arcmin}^{-2}$ over 
$18000\,{\rm deg}^2$, and a galaxy redshift distribution of $dN/dz \propto z^2 {\rm exp}[ 
-(z/0.28)^{0.9} ]$~\cite{LSSTDarkEnergyScience:2018jkl}.} to have a number density of 
$4\times10^{-3}\,{\rm Mpc}^{-3}$ for $0.5<z<2$. Assuming that this sample is used for the 
electron template but that the velocity template comes from a sample with $\bar{n}_{\rm g} 
\sim 10^{-4}\,{\rm Mpc}^{-3}$, we infer from Figs.~\ref{fig:R_gv2}-\ref{fig:cl_ksz_ratio_beta1} 
that a de-kSZing efficiency of roughly $40\%$ at $\ell>3000$ may be achievable. More 
detailed forecasts will be needed to confirm this conclusion, however.
%%%%
\item
Extra weights could be applied to the input galaxy survey
to optimize the correlation of the kSZ template with the true kSZ fluctuations (in a statistical 
sense). For example, if halo mass estimates for each galaxy are available, different mass bins 
could be weighted differently (e.g.~\cite{Hamaus:2011dq,Munchmeyer:2018eey}) in order to 
better account for the mass- and redshift-dependence of the kSZ signal (recall 
Fig.~\ref{fig:cl_ksz_integrand}). The expected improvement would depend on the accuracy of 
the halo masses, which would determine the number of mass bins that could be used.
%%%%
\item Tomographic reconstruction of large-scale velocities using the kSZ 
effect~\cite{Terrana:2016xvc,Deutsch:2017ybc,Smith:2018bpn,Giri:2020pkk,Cayuso:2021ljq,Hotinli:2020csk, 
Hotinli:2019wdp,AnilKumar:2022flx,Hotinli:2022jna,Kumar:2022bly} could 
possibly be integrated into a de-kSZing procedure as a way to improve upon an external 
velocity template.
This would have the advantage of constructing the velocity template from small-scale modes 
of the galaxy density (along with small-scale CMB temperature modes), but will introduce 
higher-point biases and noise contributions that would need to be accounted for, since the 
velocity template will then be quadratic ($\sim \deltags T$) instead of linear ($\sim \deltags$). 
This procedure would be analogous to iterative delensing of CMB polarization~\cite{Smith:2010gu}.
%%%%
\item Rather than viewing de-kSZing as a subtraction procedure, it may be possible to use 
Bayesian (e.g.~\cite{Millea:2020cpw}) or machine learning (e.g.~\cite{Guzman:2021nfk,Tanimura:2022fde}) 
techniques to jointly estimate the statistics of the kSZ effect, the primary CMB, and/or other 
CMB secondaries. 
The former approach would allow for uncertainties in the procedure and data to be accounted 
for more systematically, while the latter may be able to exploit features of the kSZ signal (such as non-Gaussian information) that 
are not incorporated in our template construction. We note, however, that either approach 
would likely require significant computing power.
%%%%
\end{itemize}
We leave these possibilities to future work.

%--------------------------------------------------------------------------------------
% Acknowledgments
%--------------------------------------------------------------------------------------
\acknowledgments

We thank Neal Dalal, Matthew~C.~Johnson, Elisabeth Krause, Jessie Muir, Neelima Sehgal, Kendrick Smith, and Sihan Yuan for helpful discussions.  

Research at the Perimeter Institute is supported in part by the Government of Canada through the Department of Innovation, Science and Economic Development, and by the Province of Ontario through the Ministry of Colleges and Universities. SCH is supported by the Horizon Fellowship from Johns Hopkins University. SCH also acknowledges the support of a grant from the Simons Foundation at the Aspen Center for Physics.  This work was performed in part at the Aspen Center for Physics, which is supported by National Science Foundation grant PHY-1607611. {SCH would like to thank Imperial College High Performance Computing Service at Imperial College London (UK) for providing computational resources at various early stages of this project.}

\onecolumngrid
\appendix
%--------------------------------------------------------------------------------------
% SECTION: Shot noise contributions to kSZ template auto spectrum
%--------------------------------------------------------------------------------------
\section{Shot noise contributions to kSZ template auto spectrum} 
\label{app:shotnoise}

The kSZ template described in Sec.~\ref{sec:templates} is related to a product of filtered copies of an observed galaxy density contrast~$\deltag$, and therefore the autocorrelation of the template is sensitive to the four-point function of $\deltag$. (For simplicity, we neglect the impact of redshift-space distortions in this appendix, and therefore use $\deltag$ instead of $\deltags$.) This four-point function contains several contributions from the shot noise in $\deltag$; in Fourier space and for Poissonian shot noise, these contributions are given by (e.g.~\cite{Sugiyama:2019ike,Darwish:2020prn})
\beq
\left\la \deltag(\vk_1) \deltag(\vk_2) \deltag(\vk_3) \deltag(\vk_4) \right\ra_{\rm shot}
	= (2\pi)^3 \dirac(\vk_1+\vk_2+\vk_3+\vk_4)
	\lb S_{4} + S_{3,1} + S_{2,2} + S_{2,1,1} \rb
	\label{eq:deltag-4pt}
\eeq
where
\begin{align*}
S_4 &= \frac{1}{\bar{n}_{\rm g}^3}\ , \\
S_{3,1} &= \frac{1}{\bar{n}_{\rm g}^2} 
	\lb \Pgg(k_1) + \Pgg(k_2) + \Pgg(k_3) + \Pgg(k_4)  \rb\ , \\
S_{2,2} &= \frac{1}{\bar{n}_{\rm g}^2} 
	\lb \Pgg(|\vk_1+\vk_2|) + \Pgg(|\vk_1+\vk_3|) + \Pgg(|\vk_1+\vk_4|) \rb\ , \\
S_{2,1,1} &= \frac{1}{\bar{n}_{\rm g}} 
	\lb B_{\rm ggg}(\vk_1+\vk_2,\vk_3,\vk_4) + B_{\rm ggg}(\vk_1+\vk_3,\vk_2,\vk_4)
	+ B_{\rm ggg}(\vk_1+\vk_4,\vk_2,\vk_3) \right. \\
&\qquad\quad\left. +\, B_{\rm ggg}(\vk_2+\vk_3,\vk_1,\vk_4)
	+ B_{\rm ggg}(\vk_2+\vk_4,\vk_1,\vk_3) + B_{\rm ggg}(\vk_3+\vk_4,\vk_1,\vk_2) \rb
	\numberthis
	\label{eq:Sfuncs}
\end{align*}
and we have omitted the redshift-dependence and the arguments of the $S$ functions for brevity. Each contribution arises from a different subset of points taken to be at zero lag in the position-space correlation function: $S_4$ is when all four points are at the same spatial location; $S_{3,1}$ is when three points are at the same location and one is elsewhere; $S_{2,2}$ is when two pairs of points are each at the same location, but the pairs are not co-located; and $S_{2,1,1}$ is when two points are at the same location, while the other two points are each at different locations.

The power spectrum of the electron momentum template in Eq.~\eqref{eq:qhatr} evaluates to
\begin{align*}
\left\la \hqr(\vk_1) \hqr(\vk_2) \right\ra
	&= \left\la \int \frac{d^3\vk_1'}{(2\pi)^3} \eta(\vk_1') \epsilon(\vk_1-\vk_1') 
	\int \frac{d^3\vk_2'}{(2\pi)^3} \eta(\vk_2') \epsilon(\vk_2-\vk_2') \right\ra \\
&= 	(-1) \int \frac{d^3\vk_1'}{(2\pi)^3} \int \frac{d^3\vk_2'}{(2\pi)^3}  \mu_1' \mu_2' 
	\frac{\Pgv(k_1')}{\Pggtot(k_1')} \frac{\Pgv(k_2')}{\Pggtot(k_2')} 
	\frac{\Pge(|\vk_1-\vk_1'|)}{\Pggtot(|\vk_1-\vk_1'|)} 
	\frac{\Pge(|\vk_2-\vk_2'|)}{\Pggtot(|\vk_2-\vk_2'|)} \\
&\qquad\qquad\times 
	\left\la \deltag(\vk_1') \deltag(\vk_1-\vk_1') \deltag(\vk_2') \deltag(\vk_2'-\vk_2) \right\ra, 
	\numberthis
	\label{eq:qhatqhat1}
\end{align*}
and combined with Eqs.~\eqref{eq:deltag-4pt} and~\eqref{eq:Sfuncs}, this yields
\beq
P_{\hqr\hqr}(\vk_1) = (-1) \int \frac{d^3\vk_1'}{(2\pi)^3} \int \frac{d^3\vk_2'}{(2\pi)^3}  \mu_1' \mu_2' 
	\frac{\Pgv(k_1')}{\Pggtot(k_1')} \frac{\Pgv(k_2')}{\Pggtot(k_2')} 
	\frac{\Pge(|\vk_1-\vk_1'|)}{\Pggtot(|\vk_1-\vk_1'|)} 
	\frac{\Pge(|\vk_1+\vk_2'|)}{\Pggtot(|\vk_1+\vk_2'|)}
	\lb \hat{S}_4 + \hat{S}_{3,1} + \hat{S}_{2,2} + \hat{S}_{2,1,1} \rb\ ,
	\label{eq:Pqhatqhat}
\eeq
where
\begin{align*}
\hat{S}_4 &= \frac{1}{\bar{n}_{\rm g}^3}\ , \\
\hat{S}_{3,1} &= \frac{1}{\bar{n}_{\rm g}^2} 
	\lb \Pgg(k_1') + \Pgg(|\vk_1-\vk_1'|) + \Pgg(k_2') + \Pgg(|\vk_1+\vk_2'|)  \rb\ , \\
\hat{S}_{2,2} &= \frac{1}{\bar{n}_{\rm g}^2} 
	\lb \Pgg(k_1) + \Pgg(|\vk_1'+\vk_2'|) + \Pgg(|\vk_1'-\vk_1-\vk_2'|) \rb\ , \\
\hat{S}_{2,1,1} &= \frac{1}{\bar{n}_{\rm g}} 
	\lb 
	B_{\rm ggg}(\vk_1,\vk_2',-\vk_1-\vk_2') 
	+ B_{\rm ggg}(\vk_1'+\vk_2',\vk_1-\vk_1',-\vk_1-\vk_2')
	+ B_{\rm ggg}(\vk_1'-\vk_1-\vk_2',\vk_1-\vk_1',\vk_2') \right. \\
&\qquad\quad\left. 
	+\, B_{\rm ggg}(\vk_1-\vk_1'+\vk_2',\vk_1',-\vk_1-\vk_2')
	+ B_{\rm ggg}(-\vk_1'-\vk_2',\vk_1',\vk_2') 
	+ B_{\rm ggg}(-\vk_1,\vk_1',\vk_1-\vk_1') \rb\ .
	\numberthis
	\label{eq:Shatfuncs}
\end{align*}
In the squeezed limit ($k_1', k_2' \ll k_1$), many of these terms will integrate to zero in Eq.~\eqref{eq:Pqhatqhat}, because they have no dependence on either $\mu_1'$ or $\mu_2'$ but are integrated against $\mu_1' \mu_2'$. (Physically, this reflects the fact that the kSZ effect is sensitive to the line-of-sight component of the large-scale velocity, and the angular average of this is zero.) However, terms that involve the sum $\vk_1'+\vk_2'$ may contribute non-negligibly. In particular, such terms involve power at larger scales than one might naively expect, because it is possible to have $|\vk_1'+\vk_2'| \ll k_1', k_2'$ if $\vk_1'$ and $\vk_2'$ are similar in magnitude and anti-aligned.\footnote{In Ref.~\cite{Darwish:2020prn}, these terms were found to significantly affect the precision with which long-wavelength modes of $\deltag$ can be reconstructed with a quadratic estimator. \\}
% Ugly hack required to make footnote single-column

\begin{figure*}[t]
\includegraphics[width=\textwidth, trim={0 15 0 0}]{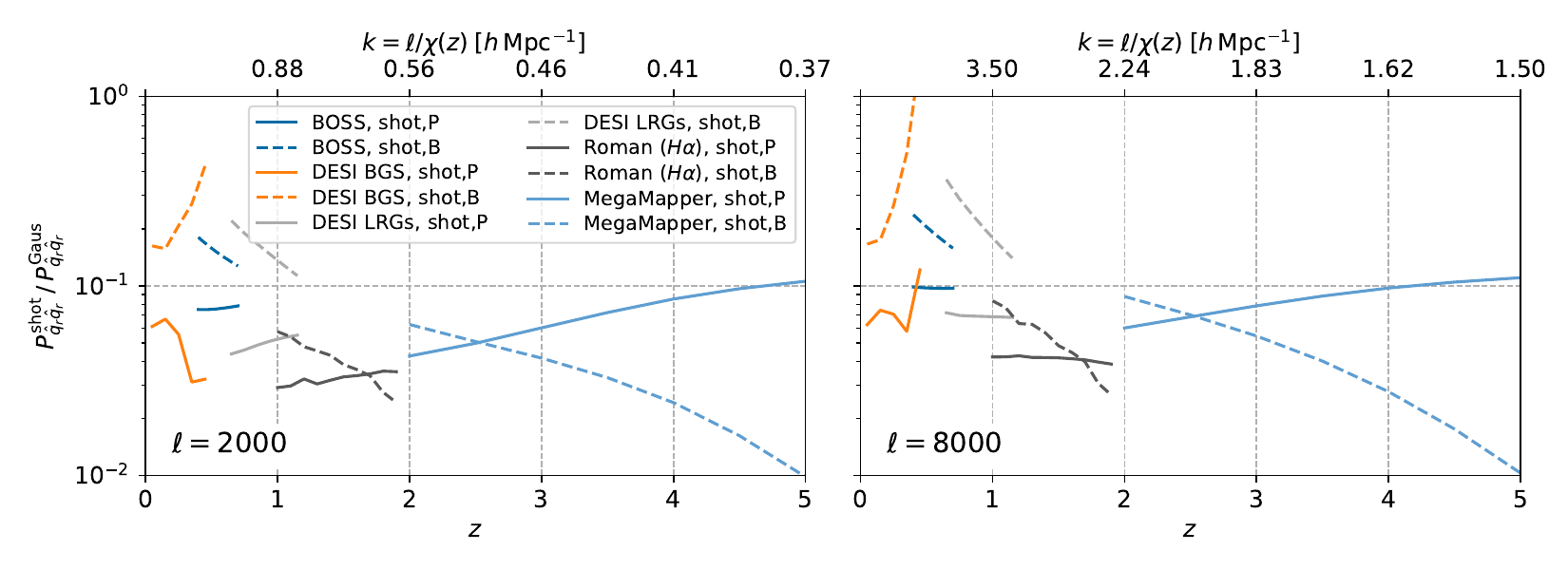}
\caption{%
The two dominant shot-noise contributions to the auto power spectrum of the electron momentum template $\hqr$, plotted as ratios to the Gaussian (i.e.\ reconstructed kSZ signal) contribution to $P_{\hqr\hqr}$. {\em Solid lines} denote the contribution that involves the galaxy power spectrum (Eq.~\ref{eq:Phqrhqr-shotP}), while {\em dashed lines} correspond to the contribution related to the galaxy bispectrum (Eq.~\ref{eq:Phqrhqr-shotB}). Different colors denote different spectroscopic surveys, described in Sec.~\ref{sec:specsurveys}. With the exception of DESI BGS, these shot-noise contributions are less than $\sim$25\% of the template power at $\ell=2000$ and less than $\sim$40\% at $\ell=8000$. While subdominant to the desired signal in the template, this shot-noise-induced power should be accounted for in future de-kSZing analyses.
}
\label{fig:shot_ratio_specsurveys}
\end{figure*}

In Fig.~\ref{fig:shot_ratio_specsurveys}, we evaluate the two dominant terms of this type in the squeezed limit, and compare them to the Gaussian contribution to $P_{\hqr\hqr}$. Specifically, we compute
\beq
P_{\hqr\hqr}^{{\rm shot},P}(k) = (-1) \int \frac{d^3\vk_1'}{(2\pi)^3} \int \frac{d^3\vk_2'}{(2\pi)^3}  \mu_1' \mu_2' 
	\frac{\Pgv(k_1')}{\Pggtot(k_1')} \frac{\Pgv(k_2')}{\Pggtot(k_2')} 
	\lp \frac{\Pge(k)}{\Pggtot(k)} \rp^2
	\frac{1}{\bar{n}_{\rm g}^2} 
	\Pgg(|\vk_1'+\vk_2'|)
	\label{eq:Phqrhqr-shotP}
\eeq
and
\beq
P_{\hqr\hqr}^{{\rm shot},B}(k) = (-1) \int \frac{d^3\vk_1'}{(2\pi)^3} \int \frac{d^3\vk_2'}{(2\pi)^3}  \mu_1' \mu_2' 
	\frac{\Pgv(k_1')}{\Pggtot(k_1')} \frac{\Pgv(k_2')}{\Pggtot(k_2')} 
	\lp \frac{\Pge(k)}{\Pggtot(k)} \rp^2
	\frac{1}{\bar{n}_{\rm g}} 
	B_{\rm ggg}(-\vk_1'-\vk_2',\vk_1',\vk_2')\ ,
	\label{eq:Phqrhqr-shotB}
\eeq
and compare with Eq.~\eqref{eq:Phqrqr}, ignoring redshift-space distortions:
\beq
P_{\hqr \qr}^{\rm Gaus}(k)
	=\frac{1}{6\pi^2}
	\lp \int d\kL\, \kL^2 \frac{\Pgv(\kL)^2}{\Pggtot(\kL)} \rp
	\frac{\Pge(k)^2}{\Pggtot(k)}\ .
\eeq
We evaluate the various power spectra using the halo model framework described in the main text and Appendix~\ref{app:halomodel}, for each of the spectroscopic surveys considered in Sec.~\ref{sec:specsurveys}.
We numerically compute Eqs.~\eqref{eq:Phqrhqr-shotP} and~\eqref{eq:Phqrhqr-shotB} using the Monte Carlo integration algorithm described in Ref.~\cite{Lepage:2020tgj}, as implemented in the \texttt{vegas} Python package~\cite{peter_lepage_2022_5893494}. 

For most of the surveys we consider, these shot noise contributions are below 25\% of the Gaussian contribution at $\ell=2000$ and below 40\% at $\ell=8000$. (A noteworthy exception is the ``shot,$B$" contribution for the DESI BGS sample, which strongly increases at the higher-redshift edge of the BGS sample, $z\approx 0.4$, at $\ell=8000$; this is caused by the decreasing number density of the BGS sample at this redshift, combined with the overall higher amplitude of the galaxy bispectrum in the BGS redshift range compared to the higher redshifts covered by the other surveys.) These numbers can be interpreted as upper bounds on the amount of power in a kSZ template that arises from shot noise in the galaxy sample instead of reconstructed kSZ signal. Future modelling related to applications of de-kSZing will need to account for this.

%--------------------------------------------------------------------------------------
% SECTION: Halo model
%--------------------------------------------------------------------------------------
\section{Details of halo model}
\label{app:halomodel}

In this appendix, we provide more details of the halo model approach we use to model quantities related to spectroscopic surveys. The galaxy power spectrum, electron power spectra, integrand of the kSZ angular power spectrum (see Eq.~\ref{eq:clksz-integrand}), and galaxy bispectrum are discussed in Sec.~\ref{sec:halomodel-gg}, \ref{sec:halomodel-e}, \ref{sec:clksz-integrand}, and \ref{sec:halomodel-ggg} respectively. For brevity, we have omitted redshift arguments in Sec.~\ref{sec:halomodel-gg}, \ref{sec:halomodel-e}, and \ref{sec:halomodel-ggg}.

\subsection{Galaxy power spectrum}
\label{sec:halomodel-gg}

Our formalism is based on Ref.~\cite{Smith:2018bpn}.
The galaxy power spectrum is a sum of two-halo, one-halo, and shot noise terms,
\beq
\Pgg(k, \mu) = \Pgg^{\rm 2h}(k, \mu) + \Pgg^{\rm 1h}(k) + \Pgg^{\rm shot}\ .
\eeq
As stated in the main text, we do not include Finger of God damping in our computations, and therefore the one-halo term has no $\mu$-dependence.\footnote{See Ref.~\cite{Schaan:2021gzb} for a formalism that consistently includes the Finger of God effect in a halo model framework.}
The two-halo term is given by 
\beq
\label{eq:Pgg-2h}
\Pgg^{\rm 2h}(k, \mu) = \lb b_{\rm g}(k) + f \mu^2 \rb^2 P_{\rm lin}(k)\ ,
\eeq
where $P_{\rm lin}$ is the linear matter power spectrum and the prefactor incorporates the leading effect of redshift space distortions at large scales. The linear bias $b_{\rm g}$ is given by
\beq
b_{\rm g}(k) = \frac{1}{\bar{n}_{\rm g}}
	\int d\mh\, n(\mh) b_{\rm h}(\mh) \lb \bar{N}_{\rm c}(\mh) 
	+ \bar{N}_{\rm s}(\mh) u_{\rm s}(k, \mh)  \rb
	\label{eq:beff-g}
\eeq
where $n(\mh)$ is the halo mass function, $b_{\rm h}(\mh)$ is the halo bias, $\bar{N}_{\rm c}$ and $\bar{N}_{\rm s}$ are the mean central and satellite occupation numbers, $u_{\rm s}$ is the Fourier transform of the assumed profile of satellite galaxies, and the mean galaxy number density $\bar{n}_{\rm g}$ is
\beq
\bar{n}_{\rm g} = \int d\mh\, n(\mh) \lb \bar{N}_{\rm c}(\mh) + \bar{N}_{\rm s}(\mh) \rb\ .
\label{eq:ng-halomodel}
\eeq

The one-halo term is 
\beq
\Pgg^{\rm 1h}(k) = \frac{1}{\bar{n}_{\rm g}^2}
	\int d\mh\, n(\mh) 
	\lb 2 \left\la N_{\rm c}(\mh) N_{\rm s}(\mh) \right\ra u_{\rm s}(k, \mh)
	+ \left\la N_{\rm s}(\mh) [N_{\rm s}(\mh) - 1] \right\ra 
	u_{\rm s}(k, \mh)^2  \rb\ .
	\label{eq:Pgg-1h}
\eeq
The expectation values of halo occupation numbers depend on what is assumed about correlations between numbers of central and satellite galaxies within a given halo. We use the ``minimally correlated" assumption from Ref.~\cite{Smith:2018bpn}, in which the central and satellite occupation numbers are independent Poisson random variables; this yields
\beq
\left\la N_{\rm c}(\mh) N_{\rm s}(\mh) \right\ra = \bar{N}_{\rm c}(\mh) \bar{N}_{\rm s}(\mh)\ ,
\qquad
\left\la N_{\rm s}(\mh) [N_{\rm s}(\mh) - 1] \right\ra = \bar{N}_{\rm s}(\mh)^2\ .
\eeq
In specific HOD models where a central-satellite correlation is assumed, we implement this correlation by including $\bar{N}_{\rm c}(\mh)$ as a prefactor in the model for $\bar{N}_{\rm s}(\mh)$, ensuring that $\bar{N}_{\rm s}=0$ if $\bar{N}_{\rm c}=0$ at a given halo mass.

Finally, the shot noise term is given by
\beq
\Pgg^{\rm shot} = \frac{1}{\bar{n}_{\rm g}}\ .
\eeq

\subsection{Electron power spectra}
\label{sec:halomodel-e}

In this paper, as in Ref.~\cite{Smith:2018bpn}, we assume that free electrons trace the gas within halos, such that we can describe electron distributions using the standard halo model for matter, but with the matter density profile replaced with the gas profile. The electron density power spectrum is then given by
\beq
\Pee(k) = \Pee^{\rm 2h}(k) + \Pee^{\rm 1h}(k)\ ,
\eeq
with
\beq
\label{eq:Pee-2h}
\Pee^{\rm 2h}(k) 
	= 
	\lb  \int d\mh\, n(\mh) b_{\rm h}(\mh) \frac{\mh}{\bar{\rho}_{\rm m}} u_{\rm gas}(k, \mh)  \rb^2
	P_{\rm lin}(k)
\eeq
and
\beq
\Pee^{\rm 1h}(k) = 
	\int d\mh\, n(\mh) 
	\lp \frac{\mh}{\bar{\rho}_{\rm m}} \rp^2 |u_{\rm gas}(k, \mh)|^2 \ .
\label{eq:Pee-1h}
\eeq
Similarly, the galaxy-electron cross power spectrum is
\beq
\Pge(k, \mu) = \Pge^{\rm 2h}(k, \mu) + \Pge^{\rm 1h}(k)\ ,
\eeq
with 
\beq
\label{eq:Pge-2h}
\Pge^{\rm 2h}(k, \mu) 
	=  \lb b_{\rm g}(k) + f \mu^2 \rb
	\lb  \int d\mh\, n(\mh) b_{\rm h}(\mh) \frac{\mh}{\bar{\rho}_{\rm m}} u_{\rm gas}(k, \mh)  \rb
	P_{\rm lin}(k)
\eeq
and
\beq
\Pge^{\rm 1h}(k) = 
	\int d\mh\, n(\mh) 
	\frac{\mh}{\bar{\rho}_{\rm m}} u_{\rm gas}(k, \mh) 
	\frac{\bar{N}_{\rm c}(\mh) + \bar{N}_{\rm s}(\mh) u_{\rm s}(k, \mh)}
		{\bar{n}_{\rm g}}\ .
	\label{eq:Pge-1h}
\eeq
Note that these quantities refer to the correlation between the observed galaxy density (which is affected by redshift space distortions) and the ``true" electron density (which has no RSD contribution).

\subsection{Integrand of kSZ angular power spectrum}
\label{sec:clksz-integrand}

It's helpful to rewrite $C_\ell^{\rm kSZ}$ as an integral over redshift and halo mass, so that we can compare the relative contributions from different ranges of these two quantities (see Fig.~\ref{fig:cl_ksz_integrand}).
Combining Eqs.~\eqref{eq:clksz} and~\eqref{eq:Pqrqr}, and changing variables from $\chi$ to $z$, we can write
\beq
C_\ell^{\rm kSZ} = \int_0^{z_*} dz \frac{d\chi}{dz} \frac{1}{\chi[z]^2} \tilde{K}(z)^2 
	\frac{1}{6\pi^2} \lb \int d\kL\, \kL^2 \Pvv(\kL; z) \rb \Pee\!\lp \frac{\ell+1/2}{\chi[z]}; z \rp\ .
\eeq
We can further rewrite $\Pee$ as
\beq
\Pee(k; z) = \int d\log\mh\, \mh \lp \frac{d\Pee^{\rm 2h}(k;z)}{d\mh} + \frac{d\Pee^{\rm 1h}(k;z)}{d\mh} \rp
\eeq
with
\begin{align}
\label{eq:pee2h-igr}
\frac{d\Pee^{\rm 2h}(k;z)}{d\mh} 
	&= 2 n(\mh) b_{\rm h}(\mh) \frac{\mh}{\bar{\rho}_{\rm m}} u_{\rm gas}(k, \mh)
	\lb \int_{m_{\rm h,min}}^{\mh} d\mh'\, n(\mh') b_{\rm h}(\mh') 
	\frac{\mh'}{\bar{\rho}_{\rm m}} u_{\rm gas}(k, \mh')  \rb
	P_{\rm lin}(k)\ , \\
\label{eq:pee1h-igr}
\frac{d\Pee^{\rm 1h}(k;z)}{d\mh} 
	&= n(\mh) \lp \frac{\mh}{\bar{\rho}_{\rm m}} \rp^2 |u_{\rm gas}(k, \mh)|^2\ .
\end{align}
Thus, we can finally write
\beq
C_\ell^{\rm kSZ} = \int_0^{z_*} dz \int d\log \mh \frac{d^2C_\ell^{\rm kSZ}}{dz\, d\log \mh}
\eeq
with
\beq
\frac{d^2C_\ell^{\rm kSZ}}{dz\, d\log \mh}
	= \frac{d\chi}{dz} \frac{1}{\chi[z]^2} \tilde{K}(z)^2 
	\frac{1}{6\pi^2} \lb \int d\kL\, \kL^2 \Pvv(\kL; z) \rb
	\mh \lp \frac{d\Pee^{\rm 2h}(k;z)}{d\mh} + \frac{d\Pee^{\rm 1h}(k;z)}{d\mh} \rp\ ,
\eeq
and the terms in parentheses given by Eqs.~\eqref{eq:pee2h-igr}-\eqref{eq:pee1h-igr}.

\subsection{Galaxy bispectrum}
\label{sec:halomodel-ggg}

To compute the shot noise contribution to the power spectrum of a kSZ template, we also need to compute the galaxy bispectrum (see Appendix~\ref{app:shotnoise}). This is a function of three 3d wavevectors $(\vk_1, \vk_2, \vk_3)$ constrained such that $\vk_1+\vk_2+\vk_3=0$; in the absence of redshift space distortions (which we ignore for simplicity), the resulting triangle can be specified by 3 numbers, which we choose to be the magnitudes of each wavevector. The halo model description of the galaxy bispectrum can then be written as (e.g.~\cite{Yamamoto:2016anp})
\beq
B_{\rm ggg}(k_1, k_2, k_3) 
	= B_{\rm ggg}^{\rm 3h}(k_1, k_2, k_3)  
	+ B_{\rm ggg}^{\rm 2h}(k_1, k_2, k_3) 
	+ B_{\rm ggg}^{\rm 1h}(k_1, k_2, k_3) \ .
\eeq
There is also a shot noise contribution, but we do not require this for our calculations in Appendix~\ref{app:shotnoise}, so we have not written it above.

The three-halo term is
\beq
B_{\rm ggg}^{\rm 3h}(k_1, k_2, k_3) 
	= b_{\rm eff}(k_1) b_{\rm eff}(k_2) b_{\rm eff}(k_3)
	B_{\rm tree}(k_1, k_2, k_3)\ ,
\eeq
where the matter bispectrum at leading order (``tree-level") in perturbation theory is
\beq
B_{\rm tree}(\vk_1, \vk_2, \vk_3) = 2F_2(\vk_1, \vk_2) P_{\rm lin}(k_1) P_{\rm lin}(k_2) + \text{2 perms}
\eeq
and the symmetrized $F_2$ kernel is
\beq
F_2(\vk_1, \vk_2) = 
	\frac{5}{7} 
	+ \frac{1}{2} \hat{\vk}_1 \cdot \hat{\vk}_2 \lp \frac{k_1}{k_2} + \frac{k_2}{k_1} \rp
	+ \frac{2}{7} \lp \hat{\vk}_1 \cdot \hat{\vk}_2 \rp^2\ .
\eeq
The two-halo term is
\begin{align*}
B_{\rm ggg}^{\rm 2h}(k_1,k_2,k_3) 
	&= 
	\Bigg[ \frac{1}{\bar{n}_{\rm g}^2} \int d\mh \, n(\mh) b_{\rm h}(\mh) \bigg\{ 
	\left\la N_{\rm c}(\mh) N_{\rm s}(\mh) \right\ra \lp u_{\rm s}(k_1, \mh) + u_{\rm s}(k_2, \mh) \rp   \\
&\qquad\qquad\qquad\quad
	+ \left\la N_{\rm s}(\mh) \lb N_{\rm s}(\mh) - 1 \rb \right\ra u_{\rm s}(k_1, \mh) u_{\rm s}(k_2, \mh) \bigg\}
	\Bigg] b_{\rm eff}(k_3) P_{\rm lin}(k_3) + \text{2 perms}\ ,
	\numberthis
	\label{eq:Bggg2h}
\end{align*}
while the one-halo term is
\begin{align*}
B_{\rm ggg}^{\rm 1h}(k_1,k_2,k_3) 
	&= \frac{1}{\bar{n}_{\rm g}^3}
	\int d\mh\, n(\mh) \bigg\{ 
	\left\la N_{\rm c}(\mh) N_{\rm s}(\mh) (N_{\rm s}(\mh)-1) \right\ra 
	\lp u_{\rm s}(k_1, \mh) u_{\rm s}(k_2, \mh) + \text{2 perms} \rp \\
&\qquad\qquad\qquad\qquad
	+\, \left\la N_{\rm s}(\mh) (N_{\rm s}(\mh)-1) (N_{\rm s}(\mh)-2) \right\ra
	u_{\rm s}(k_1, \mh) u_{\rm s}(k_2, \mh) u_{\rm s}(k_3, \mh) \bigg\}\ .
	\numberthis
	\label{eq:Bggg1h}
\end{align*}
For independent Poisson-distributed numbers of centrals and satellites, the expectation values in Eq.~\eqref{eq:Bggg1h} are given by
\beq
\left\la N_{\rm c}(\mh) N_{\rm s}(\mh) (N_{\rm s}(\mh)-1) \right\ra 
	= \bar{N}_{\rm c}(\mh) \bar{N}_{\rm s}(\mh)^2\ ,
\qquad
\left\la N_{\rm s}(\mh) (N_{\rm s}(\mh)-1) (N_{\rm s}(\mh)-2) \right\ra 
	= \bar{N}_{\rm s}(\mh)^3\ .
\eeq

%--------------------------------------------------------------------------------------
% SECTION: Halo occupation modelling
%--------------------------------------------------------------------------------------
\section{Halo occupation modelling}
\label{app:hod}

In this section, we record the details of the HOD models used for the survey-specific forecasts in Sec.~\ref{sec:forecasts}. Note that we write parameters with dimension of mass either in $h^{-1}M_\odot$ or $M_\odot$ units, depending on what was used in the original reference, but we translate all parameters to $M_\odot$ units in our numerical computations in this work.

%----------------------
% BOSS
%----------------------
\subsection{BOSS}

For BOSS, we use the following 5-parameter HOD from Ref.~\cite{Walsh:2019luq}:
\begin{align}
\bar{N}_{\rm c}(\mh) &= \frac{1}{2} \lb 1 + 
	{\rm erf}\!\lp \frac{\log \mh - \log M_{\rm min}}{\sigma_{\log M}} \rp\rb\ , \\
\bar{N}_{\rm s}(\mh) &= \bar{N}_{\rm c}(\mh) \lp \frac{M}{M_1} \rp^\alpha e^{-M_{\rm cut}/\mh}\ ,
\end{align}
with the best-fit parameter values from the joint fit to the projected correlation function ($w_p$) and void probability function ($P_0$):
\beq
M_{\rm min} = 10^{13.18}\, h^{-1} M_\odot\ , \quad
\sigma_{\log M} = 0.55\ , \quad
M_1 = 10^{14.28}\, h^{-1} M_\odot\ , \quad
\alpha = 1.12\ , \quad
M_{\rm cut} = 10^{4.87}\, h^{-1} M_\odot\ .
\eeq

%----------------------
% DESI BGS
%----------------------
\subsection{DESI BGS}
\label{app:hod-desibgs}

For the DESI BGS sample, we use the HOD from Ref.~\cite{Smith:2017tzz} with the assumption that the scatter set by the $\sigma_{\log M}$ parameter is Gaussian:
\begin{align}
\bar{N}_{\rm c}(\mh) 
	&= \frac{1}{2} \lb 1 + {\rm erf}\!\lp \frac{\log \mh - R \log M_{\rm min}}{\sigma_{\log M}} \rp\rb\ , \\
\bar{N}_{\rm s}(\mh) 
	&= \bar{N}_{\rm c}(\mh)  \lp \frac{\mh - M_0}{(M_1')^R} \rp^\alpha\ ,
\end{align}
with fitted parameter values corresponding to absolute $r$-band magnitude ${}^{0.1} M_r - 5\log h = -20.5$:
\beq
M_{\rm min} = 10^{12.2}\, h^{-1} M_\odot\ , \quad
\sigma_{\log M} = 0.15\ , \quad
M_1' = 10^{13.5}\, h^{-1} M_\odot\ , \quad
\alpha = 1.05\ , \quad
M_0 = 10^{12}\, h^{-1} M_\odot\ .
\eeq
We fix $R$ by demanding that the $\bar{n}_{\rm g}$ expression in Eq.~\eqref{eq:ng-halomodel} evaluates to the value computed from Table 2.5 of Ref.~\cite{DESI:2016fyo} over $0.05\leq z \leq 0.45$. The resulting $R$ values are well fit by $R(z) = 0.88 + 0.23z + 0.76z^2$.

%----------------------
% DESI LRG
%----------------------
\subsection{DESI LRG}

For the DESI LRG sample, we use the HOD from Ref.~\cite{Yuan:2022rsc}:
\begin{align}
\bar{N}_{\rm c}(\mh) 
	&= \frac{f_{\rm ic}}{2} {\rm erfc}\!\lp \frac{R \log M_{\rm cut} - \log \mh}{\sqrt{2}\sigma_{\log M}} \rp\ , \\
\bar{N}_{\rm s}(\mh) 
	&= \bar{N}_{\rm c}(\mh)  \lp \frac{\mh - \kappa M_{\rm cut}}{M_1^R} \rp^\alpha\ ,
\end{align}
with
\beq
M_{\rm cut} = 10^{12.7}\, h^{-1} M_\odot\ , \quad
\sigma_{\log M} = 0.2\ , \quad
M_1 = 10^{13.6}\, h^{-1} M_\odot\ , \quad
\alpha = 1.15\ , \quad
\kappa = 0.08\ , \quad
f_{\rm ic} = 0.8\ .
\eeq
We fix $R$ such that $\bar{n}_{\rm g}$ evaluates to the target density of $5\times 10^{-4}\,{\rm Mpc}^{-3}$ quoted in Ref.~\cite{Yuan:2022rsc} over $0.6\leq z \leq 1.1$. The resulting $R$ values are well fit by $R(z)=1.05 - 0.033z$.

%----------------------
% DESI ELG
%----------------------
\subsection{DESI ELG}

For the DESI ELG sample, we use the HOD from Ref.~\cite{Yuan:2022rsc}:
\begin{align}
\bar{N}_{\rm c}(\mh) &= R \left\{ 2A \phi(\log \mh) \Phi(\log \mh)
	+ \frac{1}{2Q} \lb 1 + {\rm erf}\!\lp \frac{\log \mh - \log M_{\rm cut}}{0.01} \rp \rb \right\} \ , \\
\bar{N}_{\rm s}(\mh) 
	&= R \lp \frac{\mh - \kappa M_{\rm cut}}{M_1} \rp^\alpha\ ,
\end{align}
where
\begin{align}
\phi(\log \mh) &= \frac{1}{\sqrt{2\pi \sigma_{\rm M}^2} }
	\exp\!\lb -\frac{(\log \mh - \log M_{\rm cut})^2}{2\sigma_{\rm M}^2} \rb\ , \\
\Phi(\log\mh) &= \frac{1}{2} \lb 1 + {\rm erf}\lp 
	\gamma \frac{\log\mh - \log M_{\rm cut}}{\sqrt{2} \sigma_{\rm M}} 
	\rp \rb\ , \\
A &= p_{\rm max} - \frac{1}{Q}
\end{align}
and
\beq
p_{\rm max} = 0.075\ , \;
Q = 95\ , \;
M_{\rm cut} = 10^{11.9}\, h^{-1} M_\odot\ , \;
\sigma_{M} = 0.5\ , \;
\gamma = 5\ , \;
M_1 = 10^{14.2}\, h^{-1} M_\odot\ , \;
\alpha = 0.65\ , \;
\kappa = 1.35\ .
\eeq
We use $R=1$ for our baseline modelling, but in cases where we wish to ensure that the predicted number density matches a desired input, we fix $R$ appropriately at each redshift.

%----------------------
% DESI QSO
%----------------------
\subsection{DESI QSO}

For the DESI QSO sample, we use the error-function--based HOD from Ref.~\cite{Alam:2019pwr}:
\begin{align}
\bar{N}_{\rm c}(\mh) 
	&= \frac{1}{2} p_{\rm max} {\rm erfc}\!\lp \frac{R \log M_{\rm c} - \log \mh}{\sqrt{2} \log(e) \sigma_{M}} \rp\ , \\
\bar{N}_{\rm s}(\mh) 
	&= \lp \frac{\mh - \kappa M_{\rm c}}{(M_1')^R} \rp^\alpha\ ,
\end{align}
with
\beq
M_{\rm c} = 10^{12.21}\, h^{-1} M_\odot\ , \quad
\sigma_{M} = 0.6\ , \quad
M_1 = 10^{14.09}\, h^{-1} M_\odot\ , \quad
\kappa = 1.0\ , \quad
\alpha = 0.39\ , \quad
p_{\rm max} = 0.033\ .
\eeq
We fix $R$ such that $\bar{n}_{\rm g}$ evaluates to the QSO density given in Ref.~\cite{DESI:2016fyo} over $0.65\leq z \leq 1.85$. The resulting $R$ values are well fit by $R(z)=1.16 - 0.040z$.

%----------------------
% Roman
%----------------------
\subsection{Roman}

For the H$\alpha$ sample of ELGs from the Roman High Latitude Spectroscopic Survey, we fit the following parameterized model to measurements from the mock catalog in Ref.~\cite{Zhai:2021pja}:
\begin{align}
\nonumber
\bar{N}_{\rm c}(\mh) 
	&= R \left\{ A_1 \exp\!\lb -\frac{(\log\mh - \log M_{\rm c})^2}{2\sigma_{\log M}^2} \rb
	+ A_2 \lb 1 + {\rm erf}\!\lp \frac{\log\mh - \log M_{\rm c}}{\sigma_{\log M}} \rp \rb \right\} \ , \\
\bar{N}_{\rm s}(\mh) &=  \left\{
	\begin{array}{ll}
	RA_2 \lp \frac{\mh}{M_1} \rp^{\alpha_1}\ , & \mh < M_1\ , \\
	RA_2 \lp  \frac{\mh}{M_1} \rp^{\alpha_2}\ , & \mh \geq M_1\ .
	\end{array}
	\right.
\end{align}
The central occupation is composed of a peak with amplitude $A_1$ at $\mh=M_{\rm c}$ and a smooth transition to a plateau with amplitude $2A_2$ at higher masses, while the satellite occupation is a broken power law with tilt $\alpha_1$ for $\mh < M_1$ and $\alpha_2$ for $\mh\geq M_1$. Note that the measured HODs from Ref.~\cite{Zhai:2021pja} typically have a double-peaked structure as a function of halo mass, but the higher-mass peak is typically weaker and its physical origin is somewhat unclear, so our parameterization only incorporates the lower-mass peak.  We fix the free parameters to approximately match the measurements shown in Figure 8 of Ref.~\cite{Zhai:2021pja} for $1.0<z<1.1$, $A_V=1.65$, $f_{\rm lin,1}=10^{-16}\,{\rm erg}\,{\rm s}^{-1}\,{\rm cm}^{-2}$, and $f_{\rm lin,2R}=0.5$, resulting in 
\beq
A_1=0.2\ , \quad
A_2=0.01\ , \quad
 M_{\rm c} = 11.9 M_\odot\ , \quad
  \sigma_{\log M}=0.2\ , \quad
M_1=12.3 M_\odot\ , \quad
\alpha_1=2.5\ , \quad
\alpha_2=0.5\ .
\eeq
We fix $R$ by such that the predicted galaxy number density matches that from Ref.~\cite{Wang:2021oec} corresponding to fluxes $>10^{16}\,{\rm erg}\,{\rm s}^{-1} {\rm cm}^{-2}$ and dust attenuation parameter $A_V=1.65$. The resulting $R$ values are well fit by $R(z) = 5.41 -2.77 z$.

%----------------------
% MegaMapper
%----------------------
\subsection{MegaMapper}

For the MegaMapper LBG sample, we use the ``linear HOD model" from Ref.~\cite{Harikane:2017lcw}:
\begin{align}
\bar{N}_{\rm c}(\mh) 
	&= \frac{1}{2} \lb 1 + {\rm erf}\!\lp \frac{\log \mh - \log M_{\rm min}}{\sqrt{2} \sigma_{\log M}} \rp\rb\ , \\
\bar{N}_{\rm s}(\mh) 
	&= \bar{N}_{\rm c}(\mh)  \lp \frac{\mh - M_{\rm cut}}{M_{\rm sat}} \rp^\alpha\ .
\end{align}
We use the best-fit parameters corresponding to $z\approx 3.8$ with $m_{\rm UV}^{\rm th}=24.5$:
\beq
M_{\rm min} = M_{\rm cut} =  10^{12.22} M_\odot\ , \quad
\sigma_{\log M} = 0.2\ , \quad
M_{\rm sat} = 10^{14.23} M_\odot\ , \quad
\alpha = 1.0\ .
\eeq
Note that we have chosen to set $M_{\rm cut}$ equal to $M_{\rm min}$, which differs from the choice made in Ref.~\cite{Harikane:2017lcw}, but the precise value of $M_{\rm cut}$ has a minimal impact on our results.

%--------------------------------------------------------------------------------------
% SECTION: Modelling for 21cm intensity mapping
%--------------------------------------------------------------------------------------
\section{Modelling for \tcm intensity mapping}
\label{app:21cm}

In this appendix, we provide the details of our approach to modelling \tcm observations of large-scale structure.

%----------------------
% Halo model
%----------------------
\subsection{Halo model}

\subsubsection{Formalism}

The halo model framework we use for modelling the distribution of HI generally follows that in Appendix~\ref{app:halomodel}, with a few differences adapted from the formalism in Ref.~\cite{Schaan:2021gzb}. HI halo models are often written as predictions for the statistics of the observed brightness temperature, but for consistency with our other forecasts, we write predictions for the HI overdensity instead of the brightness temperature fluctuations. In this approach, the halo occupation functions $\bar{N}_{\rm c}(\mh)$ and $\bar{N}_{\rm s}(\mh)$ are replaced by a HI mass-halo mass relation $m_{\rm HI}(\mh)$, and a HI-specific halo density profile $u_{\rm HI}$ is used, such that the linear bias from Eq.~\eqref{eq:beff-g} transforms into
\beq
b_{\rm HI}(k) = \frac{1}{\bar{\rho}_{\rm HI}}
	\int d\mh\, n(\mh) b_{\rm h}(\mh)  m_{\rm HI}(\mh)
	u_{\rm HI}(k, \mh) 
\eeq
and the one-halo term from Eq.~\eqref{eq:Pgg-1h} becomes
\beq
P_{\rm HI}^{\rm 1h}(k) = \frac{1}{\bar{\rho}_{\rm HI}^2}
	\int d\mh\, n(\mh) 
	m_{\rm HI}(\mh)^2
	u_{\rm HI}(k, \mh)^2 \ ,
\eeq
where
\beq
\bar{\rho}_{\rm HI} = \int d\mh\, n(\mh) m_{\rm HI}(\mh)\ .
\eeq
Note that the two-halo term in Eq.~\eqref{eq:Pgg-2h} retains the same form.

\subsubsection{HI mass-halo mass relation}

For the HI mass-halo mass relation $m_{\rm HI}(\mh, z)$, we consider two fitting functions from the literature. 
The first is from Ref.~\cite{Villaescusa-Navarro:2018vsg}, and has been fit to measurements of HI from the IllustrisTNG simulations:
\beq
m_{\rm HI}(\mh) = m_0  \lp \frac{\mh}{m_{\rm min}} \rp^\alpha \exp\!\lb - \lp \frac{m_{\rm min}}{\mh} \rp^{0.35} \rb\ ,
\eeq
with the best-fit ``FoF-SO" values for $\alpha$, $m_0$, and $m_{\rm min}$ listed for $z=0$ to $5$ in their Table 1. 

The second is from Ref.~\cite{Padmanabhan:2016fgy}, and has been fit to a variety of HI observations (resolved low-redshift galaxies, intensity-mapping determinations of the mean HI density, and higher-redshift damped Lyman-$\alpha$ absorbers):
\beq
m_{\rm HI}(\mh, z) = \alpha f_{{\rm H},c} \mh 
	 \lp \frac{\mh}{10^{11} h^{-1} M_\odot} \rp^\beta 
	 \exp\!\lb - \lp \frac{v_{c,0}}{v_{c}(\mh, z)} \rp^{3} \rb\ ,
\eeq
with $f_{{\rm H},c} = \Omega_{\rm b} (1 - Y_{\rm p}) / \Omega_{\rm m}$, halo virial velocity $v_c(\mh)$ given by
\beq
v_c(\mh, z) = \sqrt{\frac{G \mh}{r_{\rm vir}(\mh, z)}}\ ,
\eeq
and best-fit parameters
\beq
\alpha = 0.09\ , \quad 
\log\!\lp \frac{v_{c,0}}{{\rm km}\,{\rm s}^{-1}} \rp = 1.56\ , \quad 
\beta = -0.58\ .
\eeq

\subsubsection{HI density profile}

Likewise, we consider two forms of the HI density profile. The first, which we use for our main computations, is from Ref.~\cite{Villaescusa-Navarro:2018vsg}:
\beq
\rho_{\rm HI}(r, \mh, z) = \frac{\rho_0}{r^{\alpha_*}} \exp\!\lp -\frac{r_0}{r} \rp\ ,
\eeq
using the best-fit parameters for $z=0$ to $5$ and $m=10^9 h^{-1} M_\odot$ to $10^{15} h^{-1} M_\odot$ listed in their Table 2. We numerically evaluate a fast Fourier transform of this profile to obtain $u_{\rm HI}(k, \mh, z)$.

The second profile is from Ref.~\cite{Padmanabhan:2016fgy}:
\beq
\rho_{\rm HI}(r, \mh, z) = \rho_0(\mh, z) \exp\!\lp -\frac{r}{r_{\rm s}(\mh, z)} \rp\ ,
\eeq
with halo scale radius $r_{\rm s}(\mh, z) = r_{\rm vir}(\mh, z) / c_{\rm HI}(\mh, z)$ using an HI-specific concentration-mass relation:
\beq
c_{\rm HI}(\mh, z) = c_{\rm HI,0} \lp \frac{\mh}{10^{11} M_\odot} \rp^{-0.109}
	\frac{4}{(1+z)^\gamma}\ ,
\eeq
with
\beq
c_{\rm HI,0} = 28.65\ , \quad \gamma = 1.45\ .
\eeq
The normalization $\rho_0(\mh, z)$ is fixed so that the HI mass enclosed within the virial radius is equal to $m_{\rm HI}(\mh, z)$. In Fourier space, the normalized profile $u_{\rm HI}$ is then given by
\beq
u_{\rm HI}(k, \mh, z) = \frac{2}{[1 + k^2 r_{\rm s}(\mh, z)^2]^2}\ .
\eeq
In Appendix~\ref{app:21cm-assumptions}, we compare the results if either of these profiles is used.

%----------------------
% Instrumental noise
%----------------------
\subsection{Instrumental noise}

In \tcm intensity mapping, the dominant noise in measurements of the power spectrum is typically the instrumental noise associated with finite observing time and properties of the instrument, rather than the intrinsic shot noise of \tcm-emitting objects. Thus, we replace the shot power term in Eq.~\eqref{eq:Pggtot} with an ``effective" shot noise that includes both instrumental and Poisson contributions: 
\beq
P^{\rm shot, eff}_{\rm HI}(k_\perp, z) = \frac{P_{\rm N}(k_\perp, z)}{\bar{T}_{\rm HI}(z)^2} + \frac{1}{n_{\rm HI}(z)}\ .
\label{eq:21cmpshoteff}
\eeq
The instrumental noise $P_{\rm N}$ is associated with \tcm brightness temperature fluctuations, such that dividing it by $\bar{T}_{\rm HI}(z)^2$ converts into the noise power spectrum associated with the HI overdensity. We compute the noise power spectrum for PUMA, mean brightness temperature, and intrinsic shot noise as described in the appendices of Ref.~\cite{CosmicVisions21cm:2018rfq} and implemented in the {\sc PUMANoise} code\footnote{\url{https://github.com/slosar/PUMANoise}}. From Eq.~\ref{eq:21cmpshoteff}, we can define an effective number density $\bar{n}_{\rm eff}(k_\perp, z) \equiv P^{\rm shot, eff}_{\rm HI}(k_\perp, z)^{-1}$, which we plot in Fig.~\ref{fig:21cm_neff}.

%----------------------
% Foregrounds
%----------------------
\subsection{Foregrounds}
\label{app:21cm-foregrounds}

The need to remove bright foregrounds, dominated by Galactic synchrotron emission and extragalactic radio sources, from \tcm observations is expected to impose two restrictions on the modes that are available for cosmological analysis\footnote{
We note that observational and methodological work is underway to understand in more detail which modes will be eliminated by \tcm foreground cleaning and how to incorporate this into forecasts (e.g.~\cite{Cunnington:2020wdu}), but in this work we adopt assumptions which have been motivated by previous studies~\cite{Liu:2019awk}. We also note that reconstruction techniques are under development which could allow for the recovery of foreground-obscured modes (e.g.~\cite{Zhu:2016esh,Modi:2019hnu,Darwish:2020prn}), but such reconstructed modes will have different noise than directly-observed modes, so we only consider directly-observed modes in this work.
\\ \\ \\ \\ \\ \\ % Ugly hack required to make footnote single-column.
}:
\begin{enumerate}
\item Modes with $|k_\parallel| < \kparmin$ will be indistinguishable from smooth-spectrum foregrounds, and will therefore be filtered out. We use $\kparmin=0.03\,{\rm Mpc}$, following Ref.~\cite{Shaw:2014khi}, who found that an optimal foreground cleaning approach imposed $\kparmin \approx 0.02h^{-1}\,{\rm Mpc}$ when applied to simulations.
\item In interferometric observations, spectrally-smooth foreground power will leak beyond pure low-$\kpar$ modes into higher-$\kpar$ modes in a baseline-dependent way, creating a so-called ``foreground wedge" of contamination (e.g.~\cite{Morales:2012kf,Parsons:2012qh,Liu:2014bba}). This is described by by $|\kpar| < \beta(z) k_\perp$ where
\beq
\beta(z) \equiv \frac{\chi(z) H(z)}{c(1+z)} \sin \theta_{\rm w}(z)\ .
\eeq
The extent of the contamination is set by $\theta_{\rm w}(z)$, the maximum angle away from the receiver's phase center at which a spectrally smooth sky signal can contaminate higher-$\kpar$ modes. Following Ref.~\cite{CosmicVisions21cm:2018rfq}, we set this to 3 times the width of the PUMA primary beam, i.e.\ $\theta_{\rm w}(z) = 3 \times 1.2 \lambda(z) / D_{\rm eff}$, with $\lambda(z) \equiv 21(1+z)\,{\rm cm}$ and $D_{\rm eff}=5\,{\rm m}$. We explore our sensitivity to this choice, and the choice of $\kparmin$, in Appendix~\ref{app:21cm-assumptions}.
\end{enumerate}
These restrictions will apply separately to templates for velocity and electron density fields constructed from \tcm observations. However, the total kSZ template is constructed from the product of these templates (recall Eqs.~\ref{eq:eta-def}-\ref{eq:eps-def} and~\ref{eq:qhatr}), which becomes a convolution in Fourier space:
\beq
\hat{q}_r(\vkS, z) = \int_{\vkL} \eta(\vkL, z) \epsilon(\vkS-\vkL, z) \ .
\eeq
Thus, low-$\kpar$ modes of $\hat{q}_r(\vkS, z)$ are not obscured by foregrounds, since they can be sourced by pairs of $\eta$ and $\epsilon$ modes whose line-of-sight wavenumbers are individually much larger (i.e.\ it is possible that $|\kparS| < \kparmin$ while having $|\kparL|$ and $|\kparS-\kparL | $ both greater than $\kparmin$). We will omit redshift arguments in what follows.

To implement these foreground restrictions in computations, we start by recalling the cross power spectrum of $\hat{q}_r$ and $q_r$ from Eq.~\eqref{eq:Phqrqr1}, prior to taking the squeezed limit of $\Peeps$ (the discussion below also applies to the auto spectrum of $\hat{q}_r$):
\beq
P_{\hqr \qr}(\vkS) \approx \int \frac{d^3\vkL}{(2\pi)^3} P_{\velr \eta}(\vkL) \Peeps(\vkS-\vkL)\ .
\label{eq:Phqrqr-app1}
\eeq
We must set the integrand to zero {\em outside} of the region defined by the union of the following four conditions:
\beq
|\kparL| > \kparmin\ , \quad 
|\kparL-\kparS| > \kparmin\ , \quad
|\kparL| > \beta k_{{\rm L}\perp}\ , \quad 
|\kparS-\kparL| > \beta |\vk_{{\rm S}\perp} - \vk_{{\rm L}\perp}|\ .
\label{eq:fg-conditions}
\eeq
With $\kparL = \kL\muL$ and $k_{{\rm L}\perp} = \kL\sqrt{1-\muL^2}$ and analogously for $\vkS$, and also defining $\cos\phi_\perp \equiv \hat{\vk}_{{\rm S}\perp} \cdot \hat{\vk}_{{\rm L}\perp}$, the fourth condition can be written as $\cos\phi_\perp > \gamma(\kS, \muS; \kL, \muL)$, where
\beq
\gamma(\kS, \muS; \kL, \muL) \equiv 
	\frac{\beta^2 \lb (1-\muS^2) \kS^2 + (1-\muL^2) \kL^2 \rb - (\muS \kS - \muL \kL)^2}
	{2\beta^2 \sqrt{1-\muS^2} \sqrt{1-\muL^2} \kS \kL}\ .
\eeq
We can integrate this constraint out of Eq.~\eqref{eq:Phqrqr-app1} by approximating $\Peeps(\vkS-\vkL) \approx \Peeps(\vkS)$ and using
\beq
\mathcal{C}_\perp(\kS, \muS; \kL, \muL)
	\equiv \int_0^{2\pi} d\phi_\perp\, \Theta(\cos\phi_\perp - \gamma)
	= \left\{  \begin{array}{ll} 
		2\pi\ , & \gamma \leq -1\ , \\
		2\cos^{-1} \gamma \ , & -1 < \gamma < 1\ , \\
		0\ , & \gamma \geq 1\ ,
	\end{array} \right.
\eeq
where $\Theta(\cdots)$ is a step function. Meanwhile, the first three conditions in Eq.~\eqref{eq:fg-conditions} are equivalent to setting the integrand to zero if any of the following conditions {\em are} satisfied:
\beq
|\muL| < \frac{\kparmin}{\kL}\ , \quad 
|\kL \muL - \kS \muS| < \kparmin\ , \quad
|\muL| < \frac{\beta}{\sqrt{1+\beta^2}}\ , 
\label{eq:fg-3conditions}
\eeq
and we enforce these numerically when evaluating Eq.~\eqref{eq:Phqrqr-app1}. 

Finally, we compute $C_\ell^{\hqr \qr}$ in the Limber approximation, which requires taking the $\muS\to0$ limit of Eq.~\eqref{eq:Phqrqr-app1}:
\beq
P_{\hqr \qr}^{\rm 21cm}(\kS, 0)
	= \frac{1}{8\pi^3}
	 \int d\kL\, \kL^2 \int_{-1}^1 d\muL\, \muL^2 \frac{\Pgv(\kL, \muL)^2}{\Pggtot(\kL, \muL)} 
	\frac{\Pge(\kS, 0)^2}{\Pggtot(\kS, 0)}
	\mathcal{C}_\perp(\kS, 0; \kL, \muL) \,
	\mathcal{C}_\parallel(\kS, 0; \kL, \muL)\ ,
\eeq
where $\mathcal{C}_ \parallel(\kS, \muS; \kL, \muL)$ implements the conditions in Eq.~\eqref{eq:fg-3conditions}. Note that we take the $\muS\to0$ limit {\em after} accounting for foreground mode cuts as described above, and therefore we only consider modes of $\hqr$ that survive these cuts.

%----------------------
% 21cm Assumptions
%----------------------
\subsection{Dependence of forecasts on HI profile and foreground cuts}
\label{app:21cm-assumptions}

\begin{figure*}[t]
\includegraphics[width=0.9\textwidth, trim=0 15 0 0]{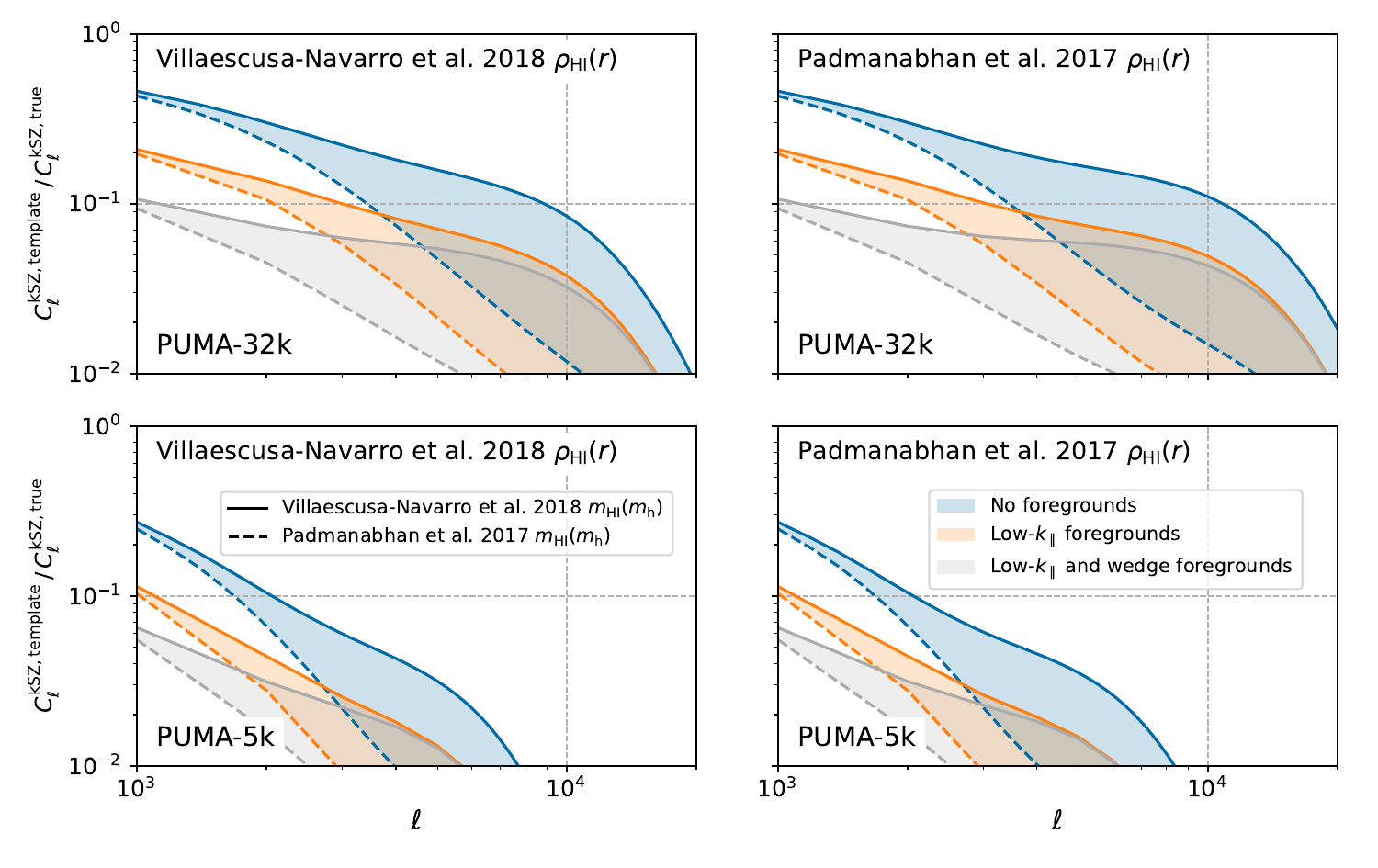}
\caption{%
Forecasts for kSZ templates constructed from \tcm intensity mapping measurements from PUMA, comparing our baseline forecasts that assume the simulation-based HI density profile from Ref.~\cite{Villaescusa-Navarro:2018vsg} ({\em left panels}) with forecasts that use the  the exponential profile from Ref.~\cite{Padmanabhan:2016fgy} ({\em right panels}). The choice of HI profile has negligible effect at lower multipoles, while it can affect the results by as much as 30\% (70\%) for PUMA-32k (PUMA-5k) at $\ell\sim 10000$.
}
\label{fig:21cm_templates_profile_comparision}
\end{figure*}

\begin{figure*}[t]
\includegraphics[width=\textwidth, trim=30 25 30 0]{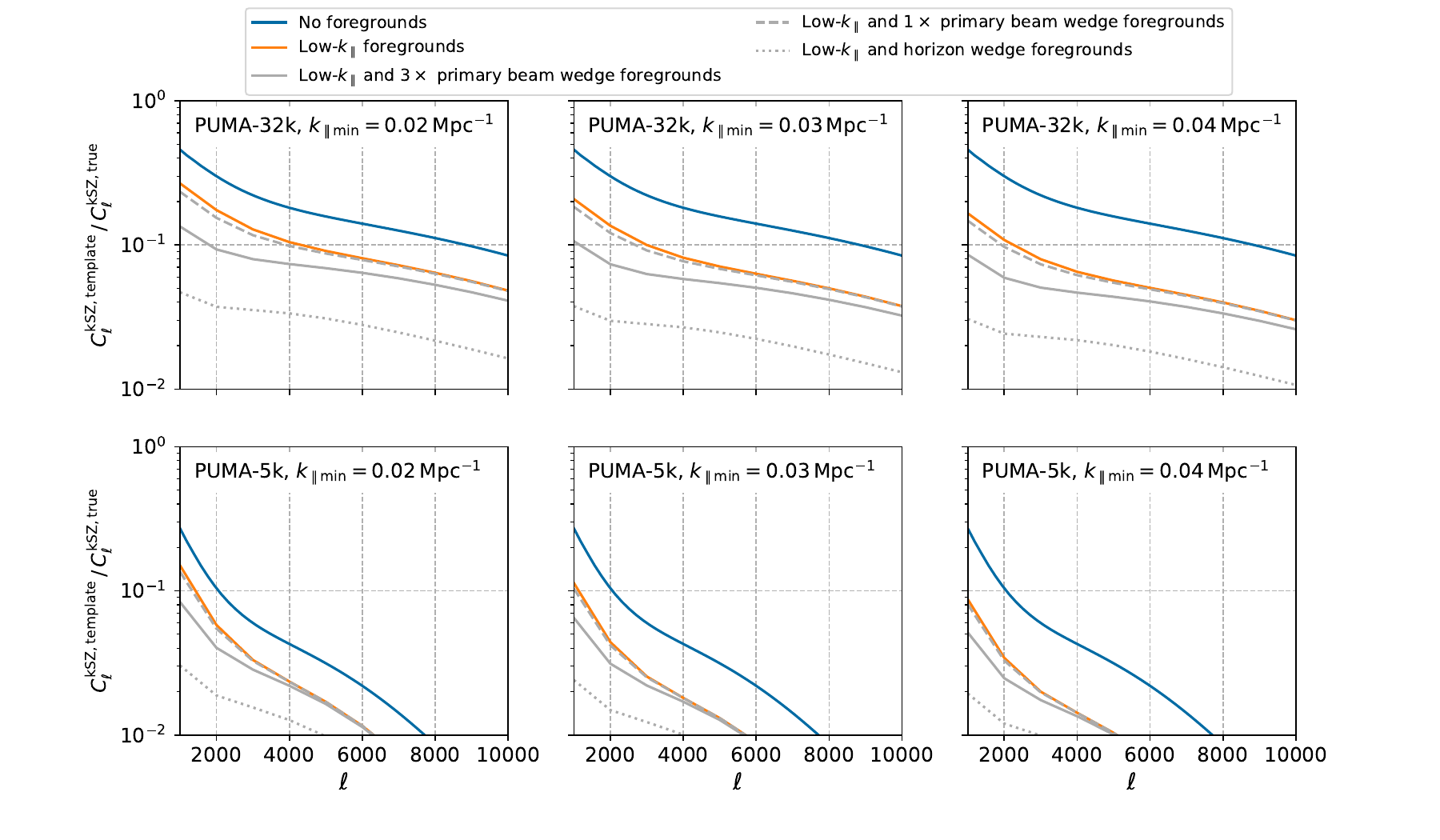}
\caption{%
Dependence of a \tcm-based kSZ template on assumptions about foreground cleaning. For simplicity, we only show results assuming the HI mass-halo mass relation and HI density profile from Ref.~\cite{Villaescusa-Navarro:2018vsg}. Different columns show different assumptions about the minimum $\kpar$ that can be cleaned of foregrounds, while different linestyles of the grey curves show different assumptions about the severity of the foreground wedge. We find a factor of 5 (at higher $\ell$) to 10 (at lower $\ell$) difference between the most optimistic and pessimistic cases that include the wedge.
}
\label{fig:21cm_templates_fg_comparision}
\end{figure*}

In Fig.~\ref{fig:21cm_templates_profile_comparision}, we compare kSZ template forecasts assuming the simulation-based HI profile from Ref.~\cite{Villaescusa-Navarro:2018vsg} with those assuming the exponential profile from Ref.~\cite{Padmanabhan:2016fgy}. We find that this choice does not qualitatively change the results: in particular, we see a negligible change at $\ell \lesssim 2000$ that becomes more important at larger multipoles, with as much as a 10\% (15\%) change in the results for PUMA-32k (PUMA-5k) at $\ell\sim 4000$, and a 30\% (70\%) change at $\ell\sim 10000$. This uncertainty in our forecasts is subdominant to that from the uncertain form of the $m_{\rm HI}(\mh)$ relation (this uncertainty is also lower at lower multipoles).

In Fig.~\ref{fig:21cm_templates_fg_comparision}, we explore how our forecasts depend on assumptions about \tcm foregrounds. Each column corresponds to a different choice of $\kparmin$, while the different linestyles for the grey curves denote different assumptions about the foreground wedge: an angular extent of 3 times the primary beam width of a single $6\,{\rm m}$ dish, 1 times the primary beam width (a more optimistic case), or the full sky visible above the horizon (a more pessimistic case). Between the most optimistic and pessimistic assumptions for the grey curves, we find roughly an order of magnitude variation at $\ell\sim 1000$ and a factor of 5 at $\ell\sim 10000$, indicating that the performance of a \tcm-based kSZ template will be very sensitive to the level of foreground cleaning.

%----------------------
% Details of CMB Forecasts
%----------------------
\section{CMB Forecasts}\label{sec:CMB_forecasts}

\begin{table}[t]
\begin{tabular}{|l|c|c|c|c|}
\hline
& \multicolumn{2}{c|}{Beam FWHM} & \multicolumn{2}{c|}{Noise RMS} \\
& \multicolumn{2}{c|}{} & \multicolumn{2}{c|}{($\mu$K-arcmin)} \\ 
\hline
& \multicolumn{1}{c|}{S4} & \multicolumn{1}{c|}{HD} &
\multicolumn{1}{c|}{S4} & \multicolumn{1}{c|}{HD} \\ \hline
39 GHz & $5.1'$ & $0.94'$ & 12.4 & 3.4 \\
93 GHz & $2.2'$ & $0.42'$ & 2.0 & 0.7 \\
145 GHz & $1.4'$ & $0.25'$ & 2.0 & 0.8 \\
225 GHz & $1.0'$ & $0.17'$ & 6.9 & 2.0 \\
280 GHz & $0.9'$ & $0.13'$ & 16.7 & 2.7 \\ \hline
\end{tabular}
\caption{Inputs to ILC noise for the CMB configurations. We have chosen the temperature noise RMS and the beam parameters to approximately match CMB-S4 and CMB-HD. We account for the effect of Earth's atmosphere by setting $\ell_{\rm knee}=100$ and $\alpha_{\rm knee}=-3$. We define the polarization noise as $\Delta_E=\Delta_B=\sqrt{2}\Delta_T$.}
\label{tab:cmb_specs}
\end{table}

We model the CMB noise including the pink and white noise components as 
\be
N_\ell=\Delta_T^2\exp\left[\ell(\ell+1)\frac{\theta_{\rm FWHM}^2}{8\log(2)}\right][1+(\ell_{\rm knee}/\ell)^{\alpha_{\rm knee}}]\label{eq:CMB_noise}
\ee
where we set the beam full-width half-maximum (FWHM), the temperature noise RMS $\Delta_T$, and the $\{\alpha_{\rm knee},\ell_{\rm knee}\}$ parameters which model the effect of the Earth's atmosphere to match upcoming experiments as given in Table~\ref{tab:cmb_specs}.

The contributions to the signal in the millimeter wavelength include clustered CIB, Poisson CIB and tSZ foregrounds, and the black-body late-time kSZ following~Ref.~\cite{Madhavacheril:2017onh} based on fits to data from Ref.~\cite{Dunkley:2013vu}. We do not include the correlation between tSZ and CIB. We also include radio sources in the 39, 93 and 145 GHz channels, using the flux-limit-dependent radio-source power model from Ref.~\cite{Lagache19}. For CMB-S4, we assume flux limits of 10, 7 and 10 mJy, respectively in those channels. For CMB-HD, we assume flux limits of 2, 1 and 1 mJy respectively. In addition, we include the lensed CMB black-body contribution which we calculate using CAMB~\cite{CAMB}, and the kSZ signal from  reionization following Ref.~\cite{Park:2013mv}. 

We calculate the total covariance between two frequency channels at each multipole $\ell$, ${\boldsymbol C}^{ij}_{\ell}$, taking into account the signal components correlated across frequencies, as well as the uncorrelated beam-deconvolved noise, indexed by $i$. The resulting minimum-variance standard ILC noise for the black-body signal (lensed CMB+kSZ) including all other contributions is then given by 
\be N_{\ell} = \Big[\sum_{ij}  \left({\boldsymbol C}^{-1}\right)^{ij}_{\ell}\Big]^{-1}\,.\ee

%----------------------
% Moving-lens equations
%----------------------
\section{Forecasts for moving-lens signal}
\label{app:movinglens}

Ref.~\citep{Hotinli:2020ntd} derives an optimal matched filter for measuring transverse velocities with the moving-lens effect, assuming a known form for the halo density profile. In Sec.~\ref{sec:movinglens}, we present a forecast that uses a modified form of this procedure to obtain binned information about the moving-lens profile (and therefore the halo density profile) itself. The impact of the moving-lens effect on the CMB has been written in Eqs.~\eqref{eq:ThetaML} to~\eqref{eq:rhos}.

In the form derived in Ref.~\citep{Hotinli:2020ntd}, an optimal filter $\tilde{\Psi}_i$ for the moving-lens signal along transverse direction $i$ is given by
\be
\tilde{\Psi}_i(\vec{\ell})={\sigma}_i\frac{\tilde{\mathcal{M}}_i(\vec{\ell})}{\tilde{C}^{TT}_\ell}\,, 
\label{eq:Psifilter}
\ee
where $i$ is either of two unit vectors, which we call $\hat{\vx}$ and $\hat{\vy}$, which span the $\vx_\perp$ plane.
We have written $\mathcal{M}_x=\hat{\vx}\cdot\boldsymbol{\mathcal{M}}$ and $\mathcal{M}_y=\hat{\vy}\cdot\boldsymbol{\mathcal{M}}$. The measurement noise $\sigma_i$ is given by
\be
\sigma_i = a_0^{-2}\!\!\int \frac{\dd^2\vell}{(2\pi)^2} {|\tilde{\Psi}_i(\vell)|^2}{\tilde{C}_\ell^{TT}}\, .
\label{eq:ML-Nrec}
\ee
In our binned forecasts, we use the same approach, but for a bin centered at $x_{\perp j}$, we suppose that we have applied a circular ring mask $\mathcal{R}_j(x_\perp)$ with outer (inner) radius $x_{\perp j}+\Delta x_\perp/2$ ($x_{\perp j}-\Delta x_\perp /2)$ around the halo center, such that $\tilde{\mathcal{M}}_i(\vell)$ in the numerator of Eq.~\eqref{eq:Psifilter} is the harmonic transform of $\mathcal{R}_j(x_\perp) \mathcal{M}_i(\vx_\perp)$ instead of $\mathcal{M}_i(\vx_\perp)$. We then compute the modified measurement noise using Eq.~\eqref{eq:ML-Nrec}, taking $\sigma \equiv \sigma_x = \sigma_y$. Our results are demonstrated in Fig.~\ref{fig:CMB_movinglens}.

%--------------------------------------------------------------------------------------
% BIBLIOGRAPHY
%--------------------------------------------------------------------------------------
\bibliography{references}

\end{document}

%% file: preamble.tex
\pdfoutput=1

\usepackage{setspace}
\usepackage{xspace}

\usepackage{graphicx}
\usepackage{epstopdf}
\usepackage{amssymb}
\usepackage{dcolumn}% Align table columns on decimal point

\usepackage{graphicx}
\usepackage{epstopdf}
\usepackage{amsmath}
\usepackage{amsfonts}
\usepackage{amssymb}
\usepackage[usenames,dvipsnames]{xcolor}
\usepackage{array}
\usepackage{nonfloat}
\usepackage[normalem]{ulem}

\usepackage{multirow}
\usepackage{psfrag}
\pdfoutput=1

\usepackage[colorlinks,bookmarks]{hyperref}
\definecolor{linkblue}{rgb}{0,0,0.8}
\definecolor{linkgreen}{rgb}{0,0.5,0}

\hypersetup{pdfpagemode=UseNone, pdfstartview=FitH, linkcolor=linkblue, 
	citecolor=linkgreen, urlcolor=linkblue}
           
\def\la{\langle}
\def\ra{\rangle}
\def\beq{\begin{equation}}
\def\eeq{\end{equation}}
\newcommand\numberthis{\addtocounter{equation}{1}\tag{\theequation}}
\newcommand{\lp}{\left(}
\newcommand{\rp}{\right)}
\newcommand{\lb}{\left[}
\newcommand{\rb}{\right]}

\newcommand{\tcm}{21$\,$cm\xspace}  % Properly spaced writing of 21 cm

\newcommand{\vk}{\boldsymbol{k}}

\newcommand{\vn}{\boldsymbol{n}}
\newcommand{\hvn}{\hat{\vn}}
\newcommand{\vecr}{\boldsymbol{r}}

\newcommand{\vx}{\boldsymbol{x}}
\newcommand{\vy}{\boldsymbol{y}}

\newcommand{\vv}{\boldsymbol{v}}
\newcommand{\vell}{\boldsymbol{\ell}}

\newcommand{\kpar}{k_\parallel}
\newcommand{\kparL}{k_{{\rm L}\parallel}}
\newcommand{\kparS}{k_{{\rm S}\parallel}}
\newcommand{\kperp}{k_\perp}
\newcommand{\kparmin}{k_{\parallel{\rm min}}}
\newcommand{\kL}{k_{\rm L}}
\newcommand{\kS}{k_{\rm S}}
\newcommand{\vkL}{\vk_{\rm L}}
\newcommand{\vkS}{\vk_{\rm S}}

\newcommand{\muS}{\mu_{\rm S}}
\newcommand{\muL}{\mu_{\rm L}}

\newcommand{\deltam}{\delta_{\rm m}}
\newcommand{\deltae}{\delta_{\rm e}}
\newcommand{\deltag}{\delta_{\rm g}}
\newcommand{\deltags}{\delta_{\rm g}^s}

\newcommand{\qr}{q_{\rm r}}
\newcommand{\hqr}{\hat{q}_{\rm r}}
\newcommand{\velr}{v_{\rm r}}
\newcommand{\TkSZ}{T_{\rm kSZ}}
\newcommand{\hTkSZ}{\hat{T}_{\rm kSZ}}
\newcommand{\hTdekSZ}{\hat{T}_{\text{de-kSZ}}}

\newcommand{\Pee}{P_{\rm ee}}
\newcommand{\Pgg}{P_{\rm gg}}
\newcommand{\Pggtot}{\Pgg^{\rm tot}}

\newcommand{\Pge}{P_{\rm ge}}
\newcommand{\Peeps}{P_{{\rm e}\epsilon}}
\newcommand{\Pgv}{P_{{\rm g}v}}
\newcommand{\Pvv}{P_{vv}}
\newcommand{\Rgv}{R_{{\rm g}v}}

\newcommand{\dirac}{\delta_{\rm D}}

\newcommand{\dd}{{\rm d}}

\newcommand{\be}{\begin{eqnarray}}

\newcommand{\ee}{\end{eqnarray}}

\newcommand{\invMpc}{\, {\rm Mpc}^{-1}}
\newcommand{\bg}{b_{\rm g}}

\newcommand{\mh}{m_{\rm h}}

\definecolor{purple}{rgb}{0.78,0.18,0.77}